\documentclass[aps,pra,amsmath,amssymb,bibnotes]{revtex4}

\usepackage{bbm}
\usepackage{amsmath}
\usepackage{mathtools}
\usepackage[english]{babel}
\usepackage[utf8]{inputenc}
\usepackage{hyperref}
\usepackage{textcomp} 
\usepackage{physics}
\usepackage{float}
\usepackage{dsfont}
\usepackage{enumitem}
\usepackage{MnSymbol}
\usepackage{comment}
\usepackage{color}
\usepackage{ulem}
\usepackage{natbib}

\newcommand{\dr}{\!\! \mathrm{d}^3 r}
\newcommand{\dx}{\!\! \mathrm{d}^2 r}

\newcommand{\el}{\Big \langle \! \Big \langle}
\newcommand{\er}{\Big \rangle \! \Big \rangle}
\newcommand{\ri}{\mathrm{i}}
\renewcommand{\Im}{\mathrm{Im}}
\renewcommand{\Re}{\mathrm{Re}}
\renewcommand{\Tr}{\mathrm{Tr}}

\newcommand{\re}{{\rm e}}

\begin{document}

\title{Generalized coupled dipole method for thermal far-field radiation}
	
\date{\today}
	
\author{Florian Herz and Svend-Age Biehs$^{*}$}
\affiliation{Institut f{\"u}r Physik, Carl von Ossietzky Universit{\"a}t, D-26111 Oldenburg, Germany}
\email{s.age.biehs@uni-oldenburg.de} 
	
\begin{abstract}
We introduce a many body theory for thermal far-field emission of dipolar dielectric and metallic nanoparticles in the vicinity of a substrate within the framework of fluctuational electrodynamics. Our theoretical model includes the possibility to define the temperatures of each nanoparticle, the substrate temperature, and the temperature of the background thermal radiation, separately. To demonstrate the versatility of our method, we apply it in an exemplary way by discussing the thermal radiation of four particle assemblies of SiC and Ag nanoparticles above a planar SiC and Ag substrate. Furthermore, we use discrete dipole approximation to determine the thermal emission of a spherical nanoparticle in free space and close to a substrate. Finally, the calculation of the thermal far-field radiation of a sharp Si tip close to a SiC substrate using the discrete dipole approximation including the near-field scattering by the tip as well as the thermal emission of the tip and the contribution of the substrate which is partially blocked by the tip serves as another example. 
\end{abstract}
\maketitle

\section{Introduction}

The theory of near- and far-field thermal radiation in many-body systems has greatly advanced in the last decade. This development arose from the generalization of Draine's work on dipolar many-body interactions~\cite{Draine1} by allowing for fluctuational and induced dipolar moments~\cite{pbaetal}. With this approach one could study thermally fluctuating fields generated by interacting dipolar objects and, thus, describe near-field as well as far-field thermal emission of an arbitrary number of dipolar objects each thermalized at its own heat bath with given temperature. Further works generalized this model rendering it possible to place dipolar objects in an environment at a given temperature~\cite{Messina} and to include magnetic dipolar contributions~\cite{Dong}. Therewith, thermal infrared radiation emitted by small metallic objects can be described more accurately since they are known to have a massive effective magnetic response due to the induction of eddy currents~\cite{wirbelM,Dedkov2,Chapuis,gold3}. It should be mentioned that also more general models were introduced in order to provide a theoretical basis for many-body systems with objects of arbitrary size and shape~\cite{KruegerEtAlTraceFormulas2012,scatt,scatt2,ZhuEtAl2018} as well as numerical methods~\cite{RodriguezEtAl2013,Pol2015,JinWEtAl2016}. 

Here, we will focus on the dipole model only, which is the theoretical workhorse of many recent works, because of its conceptional and
numerical simplicity. For example, using this model it could be shown that the heat transfer between two dipolar objects can be enhanced 
by the presence of a substrate or more general third objects supporting surface waves like phonon polaritonic media~\cite{SaaskilahtiEtAl2014,DongEtAl2018,gold1,AsheichykEtAl2017,Asheichyk2018}, graphene~\cite{gold1}, Moir\'{e} bilayer graphene~\cite{YangEtAl2021}, or hyperbolic media~\cite{ZhangEtAl2019b, OttEtAl2021}. For non-reciprocal media even more interesting many-body effects were demonstrated like persistent heat currents and fluxes~\cite{Linxiao2016,Silveirinha2017,GueEtAl2019,OttEtAl2019}, giant magnetic resistances~\cite{LatellaPBA2018,VillaEtAL2015}, Hall and anomalous Hall effect~\cite{PBA2017,OttEtAlanom2020}, circular polarized emission and thermal angular momentum and spin~\cite{Silveirinha2017,OttEtAl2018,Khandekar2019b} as well as non-reciprocal near-field diodes~\cite{Konzeptdiode10,OttSAB2020} or spin-related directional thermal emission~\cite{DongEtAl2021}. Of course, also radiative heat transfer between different many-body systems has been studied theoretically for anisotropic systems~\cite{Nikbakht2014a,IncardoneEtAl2014,Nikbakht2015,Nikbakht2017,Luo}, fractal structures~\cite{Nikbakht2017,Luo}, and lattices~\cite{PhanEtAl2013,LuoEtAl2020} as well as thermal transport in 1D, 2D, and 3D nanoparticle systems~\cite{PBAEtAlSuperdiff2013,KathmannEtAl2018,TervoEtAl2019,tervo} including topological Su-Schriefer-Heeger structures~\cite{OttSAB2020b}. Predominately, such studies are of theoretical interest, but they can also be relevant for radiative heat transport measurements in particle glasses~\cite{Retsch2021} and other systems. Reviews of most of the advances in the last ten years can be found in Refs.~\cite{SABRMB2021,SongPlagiat2021,LatellaEtAl2021}.

In this work, we provide a generalized many-body theory for the thermal far-field emission of $N$ dipolar objects in a given environment. Based on the coupled dipole method of Ben-Abdallah et al.~\cite{pbaetal}, first Edalatpour and Francoeur~\cite{EdalatpourDDA} and then Ekeroth et al.~\cite{Ekeroth} provided a model for near-field heat exchange and, subsequently, also for thermal emission in the context of discrete dipole approximation (DDA) of a macroscopic object divided into many small subvolumes, so called voxels, which act as fluctuating dipoles emitting thermal radiation. Here, we want to extend that description by including magnetic dipole moments to be able to describe metallic nanoparticles. Furthermore, we include an environment consisting of thermal radiation at a given temperature in which $N$ dipolar objects and a material part at another temperature, for instance a substrate, are embedded. Therefore, our model can be regarded as a $N$-body extension of the dipole model in Refs.~\cite{Joulain2,Jarzembski,Herz,Herz3}. Our model can be used to study thermal emission of any $N$-body assembly close to a substrate, for instance, but it can also be employed for a discrete dipole approximation of a single or several macroscopic objects close to a substrate or environment as depicted in Fig.~\ref{Fig:SketchDDA}. To demonstrate the versatility of our model, we will study the thermal emission of different combinations of four nanoparticles close to a planar substrate, the thermal emission of a macroscopic spherical particle close to a substrate using DDA, and the thermal emission and scattering of near-field thermal radiation by a sharp tip, again, using DDA. This allows us also to check the validity of our results against those of Refs.~\cite{Dong,Ekeroth,Edalatpour2}. Finally, we will also discuss the impact of different choices of the dressed polarizability to some extent. 

Our paper is organized as follows: In Sec~II. we introduce the theoretical framework and derive the general $N$-body expressions. This includes also the introduction of different definitions of the dressed polarizability. In Sec.~III we provide several numerical results including those for the thermal emission of a single nanoparticle, of a combination of four nanoparticles with different temperatures, a DDA calculation for a spherical particle, and a DDA calculation for a sharp tip. Finally, in Sec.~IV we summarize our findings.

\section{Theoretical framework}

\begin{figure}
	\centering
	\includegraphics[width=0.45\textwidth]{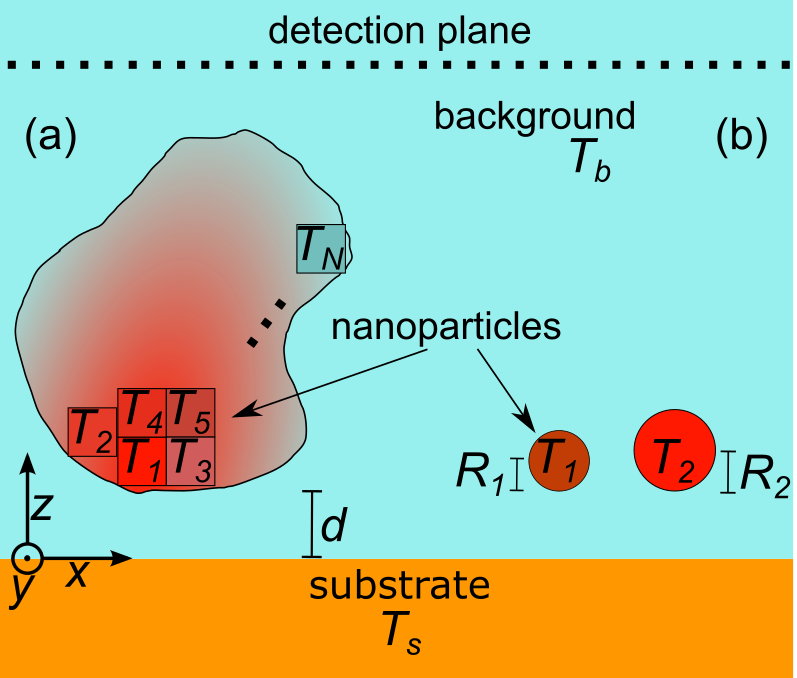}
	\caption{Sketch of a DDA of a macroscopic object by $N$ voxels (a) and of two spherical nanoparticles differing in size and material (b) each emitting heat radiation into the far-field. The substrate, the different nanoparticles, and the background radiation have, in general, different temperatures. The radiation is gathered at the detection plane lying parallel to the substrate. The assemblies are at edge-to-edge-distance $d$ to the substrate.}
	\label{Fig:SketchDDA}
\end{figure}

In this chapter we introduce the general theoretical expressions for the emitted power of a heated system that consists of $N$ nanoparticles treated as point-like dipoles in presence of a substrate as sketched in Fig.~\ref{Fig:SketchDDA}. This results in two possible applications to determine thermal emission. One is pursued by approximating macroscopic objects by voxels like in Fig.~\ref{Fig:SketchDDA}(a). In that case, we refer to this procedure as DDA. The other one describes the thermal emission of several separated subwavelength emitters as depicted in Fig.~\ref{Fig:SketchDDA}(b). Since this situation is different from DDA, we will refer to it by using the term coupled dipole method (CDM). In the CDM-cases considered in the following, these emitters will be spherical. In both cases the same formalism applies which we develop in the following. We start by determining the electric and magnetic fields generated by these dipoles in terms of the dressed polarizability for which we also discuss the different choices used in literature. Finally, we apply the fluctuation-dissipation theorem (FDT) to derive an expression for the emitted power of the $N$-body system through a ``detection plane''.

\subsection{Fluctuational fields}

Let us start by considering a configuration of $1 \leq \alpha \leq N$ nanoparticles arbitrarily distributed in front of a substrate occupying the half space at $z<0$. We ascribe different temperatures $T_\alpha$, $T_s$, and $T_b$ to the nanoparticles, substrate, and background, respectively. We assume that the nanoparticles fulfill the dipole approximation. That means their radii $R_\alpha$ obey $3 R_\alpha < d_\alpha \ll \lambda_\text{th}$ with particle-substrate distance $d_\alpha$ and thermal wavelength $\lambda_\text{th}$. Each nanoparticle and the substrate possess non-magnetic, homogeneous, and isotropic material properties, therefore, we neglect their permeabilities but we allow for eddy currents inside of the nanoparticles so that we can treat metallic nanoparticles~\cite{Dong,wirbelM,Dedkov2,Chapuis,gold3}. To include the effect of self interaction, we want to start by composing the equations for the fields in the nanoparticle $\alpha$ within the framework of macroscopic electrodynamics. For these fields we make the ansatz ($k = \text{E},\text{H}$)
\begin{align}
    \mathbf{F}_{k, \text{in}} (\mathbf{r}_\alpha) & = \mathbf{F}_{k,\text{env}} (\mathbf{r}_\alpha) + \mu_0 \omega \sum_{l \in \{\text{E,H}\}} \biggl[ \ri \int_{V_\alpha} \dr \mathds{G}_{kl} (\mathbf{r}_\alpha, \mathbf{r}) \mathbf{J}_l (\mathbf{r}) + \omega \sum_{\beta \neq \alpha} \mathds{G}_{kl} (\mathbf{r}_\alpha, \mathbf{r}_\beta) \mathbf{d}_l^\beta \biggr] .
\end{align}
Here $\mathbf{F}_\text{E} = \mathbf{E}$ is the electric field, $\mathbf{F}_\text{H} = \mathbf{H}$ the magnetic field, $\mathbf{d}_\text{E} = \mathbf{p}$ the electric dipole moment, and $\mathbf{d}_\text{H} = \mathbf{m}$ the magnetic dipole moment. $\mathbf{J}_\text{H}$ describes the above mentioned eddy currents. The index ``env'' indicates a background field consisting of the radiation of the background and substrate, $\mathds{G}(\mathbf{r}, \mathbf{r}')$ denotes the corresponding Green's function, $V_\alpha$ the volume of nanoparticle $\alpha$, $\omega$ the angular frequency, $\ri$ the imaginary unit, and $\mu_0$ the vacuum's permeability. Exploiting the long wavelength approximation (LWA) and, thus, assuming that the current density is constant inside of the minuscule nanoparticle volume, we extract the current density from the volume integral so that we can interpret it as volume average of the Green's function. After replacing the current density by the corresponding dipole moment, we obtain ($k = \text{E},\text{H}$)
\begin{align}
    \mathbf{F}_{k,\text{in}} (\mathbf{r}_\alpha) & = \mathbf{F}_{k,\text{env}} (\mathbf{r}_\alpha) + \mu_0 \omega^2 \sum_{l \in \{\text{E,H}\}} \biggl[ \expval{\mathds{G}_{kl} (\mathbf{r}_\alpha)} \mathbf{d}_l^\alpha + \sum_{\beta \neq \alpha} \mathds{G}_{kl} (\mathbf{r}_\alpha, \mathbf{r}_\beta) \mathbf{d}_l^\beta \biggr] 
\label{eq:F1}
\end{align}
with the volume averages
\begin{align}
   \expval{\mathds{G} (\mathbf{r})} & = \frac{1}{V_\alpha} \int_{V_\alpha} \dr' \mathds{G} (\mathbf{r}, \mathbf{r}') .
\end{align}
Note that this integral contains a singularity at $\mathbf{r}'=\mathbf{r}$. However, a rigorous treatment is provided by \cite{Lakhtakia, Fikioris, Yaghjian}. To obtain a closed formula for the fields, we decompose the dipole moments into a fluctuating part and an induced one, so that
\begin{align}
   \mathbf{d}_k^\alpha & = \mathbf{d}^\alpha_{k, \text{fl}} + \varepsilon_0 V_\alpha \chi_k^\alpha \mathbf{F}_{k, \text{in}} (\mathbf{r}_\alpha) .
\label{eq:d}
\end{align}
Here, $\varepsilon_0$ denotes the vacuum's permittivity and $\chi_\text{E/H}$ are the electric and magnetic susceptibilities which will be specified later. Inserting Eq. \eqref{eq:d} into Eq. \eqref{eq:F1}, we find for the fields inside of nanoparticle $\alpha$ the closed expression
\begin{align}
    \mathbf{F}_{k, \text{in}} (\mathbf{r}_\alpha) & = \mathbf{F}_{k, \text{env}} (\mathbf{r}_\alpha) + \sum_{l \in \{\text{E,H}\}} \biggl[ \mu_0 \omega^2 \expval{\mathds{G}_{kl} (\mathbf{r}_\alpha)} \mathbf{d}_{l, \text{fl}}^\alpha + k_0^2 V_\alpha \chi_l^\alpha \expval{\mathds{G}_{kl} (\mathbf{r}_\alpha)} \mathbf{F}_{l, \text{in}} (\mathbf{r}_\alpha) \notag \\
                                                  & \quad + \sum_{\beta \neq \alpha} \Bigl( \mu_0 \omega^2 \mathds{G}_{kl} (\mathbf{r}_\alpha, \mathbf{r}_\beta) \mathbf{d}_{l, \text{fl}}^\beta + k_0^2 V_\beta \chi_l^\beta \mathds{G}_{kl} (\mathbf{r}_\alpha, \mathbf{r}_\beta) \mathbf{F}_{l, \text{in}} (\mathbf{r}_\beta) \Bigr) \biggr] 
\end{align}
with the vacuum wave number $k_0 = \omega/c$. This system of $N$ vector equations for $1 \leq \alpha \leq N$ can be rewritten in a more compact way by exploiting the block matrix notation. Therefore, let us introduce the block vectors
\begin{align}
  \underline{\mathbf{F}}_k & = (\mathbf{F}_k (\mathbf{r}_1) , ... , \mathbf{F}_k (\mathbf{r}_N) )^t , \\
  \underline{\mathbf{d}}_k & = (\mathbf{d}_k^1 , ... , \mathbf{d}_k^N )^t
\end{align}
and the block matrices 
\begin{align}
  \underline{\underline{\mathds{G}}}_{kl}^{\alpha \beta} & = \delta_{\alpha \beta} \expval{\mathds{G}_{kl} (\mathbf{r}_\beta)} + (1 - \delta_{\alpha \beta}) \mathds{G}_{kl} (\mathbf{r}_\alpha, \mathbf{r}_\beta) , \\
   \underline{\underline{\mathds{X}}}_k^{\alpha \beta} & = \delta_{\alpha \beta} V_\beta \chi_k^\beta .
\end{align}
With these definitions we can express the corresponding fields by the two block matrix equations ($k = \text{E},\text{H}$)
\begin{align}
  \underline{\mathbf{F}}_{k, \text{in}} & = \underline{\mathbf{F}}_{k, \text{env}} + \sum_{l \in \{\text{E,H}\}} \Bigl[ \mu_0 \omega^2 \underline{\underline{\mathds{G}}}_{kl} \underline{\mathbf{d}}_{l, \text{fl}} + k_0^2 \underline{\underline{\mathds{G}}}_{kl} \underline{\underline{\mathds{X}}}_l \underline{\mathbf{F}}_{l, \text{in}} \Bigr] .
\label{eq:Fin1}
\end{align}
Eventually, one can interpret these two block matrix equations as one new block matrix equation introducing the block vectors
\begin{align}
  \stackrel{\rightarrow}{\mathbf{F}} & = (\underline{\mathbf{E}} , \underline{\mathbf{H}})^t , \\
  \stackrel{\rightarrow}{\mathbf{d}} & = (\underline{\mathbf{p}} , \underline{\mathbf{m}})^t 
\end{align}
and the new block matrices
\begin{align}
  \stackrel{\leftrightarrow}{\mathds{G}} & = \begin{pmatrix} \underline{\underline{\mathds{G}}}_\text{EE} & \underline{\underline{\mathds{G}}}_\text{EH} \\ \underline{\underline{\mathds{G}}}_\text{HE} & \underline{\underline{\mathds{G}}}_\text{EE} \end{pmatrix} , \\
  \stackrel{\leftrightarrow}{\mathds{X}} & = \begin{pmatrix} \underline{\underline{\mathds{X}}}_\text{E} & 0 \\ 0 & \underline{\underline{\mathds{X}}}_\text{H} \end{pmatrix} .
\end{align}
Inserting the latter into Eq. \eqref{eq:Fin1}, we obtain
\begin{align}
  \stackrel{\rightarrow}{\mathbf{F}}_\text{in} & = \Bigl[ \stackrel{\leftrightarrow}{\mathds{1}} - k_0^2 \stackrel{\leftrightarrow}{\mathds{G}} \stackrel{\leftrightarrow}{\mathds{X}} \Bigr]^{-1} \Bigl(\stackrel{\rightarrow}{\mathbf{F}}_\text{env} + \mu_0 \omega^2 \stackrel{\leftrightarrow}{\mathds{G}} \stackrel{\rightarrow}{\mathbf{d}}_\text{fl} \Bigr) .
\label{eq:Fin2}
\end{align}
This expression allows us now to derive the fields outside of the nanoparticles. By inserting Eq.~\eqref{eq:Fin2} into Eq.~\eqref{eq:d}
\begin{align}
  \stackrel{\rightarrow}{\mathbf{d}} & = \Bigl(\stackrel{\leftrightarrow}{\mathds{1}} + k_0^2 \stackrel{\leftrightarrow}{\boldsymbol{\alpha}} \stackrel{\leftrightarrow}{\mathds{G}} \Bigr) \stackrel{\rightarrow}{\mathbf{d}}_\text{fl} + \varepsilon_0 \stackrel{\leftrightarrow}{\boldsymbol{\alpha}} \stackrel{\rightarrow}{\mathbf{F}}_\text{env} ,
\label{eq:d2}
\end{align}
we obtain the closed form of the dipole moments. Here, we introduced the dressed polarizability 
\begin{align}
  \stackrel{\leftrightarrow}{\boldsymbol{\alpha}} & = \stackrel{\leftrightarrow}{\mathds{X}} \Bigl[ \stackrel{\leftrightarrow}{\mathds{1}} - k_0^2 \stackrel{\leftrightarrow}{\mathds{G}} \stackrel{\leftrightarrow}{\mathds{X}} \Bigr]^{-1} .
\label{eq:alpha}
\end{align}
A general discussion of this dressed polarizability will follow up in the next section. Here, we focus on the derivation of the fields outside of the nanoparticles which are given by the fields generated by the dipole moments and the environmental fields so that we have 
\begin{align}
   \mathbf{F}_{k, \text{out}} (\mathbf{r}) & = \mathbf{F}_{k, \text{env}} (\mathbf{r}) + \mu_0 \omega^2 \sum_{l \in \{\text{E,H}\}} \sum_\alpha \mathds{G}_{kl} (\mathbf{r}, \mathbf{r}_\alpha) \mathbf{d}_l^\alpha.
\end{align}
By inserting Eq.~\eqref{eq:d2}, we finally obtain the result for the fields outside of the nanoparticles
\begin{align}
  \mathbf{F}_{k, \text{out}} (\mathbf{r}) & = \mathbf{F}_{k, \text{env}} (\mathbf{r}) + \sum_{l,m \in \{\text{E,H}\}} \sum_{\alpha, \beta} \mathds{G}_{kl} (\mathbf{r}, \mathbf{r}_\alpha) \left( \mu_0 \omega^2 \mathds{A}_{lm}^{\alpha \beta} \mathbf{d}_{m, \text{fl}}^\beta + k_0^2 \boldsymbol{\alpha}_{lm}^{\alpha \beta} \mathbf{F}_{m, \text{env}} (\mathbf{r}_\beta) \right)
\end{align}
with
\begin{align}
  \stackrel{\leftrightarrow}{\mathds{A}} & = \stackrel{\leftrightarrow}{\mathds{1}} + k_0^2 \stackrel{\leftrightarrow}{\boldsymbol{\alpha}} \stackrel{\leftrightarrow}{\mathds{G}} .
\end{align}
Note, that this expression is very general and describes the fields generated by the fluctuational dipoles inside of the nanoparticles and the field of the background and substrate within near- and far-field regime. We will use these general expressions to compute the z-component of the mean Poynting vector integrated over a plane above the nanoparticle assembly. This will give us the spectral power emitted by the nanoparticles in vicinity of the substrate taking into account the different temperatures of the nanoparticles, substrate, and background. Hence, this power determines the emitted heat of the whole system which would also be measurable with a detector.

\subsection{dressed polarizability}

Actually, Eq.~\eqref{eq:alpha} contains four different dressed polarizabilities which are given by
\begin{align}
  \underline{\underline{\boldsymbol{\alpha}}}_{kk} & = \underline{\underline{\mathds{X}}}_k \left[ \underline{\underline{\mathds{1}}} - k_0^2 \underline{\underline{\mathds{G}}}_{kk} \underline{\underline{\mathds{X}}}_k - k_0^4 \underline{\underline{\mathds{G}}}_{k\bar{k}} \underline{\underline{\mathds{X}}}_{\bar{k}} \left[ \underline{\underline{\mathds{1}}} - k_0^2 \underline{\underline{\mathds{G}}}_{\bar{k}\bar{k}} \underline{\underline{\mathds{X}}}_{\bar{k}} \right]^{-1} \underline{\underline{\mathds{G}}}_{\bar{k}k} \underline{\underline{\mathds{X}}}_k \right]^{-1} ,
  \label{eq:akk} \\
  \underline{\underline{\boldsymbol{\alpha}}}_\text{EH/HE} & = \underline{\underline{\mathds{F}}}_\text{E/H} \underline{\underline{\boldsymbol{\alpha}}}_\text{HH/EE} = \underline{\underline{\boldsymbol{\alpha}}}_\text{EE/HH} \underline{\underline{\mathds{E}}}_\text{H/E}, 
\label{eq:aEH}
\end{align}
with 
\begin{align}
   \underline{\underline{\mathds{F}}}_\text{E/H} & = k_0^2 \underline{\underline{\mathds{X}}}_\text{E/H} \left[ \underline{\underline{\mathds{1}}} - k_0^2 \underline{\underline{\mathds{G}}}_\text{EE/HH} \underline{\underline{\mathds{X}}}_\text{E/H} \right]^{-1} \underline{\underline{\mathds{G}}}_\text{EH/HE} , \\
   \underline{\underline{\mathds{E}}}_\text{E/H} & = k_0^2 \underline{\underline{\mathds{G}}}_\text{HE/EH} \underline{\underline{\mathds{X}}}_\text{E/H} \left[ \underline{\underline{\mathds{1}}} - k_0^2 \underline{\underline{\mathds{G}}}_\text{EE/HH} \underline{\underline{\mathds{X}}}_\text{E/H} \right]^{-1}.
\end{align}
Note that the bar notation indicates $\bar{\text{E}}=\text{H}$ and $\bar{\text{H}}=\text{E}$, respectively. Additionally, the relations
\begin{align}
  \underline{\underline{\boldsymbol{\alpha}}}_\text{EE/HH} & = \frac{1}{k_0^2} \underline{\underline{\mathds{F}}}_\text{E/H} \underline{\underline{\mathds{G}}}_\text{EH/HE}^{-1} + \underline{\underline{\mathds{F}}}_\text{E/H} \underline{\underline{\boldsymbol{\alpha}}}_\text{HH/EE} \underline{\underline{\mathds{E}}}_\text{E/H} 
\label{eq:trick2}
\end{align}
hold. These relations are the general $N$-body formulas for the dressed polarizabilities. 

To show later that we can retrieve the known single-particle expressions from the general ones, we restrict ourselves to the electric polarizability by setting $\chi_\text{H} = 0$. Then the above expressions simplify to
\begin{align}
  \boldsymbol{\alpha}_\text{EE} & = \alpha_{\text{EE}, \perp} [\mathbf{e}_x \otimes \mathbf{e}_x + \mathbf{e}_y \otimes \mathbf{e}_y] + \alpha_{\text{EE}, z} \mathbf{e}_z \otimes \mathbf{e}_z , \\
  \alpha_{\text{EE}, \perp/z} & = \frac{V \chi_\text{E}}{1 - k_0^2 V \chi_\text{E} \expval{G_{\text{EE}, \perp/z}}},
\end{align}
whereas the other polarizabilities vanish. Here, the volume average of the components perpendicular and parallel to the surface normal split up in the vacuum contribution $\expval{G_{\text{EE,vac},\perp/z}}$ and the contribution of the surface reflected fields $\expval{G_{\text{EE,ref},\perp/z}}$.
To further evaluate this expression, we require the volume averages of the Green's functions. These averages depend on the shape of the nanoparticles. For pure vacuum, i.e.\ without substrate, they are well-known and long-established~\cite{Lakhtakia, Fikioris, Yaghjian}. Nonetheless, throughout the literature different forms are used which lead to different expressions for the dressed polarizability~\cite{Fikioris, Albaladejo, Ekeroth, Edalatpour1, Draine1, Goedecke, Draine2, Messina, Carminati}, especially for DDA. When considering spherical nanoparticles, for instance, one finds the expressions~\cite{Albaladejo, Messina, Carminati, Ekeroth, Lakhtakia}
\begin{equation}
	\expval{G_{\text{EE,vac},\perp/z}}_{\rm d-cdm} \approx - \frac{1}{3 V k_0^2} + \frac{\ri k_0}{6 \pi} 
\end{equation} 
which includes the so-called ``radiation correction'' term and is also called ``Draine's form'' of coupled dipole moments (D-CDM)~\cite{Lakhtakia}. On the other hand, the expression~\cite{Fikioris, Edalatpour1, Lakhtakia}
\begin{equation}
	 \expval{G_{\text{EE,vac},\perp/z}}_{\rm s-cdm} = \biggl( \frac{2}{3} \left[1 - \ri k_0 R \right] e^{\ri k_0 R} - 1\biggr) \frac{1}{V k_0^2}
\end{equation}
is used which is also called the ``strong form'' of coupled dipole moments (S-CDM)~\cite{Lakhtakia}. Note that in the limit $k_0 R \ll 1$ one has 
\begin{equation}
	\expval{G_{\text{EE,vac},\perp/z}}_{\rm s-cdm} \approx - 1/(3 k_0^2 V) = \expval{G_{\text{EE,vac},\perp/z}}_{\rm w-cdm}.
\end{equation}
That means in the small sphere limit the strong form of coupled dipole moments converges to the ``weak form'' of coupled dipole moments (W-CDM) and the only difference between the weak/strong form and Draine's form in that limit is the radiation correction term $\ri k_0/ 6 \pi$ which is typically small in the limit. As a consequence, when using the susceptibility
\begin{equation}
	\chi_\text{E} = \varepsilon - 1
\end{equation}
the polarizability for a nanoparticle in vacuum including the radiation correction is~\cite{Albaladejo, Messina, Carminati, Ekeroth, Lakhtakia}
\begin{equation}
	{\alpha_{\text{EE},\perp/z}}_{\rm d-cdm} = \frac{\alpha_{\rm w-cdm}}{1 - \alpha_{\rm w-cdm}  \frac{\ri k_0^3}{6 \pi}}
\end{equation}
with the Clausius-Mosotti-like polarizability of a spherical nanoparticle also obtained from the weak form of coupled dipole moments
\begin{equation}
	\alpha_{\rm w-cdm} = 3 V \frac{\varepsilon - 1}{\varepsilon + 2}.
\end{equation}
On the other hand, the strong form of coupled dipole moments leads to the polarizability
\begin{equation}
	{\alpha_{\text{EE},\perp/z}}_{ \rm s-cdm} = \frac{V \chi_\text{E}}{1 - \chi_\text{E} \left(\frac{2}{3} \left[1 - \ri k_0 R \right] e^{\ri k_0 R} - 1\right)} 
	\label{Eq:alphaSCDM}
\end{equation}
which coincides in the limit $k_0 R \ll 1$ with the weak form $\alpha_{\rm w-cdm}$ like it is used in Ref.~\cite{Edalatpour1}. 

Note, that these expressions are only valid without substrate. With substrate also the average of the scattered part of the Green's function needs to be added. In that case we have the three forms
\begin{align}
	{\alpha_{\text{EE},\perp/z}}_{\rm d-cdm} &= \frac{\alpha_{\rm w-cdm}}{1 - \alpha_{\rm w-cdm}  \frac{\ri k_0^3}{6 \pi} - k_0^2 \alpha_{\rm w-cdm} \expval{G_{\text{EE,ref},\perp/z}} },  \\
	{\alpha_{\text{EE},\perp/z}}_{\rm s-cdm} &= \frac{V \chi_\text{E}}{1 - \chi_\text{E} \left(\frac{2}{3} \left[1 - \ri k_0 R \right] e^{\ri k_0 R} - 1\right) - k_0^2 V \chi_E  \expval{G_{\text{EE,ref},\perp/z}}}, \label{Eq:alphascdmvac} \\
        {\alpha_{\text{EE},\perp/z}}_{\rm w-cdm} &= \frac{\alpha_{\rm w-cdm}}{1  - k_0^2 \alpha_{\rm w-cdm} \expval{G_{\text{EE,ref},\perp/z}}}. 
\end{align}
The weak form is the one used in Ref.~\cite{Joulain2} together with an approximated Mie expression for the dipolar polarizabilities in place of $\alpha_{\rm w-cdm}$. 
This combination of the weak form together with Mie expressions is not unusual and has been applied to spherical nanoparticles for the electric and magnetic part~\cite{Joulain2}, because it might also be applicable to spherical particles of larger size as long as the dipole contributions are dominant. It corresponds to setting $\expval{G_{\text{EE,vac},\perp/z}} = 0$ and using 
\begin{align}
	V_\alpha \chi_\text{E}^\alpha & = \frac{9 V_\alpha\ri}{2 x_\alpha^3} a_1, \\
	V_\alpha \chi_\text{H}^\alpha & = \frac{9 V_\alpha\ri}{2 x_\alpha^3} \frac{\mu_0}{\epsilon_0} b_1
\end{align}
where~\cite{bohren}
\begin{align}
	a_1 & =  \frac{\varepsilon_\alpha j_1(y_\alpha) [x_\alpha j_1(x_\alpha)]' - j_1(x_\alpha) [y_\alpha j_1(y_\alpha)]'}{\varepsilon_\alpha j_1(y_\alpha) [x_\alpha h_1^{(1)}(x_\alpha)]' - h_1^{(1)}(x_\alpha) [y_\alpha j_1(y_\alpha)]'} , \label{ChiE} \\
        b_1 & = \frac{j_1(y_\alpha) [x_\alpha j_1(x_\alpha)]' - j_1(x_\alpha) [y_\alpha j_1(y_\alpha)]'}{j_1(y_\alpha) [x_\alpha h_1^{(1)}(x_\alpha)]' - h_1^{(1)}(x_\alpha) [y_\alpha j_1(y_\alpha)]'} \label{ChiH}
\end{align}
are the dipolar Mie coefficients with $x_\alpha = k_0 R_\alpha$, $y_\alpha = \sqrt{\epsilon_\alpha} x_\alpha$, and the spherical Bessel and Hankel functions of the first kind \cite{Dong}. The primes denote the derivatives with respect to the function's argument. $\varepsilon_\alpha$ denotes the permittivity of nanoparticle $\alpha$. In the limit $x_\alpha,y_\alpha \ll 1$ the susceptibilities simplify to 
\begin{align}
  V_\alpha \chi_\text{E}^\alpha & = 3 V_\alpha \frac{\varepsilon_\alpha - 1}{\varepsilon_\alpha + 2} , \label{ChiEappr}\\
	V_\alpha \chi_\text{H}^\alpha & = \frac{\mu_0}{\epsilon_0} \frac{V_\alpha}{10} x_\alpha^2 (\varepsilon_\alpha - 1). \label{ChiHappr}
\end{align}
Note that in this limit $V_\alpha \chi_\text{E}^\alpha$ equals $\alpha_{\rm w-cdm}$ so that the electrical part retrieves the Clausius-Mosotti form of the polarizability. This mixed approach of the weak form with the Mie coefficients is called the ``weak Mie form'' (W-Mie-CDM) in the following.

In the numerical evaluations of thermal radiation of one or several nanoparticles within the CDM we will use the expressions for spherical nanoparticles. On the other hand, for DDA simulations we will use cubic volume elements of a single unit size to replace the macroscopic object by a assembly of voxels. More elaborated DDA simulations with rectangular voxels or voxels of different sizes and shapes for better approximation of the shape of the macroscopic object can, of course, also be used~\cite{Yurkin2,Edalatpour3}. An overview on the DDA method in general can be found in Ref.~\cite{Yurkin1}, for instance. A more detailed discussion on finite size effects of some of the considered polarizabilities which are important for large radii can be found in Ref.~\cite{Yurkin3}.

\subsection{Total emitted heat radiation}

Now, we determine the total heat flux through a plane parallel to the substrate's surface above the nanoparticle assembly. This heat flux is equivalent to the power emitted into the far-field. As mentioned before, we need the Poynting vector which is computed by means of the FDT~\cite{Rytov}. To this end, we assume that the background fields as well as the electric and magnetic dipole moments are uncorrelated. We obtain
\begin{align}
   P_z & = 2 \Re \int \dx \epsilon_{ijz} \el \mathbf{E}_\text{out} (\mathbf{r}) \otimes \mathbf{H}_\text{out}^\dagger (\mathbf{r}) \er^\omega_{ij} \notag \\
      & = P_\text{s} + P_\text{np} + P_\text{env}
	\label{Eq:PsPnpPenv}
\end{align}
with the power solely emitted by the background fields
\begin{align}
   P_\text{s} & = 2 \Re \int \dx \epsilon_{ijz} \el \mathbf{E}_\text{env} (\mathbf{r}) \otimes \mathbf{H}_\text{env}^\dagger (\mathbf{r}) \er^\omega_{ij},
\end{align}
the power emanated from the nanoparticles
\begin{align}
   P_\text{np} & = 2 \mu_0^2 \omega^4 \sum_{k,l \in \{\text{E,H}\}} \sum_{\alpha, \beta, \gamma, \delta} \Re \int \dx \epsilon_{ijz} \Bigl[ \mathds{G}_{\text{E}k} (\mathbf{r}, \mathbf{r}_\alpha) \mathds{A}_{k\text{E}}^{\alpha \beta} \el \mathbf{p}_\text{fl}^\beta \otimes \mathbf{p}_{\text{fl}}^\delta \er^\omega \mathds{A}_{l\text{E}}^{\gamma \delta \dagger} \mathds{G}_{\text{H}l}^\dagger (\mathbf{r}, \mathbf{r}_\gamma) \notag \\
               & \quad + \mathds{G}_{\text{E}k} (\mathbf{r}, \mathbf{r}_\alpha) \mathds{A}_{k\text{H}}^{\alpha \beta} \el \mathbf{m}_\text{fl}^\beta \otimes \mathbf{m}_{\text{fl}}^\delta \er^\omega \mathds{A}_{l\text{H}}^{\gamma \delta \dagger} \mathds{G}_{\text{H}l}^\dagger (\mathbf{r}, \mathbf{r}_\gamma) \Bigr]_{ij} ,
\end{align}
and the power stemming from interactions between substrate and nanoparticles
\begin{align}
   P_\text{env} & = 2 k_0^2 \sum_{k,l \in \{\text{E,H}\}} \sum_{\alpha, \beta} \Re \int \dx \epsilon_{ijz} \Bigl[ \el \mathbf{E}_\text{env} (\mathbf{r}) \otimes \mathbf{F}_{l, \text{env}}^\dagger (\mathbf{r}_\beta) \er^\omega \boldsymbol{\alpha}_{kl}^{\alpha \beta \dagger} \mathds{G}_{\text{H}k}^\dagger (\mathbf{r}, \mathbf{r}_\alpha) \notag \\
                & \quad + \mathds{G}_{\text{E}k} (\mathbf{r}, \mathbf{r}_\alpha) \boldsymbol{\alpha}_{kl}^{\alpha \beta} \el \mathbf{F}_{l, \text{env}} (\mathbf{r}_\beta) \otimes \mathbf{H}_\text{env}^\dagger (\mathbf{r}) \er^\omega \notag \\
                & \quad + k_0^2 \sum_{m,n \in \{\text{E,H}\}} \sum_{\gamma, \delta} \mathds{G}_{\text{E}k} (\mathbf{r}, \mathbf{r}_\alpha) \boldsymbol{\alpha}_{kl}^{\alpha \beta} \el \mathbf{F}_{l, \text{env}} (\mathbf{r}_\beta) \otimes \mathbf{F}_{n, \text{env}}^\dagger (\mathbf{r}_\delta) \er^\omega \boldsymbol{\alpha}_{mn}^{\gamma \delta \dagger} \mathds{G}_{\text{H}m}^\dagger (\mathbf{r}, \mathbf{r}_\gamma) \Bigr]_{ij}.
\end{align}
Additionally, $\epsilon_{ijq}$ denotes the Levi-Civita symbol, $\,\,\dx$ the infinitesimal surface element of the x-y plane. The $\langle \! \langle \cdot \rangle \! \rangle$ notation indicates an ensemble average over thermally fluctuating currents and fields. For the first term $P_s$ in Eq.~(\ref{Eq:PsPnpPenv}) we obtain 
\begin{align}
  P_\text{s} & = A \hbar \omega \bigl( n_\text{s} - n_\text{b} \bigr) \int_0^{k_0} \frac{\text{d} k_\perp}{2 \pi} k_\perp (2 - |r_\text{E}|^2 - |r_\text{H}|^2).
\label{eq:P_s}
\end{align}
It conveys no additional information about the scattering process but describes the power transferred between substrate and background due to the different temperatures of the substrate and background~\cite{Polder}. Here, $A$ denotes an area in the x-y plane, $\hbar$ the reduced Planck constant, and $r_\text{E,H}$ are the Fresnel amplitude reflection coefficients for parallel and perpendicular polarized light, respectively (see appendix \ref{appA}). $n_j$ is the Bose-Einstein occupation probability 
\begin{align}
   n_j & = \left[ e^{\frac{\hbar \omega}{k_B T_j}} - 1 \right]^{-1}
\end{align}
with the Boltzmann constant $k_B$. The plane area $A$ diverges when considering the whole x-y plane, so that we will later set $A$ to be the geometrical cross section of the considered tip or nanoparticle. However, in an experiment the choice of $A$ would correspond to the collection area of a detector. Nowadays, many experiments usually operate in tapping mode with a lock-in technique~\cite{Wang, Komiyama, Kajihara2, DeWilde, Babuty, Jones, O'Callahan}, i.e. the distance between substrate surface and nanoparticle assembly, e.g. the tip of an atomic force or scanning tunnelling microscope, oscillates during measurements. Then, because $P_\text{s}$ is constant with respect to the distance between nanoparticles and substrate surface, this quantity will average out and does not contribute to the detected signal.

For the evaluation of $P_\text{np}$ we need two intermediate steps. At first, we exploit Eq. \eqref{eq:aEH} and Eq. \eqref{eq:trick2} to rewrite $\mathds{A}_\text{EH/HE}^{\alpha \beta}$ in terms of $\mathds{A}_\text{EE/HH}^{\alpha \beta}$ by
\begin{align}
   \mathds{A}_\text{EH/HE}^{\alpha \beta} & = \sum_{\gamma=1}^N \mathds{F}_\text{E/H}^{\alpha \gamma} \mathds{A}_\text{HH/EE}^{\gamma \beta} .
\end{align}
Besides, we have to evaluate the correlation function which is done by using the FDT related to~\cite{Herz3}:
\begin{align}
   \el \mathbf{d}_{k,\text{fl}} \otimes \mathbf{d}_{k,\text{fl}}^\dagger \er^\omega_{\alpha \beta} & = \frac{2 \hbar}{\mu_0 \omega^2} \left( n_\alpha + \frac{1}{2} \right) \sum_{\gamma,\delta} \mathds{A}_{kk}^{\alpha \gamma -1} \boldsymbol{\chi}^k_{\gamma \delta} \mathds{A}_{kk}^{\delta \beta -1 \dagger} \delta_{\beta \alpha}
\end{align}
with the generalized susceptibilities
\begin{align}
  \boldsymbol{\chi}^k_{\alpha \beta} & = k_0^2 \frac{\boldsymbol{\alpha}_{kk}^{\alpha \beta} - \boldsymbol{\alpha}_{kk}^{\beta \alpha \dagger}}{2 \ri} - k_0^4 \sum_{\gamma, \delta} \boldsymbol{\alpha}_{kk}^{\alpha \gamma} \biggl( \frac{\mathds{G}_{kk}^{\gamma \delta} - \mathds{G}_{kk}^{\delta \gamma \dagger}}{2 \ri} + \sum_\epsilon \frac{\mathds{E}_{\bar{k}}^{\gamma \epsilon} \mathds{G}_{\bar{k}k}^{\epsilon \delta} - \mathds{G}_{\bar{k}k}^{\epsilon \gamma \dagger} \mathds{E}_{\bar{k}}^{\delta \epsilon \dagger}}{2 \ri} \biggr) \boldsymbol{\alpha}_{kk}^{\beta \delta \dagger} .
\end{align}
These general susceptibilities used in the FDT are known for single arbitrary objects~\cite{KruegerEtAlTraceFormulas2012,Herz2} and in particular for a single dipolar object in a given environment as well as for many of them ~\cite{OttSAB2020, Herz3, ottthgg, Yurkin3, SABRMB2021} where the details of the derivation for arbitrary objects can be found in~\cite{KruegerEtAlTraceFormulas2012, Herz2} and for dipoles in~\cite{Yurkin3}.
A more detailed justification for this evaluation will follow at the end of this paragraph. With that we obtain for $P_\text{np}$
\begin{align}
   P_\text{np} & = 4 \hbar \mu_0 \omega^2 \Re \int \dx \epsilon_{ijz} \sum_{\alpha,\beta} \sum_{k \in \{\text{E,H}\}} \Biggl( \mathds{G}_{\text{E}k} (\mathbf{r}, \mathbf{r}_\alpha) \notag \\
               & \quad \times \Bigg \{ \left[ \left( n(T_{\alpha}, \omega) + \frac{1}{2} \right) \boldsymbol{\chi}^k_{\alpha \beta} + \sum_{\gamma,\delta} \left( n(T_{\gamma}, \omega) + \frac{1}{2} \right) \mathds{F}_k^{\alpha \gamma} \boldsymbol{\chi}^{\bar{k}}_{\gamma \delta} \mathds{F}_k^{\beta \delta \dagger} \right] \mathds{G}_{\text{H}k}^\dagger (\mathbf{r}, \mathbf{r}_\beta) \notag \\
               & \quad + \sum_\gamma \left[ \left( n(T_{\gamma}, \omega) + \frac{1}{2} \right) \mathds{F}_k^{\alpha \gamma} \boldsymbol{\chi}^{\bar{k}}_{\gamma \beta} + \left( n(T_{\alpha}, \omega) + \frac{1}{2} \right) \boldsymbol{\chi}^k_{\alpha \gamma} \mathds{F}_{\bar{k}}^{\beta \gamma \dagger} \right] \mathds{G}_{\text{H} \bar{k}}^\dagger (\mathbf{r}, \mathbf{r}_\beta) \Bigg \} \Biggr)_{ij} .
\end{align}

If the radiative heat sources in the substrate and background are uncorrelated, we can safely decompose the correlation function concerning the fields of the whole environment containing substrate and background into separated correlation functions dealing with either the substrate or the background. Then, we can follow the procedure outlined in Ref.~\cite{Herz3} providing
\begin{align}
   \el \mathbf{F}_k (\mathbf{r}_\alpha) \otimes \mathbf{F}_l^\dagger (\mathbf{r}_\beta) \er^\omega & = \el \mathbf{F}_k (\mathbf{r}_\alpha) \otimes \mathbf{F}_l^\dagger (\mathbf{r}_\beta) \er^\omega_\text{eq}  + \el \mathbf{F}_k (\mathbf{r}_\alpha) \otimes \mathbf{F}_l^\dagger (\mathbf{r}_\beta) \er^\omega_\text{leq}
\end{align}
for the correlation function of the whole environment with
\begin{align}
  \el \mathbf{F}_{k, \text{env}} (\mathbf{r}_\alpha) \otimes \mathbf{F}_{l, \text{env}}^\dagger (\mathbf{r}_\beta) \er^\omega_\text{eq} & = 2 \hbar \mu_0 \omega^2 \left( n_\text{b} + \frac{1}{2} \right) \frac{\mathds{G}_{kl}^{\alpha \beta} - \mathds{G}_{lk}^{\beta \alpha \dagger}}{2 \ri}
\label{eq:eq}
\end{align}
and
\begin{align}
  \el \mathbf{F}_{k, \text{env}} (\mathbf{r}_\alpha) \otimes \mathbf{F}_{l, \text{env}}^\dagger (\mathbf{r}_\beta) \er^\omega_\text{leq} & = 2 \hbar \mu_0 \omega^2 k_0^2 \left( n_\text{s} - n_\text{b} \right) \Im(\varepsilon_\text{s}) \notag \\
                            & \quad \times \int_{V_\text{s}} \dr \mathds{G}_{k\text{E}}^\text{s} (\mathbf{r}_\alpha, \mathbf{r}) \mathds{G}_{l\text{E}}^{\text{s} \dagger} (\mathbf{r}_\beta, \mathbf{r}).
\label{eq:leq}
\end{align}
In the latter equation, $V_\text{s}$ denotes the substrate's volume, $\varepsilon_\text{s}$ its permittivity, and $\mathds{G}^\text{s}$ describes the Green's function for sources inside of the substrate. If one refers to spatial arguments which do not correspond to particle positions, the superscripts in the Green's function in Eq.~\eqref{eq:eq} are replaced by the usual argument notation with coordinates of the observation point and the source point. The first correlation function stems from forcing both fields to fulfill the global equilibrium correlation function Ref.~\cite{Eckhardt} when they are at the same temperature. The second one, however, results from evaluating the correlation functions in local equilibrium for both fields separately. Using only the first correlation function in Eq.~\eqref{eq:eq} and the relations between the dressed polarizabilities ($k = {\rm E, H}$)
\begin{align}
   \underline{\underline{\boldsymbol{\alpha}}}_{kk}^t & = \underline{\underline{\boldsymbol{\alpha}}}_{kk} , 
   \label{eq:trick3} \\
   \underline{\underline{\boldsymbol{\alpha}}}_\text{EH}^t & = - \underline{\underline{\boldsymbol{\alpha}}}_\text{HE} ,
  \label{eq:trick4}
\end{align}
we obtain for the global equilibrium contribution of $P_\text{env} = P_\text{env,eq} + P_\text{env,leq}$
\begin{align}
  P_\text{env,eq} & = - 4 \hbar \mu_0 \omega^2 \left( n_\text{b} + \frac{1}{2} \right) \Re \int \dx \epsilon_{ijz} \sum_{\alpha,\beta} \sum_{k \in \{\text{E,H}\}} \notag \\
                  & \quad \times \Biggl( \mathds{G}_{\text{E}k} (\mathbf{r}, \mathbf{r}_\alpha) \left[ \boldsymbol{\chi}^k_{\alpha \beta} + \sum_{\gamma,\delta} \mathds{F}_k^{\alpha \gamma} \boldsymbol{\chi}^{\bar{k}}_{\gamma \delta} \mathds{F}_k^{\beta \delta \dagger} \right] \mathds{G}_{\text{H}k}^\dagger (\mathbf{r}, \mathbf{r}_\beta) \notag \\
                  & \quad + \sum_\gamma \mathds{G}_{\text{E}k} (\mathbf{r}, \mathbf{r}_\alpha) \left[ \mathds{F}_k^{\alpha \gamma} \boldsymbol{\chi}^{\bar{k}}_{\gamma \beta} + \boldsymbol{\chi}^k_{\alpha \gamma} \mathds{F}_{\bar{k}}^{\beta \gamma \dagger} \right] \mathds{G}_{\text{H} \bar{k}}^\dagger (\mathbf{r}, \mathbf{r}_\beta) \Biggr)_{ij} .
\end{align}
Comparing $P_\text{np}$ and $P_\text{env,eq}$, they only differ in sign and temperature. Therefore, we can merge them to one contribution to the whole system describing the heat transfer between all nanoparticles and the background. This also describes detailed balance between background and nanoparticles and also serves as justification for the evaluation of the dipole moment correlation functions a posteriori because this part vanishes when $T_p = T_b$ as it should be. Together with the matrix abbreviations
\begin{align}
  \mathds{M}_k^{\alpha \beta} & := \left( n_\alpha - n_b \right) \boldsymbol{\chi}^k_{\alpha \beta} + \sum_{\gamma,\delta} \left( n_\gamma - n_b \right) \mathds{F}_k^{\alpha \gamma} \boldsymbol{\chi}^{\bar{k}}_{\gamma \delta} \mathds{F}_k^{\beta \delta \dagger}
\end{align}
and
\begin{align}
   \mathds{N}_\text{c}^{\alpha \beta} & := \sum_\gamma \left[ \left( n_\alpha - n_b \right) \boldsymbol{\chi}^\text{E}_{\alpha \gamma} \mathds{F}_\text{H}^{\beta \gamma \dagger} + \left( n_\gamma - n_b \right) \mathds{F}_\text{E}^{\alpha \gamma} \boldsymbol{\chi}^\text{H}_{\gamma \beta} \right] 
\end{align}
we simplify the merging of $P_\text{np}$ and $P_\text{env,eq}$ to
\begin{align}
  P_\text{dir} & = P_\text{np} + P_\text{env,eq} \notag \\
               & = 4 \hbar \mu_0 \omega^2 \Re \int \dx \epsilon_{ijz} \sum_{\alpha,\beta} \biggl( \sum_{k \in \{\text{E,H}\}} \mathds{G}_{\text{E}k} (\mathbf{r}, \mathbf{r}_\alpha) \mathds{M}_k^{\alpha \beta} \mathds{G}_{\text{H}k}^\dagger (\mathbf{r}, \mathbf{r}_\beta) \notag \\
               & \quad + \mathds{G}_\text{EE} (\mathbf{r}, \mathbf{r}_\alpha) \mathds{N}_\text{c}^{\alpha \beta} \mathds{G}_\text{HH}^\dagger (\mathbf{r}, \mathbf{r}_\beta)  + \mathds{G}_\text{EH} (\mathbf{r}, \mathbf{r}_\alpha) \mathds{N}_\text{c}^{\beta \alpha \dagger} \mathds{G}_\text{HE}^\dagger (\mathbf{r}, \mathbf{r}_\beta) \biggr)_{ij} .
\end{align}
Analytically simplifying the integrals as far as possible, we end up with
\begin{align}
   P_\text{dir} & = \hbar \omega k_0 \Re \Tr \biggl( \underline{\underline{\mathds{M}}}_\text{E} \underline{\underline{\mathds{I}}}_\text{E} + \frac{\epsilon_0}{\mu_0} \underline{\underline{\mathds{M}}}_\text{H} \underline{\underline{\mathds{I}}}_\text{H} + 2 \sqrt{\frac{\varepsilon_0}{\mu_0}} \underline{\underline{\mathds{N}}}_\text{c} \underline{\underline{\mathds{I}}}_\text{c} \biggr)
\end{align}
The components of the tensors $\mathds{I}$ are integral expressions listed in appendix~\ref{appB}. 

The remaining ``local equilibrium contribution'' (LEQC) consists of two parts
\begin{align}
  P_\text{env,leq} & = P_\text{abs} + P_\text{scat} 
\end{align}
with
\begin{align}
  P_\text{abs} & = 4 k_0^4 \hbar \mu_0 \omega^2 \left( n_\text{s} - n_\text{b} \right) \sum_{k,l \in \{\text{E,H}\}} \sum_{\alpha, \beta} \Re \int \dx \int_{V_\text{s}} \dr' \epsilon_{ijz} \Im(\varepsilon_\text{s}) \notag \\
               & \quad \times \Bigl[ \mathds{G}_{\text{EE}}^\text{s} (\mathbf{r}, \mathbf{r}') \mathds{G}_{l\text{E}}^{\text{s} \dagger} (\mathbf{r}_\beta, \mathbf{r}') \boldsymbol{\alpha}_{kl}^{\alpha \beta \dagger} \mathds{G}_{\text{H}k}^\dagger (\mathbf{r}, \mathbf{r}_\alpha) \notag \\
               & \quad + \mathds{G}_{\text{E}k} (\mathbf{r}, \mathbf{r}_\alpha) \boldsymbol{\alpha}_{kl}^{\alpha \beta} \mathds{G}_{l\text{E}}^\text{s} (\mathbf{r}_\beta, \mathbf{r}') \mathds{G}_{\text{HE}}^{\text{s} \dagger} (\mathbf{r}, \mathbf{r}') \Bigr]_{ij}.
\end{align}
and
\begin{align}
  P_\text{scat} & = 4 k_0^6 \hbar \mu_0 \omega^2 \left( n_\text{s} - n_\text{b} \right) \sum_{k,l,m,n \in \{\text{E,H}\}} \sum_{\alpha, \beta, \gamma, \delta} \Re \int \dx \int_{V_\text{s}} \dr' \epsilon_{ijz} \Im(\varepsilon_\text{s}) \notag \\
              & \quad \times \Bigl[ \mathds{G}_{\text{E}k} (\mathbf{r}, \mathbf{r}_\alpha) \boldsymbol{\alpha}_{kl}^{\alpha \beta} \mathds{G}_{l\text{E}}^\text{s} (\mathbf{r}_\beta, \mathbf{r}') \mathds{G}_{n\text{E}}^{\text{s} \dagger} (\mathbf{r}_\delta, \mathbf{r}') \boldsymbol{\alpha}_{mn}^{\gamma \delta \dagger} \mathds{G}_{\text{H}m}^\dagger (\mathbf{r}, \mathbf{r}_\gamma) \Bigr]_{ij}.
\end{align}
Analytically simplifying the integrals as far as possible, we obtain
\begin{align}
P_\text{abs} & = \left(n_\text{b} - n_\text{s} \right) \hbar \omega k_0^3 \Im \Tr \biggl( \underline{\underline{\boldsymbol{\alpha}}}_\text{EE} \underline{\underline{\mathds{R}}}_\text{E} + \frac{\epsilon_0}{\mu_0} \underline{\underline{\boldsymbol{\alpha}}}_\text{HH} \underline{\underline{\mathds{R}}}_\text{H} + 2 \sqrt{\frac{\varepsilon_0}{\mu_0}} \underline{\underline{\boldsymbol{\alpha}}}_\text{HE} \underline{\underline{\mathds{R}}}_\text{c} \biggr)
\end{align}
and
\begin{align}
  P_\text{scat} & = \frac{1}{4} \left(n_\text{s} - n_\text{b} \right) \hbar \omega k_0^6 \Re \Tr \biggl( \biggl[ \underline{\underline{\boldsymbol{\alpha}}}_\text{EE} \underline{\underline{\mathbf{\Gamma}}}_\text{E} \underline{\underline{\boldsymbol{\alpha}}}_\text{EE}^\dagger + \frac{\epsilon_0}{\mu_0} \underline{\underline{\boldsymbol{\alpha}}}_\text{EH} \underline{\underline{\mathbf{\Gamma}}}_\text{H} \underline{\underline{\boldsymbol{\alpha}}}_\text{EH}^\dagger \notag \\
              & \quad + 2 \sqrt{\frac{\varepsilon_0}{\mu_0}} \underline{\underline{\boldsymbol{\alpha}}}_\text{EE} \underline{\underline{\mathbf{\Gamma}}}_\text{c} \underline{\underline{\boldsymbol{\alpha}}}_\text{EH}^\dagger \biggr] \underline{\underline{\mathds{I}}}_\text{E} + \frac{\varepsilon_0}{\mu_0} \biggl[ \underline{\underline{\boldsymbol{\alpha}}}_\text{HE} \underline{\underline{\mathbf{\Gamma}}}_\text{E} \underline{\underline{\boldsymbol{\alpha}}}_\text{HE}^\dagger + \frac{\epsilon_0}{\mu_0} \underline{\underline{\boldsymbol{\alpha}}}_\text{HH} \underline{\underline{\mathbf{\Gamma}}}_\text{H} \underline{\underline{\boldsymbol{\alpha}}}_\text{HH}^\dagger \notag \\
             & \quad + 2 \sqrt{\frac{\varepsilon_0}{\mu_0}} \underline{\underline{\boldsymbol{\alpha}}}_\text{HE} \underline{\underline{\mathbf{\Gamma}}}_\text{c}^\dagger \underline{\underline{\boldsymbol{\alpha}}}_\text{HH}^\dagger \biggr] \underline{\underline{\mathds{I}}}_\text{H} + 2 \sqrt{\frac{\varepsilon_0}{\mu_0}} \biggl[ \underline{\underline{\boldsymbol{\alpha}}}_\text{EE} \left( \underline{\underline{\mathbf{\Gamma}}}_\text{E} + \underline{\underline{\mathds{K}}}_\text{E} \right) \underline{\underline{\boldsymbol{\alpha}}}_\text{HE}^\dagger + \frac{\epsilon_0}{\mu_0} \underline{\underline{\boldsymbol{\alpha}}}_\text{EH} \left( \underline{\underline{\mathbf{\Gamma}}}_\text{H} + \underline{\underline{\mathds{K}}}_\text{H} \right) \underline{\underline{\boldsymbol{\alpha}}}_\text{HH}^\dagger \notag \\
             & \quad + \sqrt{\frac{\varepsilon_0}{\mu_0}} \underline{\underline{\boldsymbol{\alpha}}}_\text{EE} \left( \underline{\underline{\mathbf{\Gamma}}}_\text{c} + \underline{\underline{\mathds{K}}}_\text{c} \right) \underline{\underline{\boldsymbol{\alpha}}}_\text{HH}^\dagger + \sqrt{\frac{\varepsilon_0}{\mu_0}} \underline{\underline{\boldsymbol{\alpha}}}_\text{EH} \left( \underline{\underline{\mathbf{\Gamma}}}_\text{c}^\dagger - \underline{\underline{\mathds{K}}}_\text{c}^\dagger \right) \underline{\underline{\boldsymbol{\alpha}}}_\text{HE}^\dagger \biggr] \underline{\underline{\mathds{I}}}_\text{c} \biggr).
\end{align}
The integrals $\mathds{R}$ and $\mathbf{\Gamma}$ are listed in appendix~\ref{appB}. 

The total emitted spectral power is 
\begin{equation}
  P_{\rm tot} = P_{\rm dir} + P_{\rm abs} + P_{\rm s} + P_{\rm scat}
\end{equation}
where $P_{\rm dir}$ gives the direct thermal emission of the nanoparticles and $P_\text{abs}$ describes heat flux between substrate and background that is absorbed in the nanoparticles which is encoded by the imaginary part. This power can become negative due to the choice of temperatures in the Bose-Einstein factors which is opposite to the one in $P_\text{scat}$. The latter describes the power transferred to the background by being scattered at the nanoparticles. The diagonal entries of the $\Gamma$-integrals indicating identical particle indices coincide with the integrals used to describe the heat flux between a small sphere and a semi-infinite half space without multiple reflections \cite{Herz4, KruegerEtAlTraceFormulas2012}. Note, that in principle $P_\text{abs}$ and $P_\text{s}$ need to be considered together, because $P_\text{abs}$ corresponds to the power which is absorbed from $P_\text{s}$ due to the presence of the nanoparticles when $T_\text{s} > T_\text{b}$. Hence, when choosing $A$ equal to the cross section of all nanoparticles the sum of $P_\text{abs}$ and $P_\text{s}$ would be positive even though $P_\text{abs}$ itself would be negative if the dipole approximation is fulfilled.

\section{Numerical Results}

\subsection{A single nanoparticle}\label{Sec:SingleParticle}

To check for plausibility, we want to retrieve the result in \cite{Herz3} for the single nanoparticle case $N=1$. Then, all $3 N \times 3N$ block matrices reduce to ordinary $3 \times 3$ matrices corresponding to the $11$-entry of the former block matrix. For example, now, the Green's functions are fully described by the volume averages. This yields the dressed polarizabilities ($k = \text{E}, \text{H}$)
\begin{align}
  \boldsymbol{\alpha}_{kk} & = \frac{V_p \chi_k [\mathbf{e}_x \otimes \mathbf{e}_x + \mathbf{e}_y \otimes \mathbf{e}_y]}{1 - k_0^2 V_p \chi_k \expval{G_{kk,\perp}} + \frac{k_0^4 V_p^2 \chi_\text{E} \chi_\text{H} \expval{G_\text{HE}}^2}{1 - k_0^2 V_p \chi_{\bar{k}} \expval{G_{\bar{kk},\perp}}}} + \frac{V_p \chi_k \mathbf{e}_z \otimes \mathbf{e}_z}{1 - k_0^2 V_p \chi_k \expval{G_{kk,z}}} , 
\label{eq:aEE_approx} \\
  \boldsymbol{\alpha}_\text{EH} & = \boldsymbol{\alpha}_\text{HE} = \frac{k_0^2 V_p^2 \chi_\text{E} \chi_\text{H} \expval{G_\text{HE}} [\mathbf{e}_x \otimes \mathbf{e}_y - \mathbf{e}_y \otimes \mathbf{e}_x]}{\left(1 - k_0^2 V_p \chi_\text{E} \expval{G_{\text{EE},\perp}}\right) \left(1 - k_0^2 V_p \chi_\text{H} \expval{G_{\text{HH},\perp}} \right) + k_0^4 V_p^2 \chi_\text{E} \chi_\text{H} \expval{G_\text{HE}}^2}
\label{eq:aEH_approx}
\end{align}
which coincide with the dressed polarizabilities in \cite{Herz3}. The matrices $\mathds{F}$, $\mathds{E}$, and $\mathds{N}$ follow the structure of Eq. \eqref{eq:aEH_approx}, meaning that the matrices only possess a xy and yx entry. The matrices $\boldsymbol{\chi}$ and $\mathds{M}$, however, look like Eq. \eqref{eq:aEE_approx} having three entries with two identical ones accounting for one direction parallel and two directions perpendicular to the substrate's surface normal. Then, we obtain the identical results as in Ref.~\cite{Herz3} for the three contributions 
\begin{align}
  P_\text{dir} & = \hbar \omega k_0 \Re \biggl( \sum_{j \in \{\perp, z \}} [M_{\text{E},j} I_{\text{E},j} + \frac{\epsilon_0}{\mu_0} M_{\text{H},j} I_{\text{H},j}] + 2 \sqrt{\frac{\varepsilon_0}{\mu_0}} N_\text{c} I_\text{c} \biggr) , 
\label{eq:Pdir_snp} \\
  P_\text{abs} & = \left(n_\text{b} - n_\text{s} \right) \hbar \omega k_0^3 \Im \biggl( \sum_{j \in \{\perp, z \}} [\alpha_{\text{EE},j} R_{\text{E},j} + \frac{\epsilon_0}{\mu_0} \alpha_{\text{HH},j} R_{\text{H},j}] + 2 \sqrt{\frac{\varepsilon_0}{\mu_0}} \alpha_\text{HE} R_\text{c} \biggr),
\end{align}
and 
\begin{align}
  P_\text{scat} & = \frac{1}{8} \left(n_\text{s} - n_\text{b} \right) \hbar \omega k_0^6 \biggl( \biggl(|\alpha_{\text{EE},\perp}|^2 \Gamma_{\text{E},\perp} + \frac{\epsilon_0}{\mu_0} |\alpha_\text{HE}|^2 \Gamma_{\text{H},\perp} + 2 \sqrt{\frac{\varepsilon_0}{\mu_0}} \Re \left(\alpha_{\text{EE},\perp} \alpha_\text{HE}^{*} \Gamma_\text{c}\right) \biggr) I_{\text{E},\perp} \notag \\
                & \quad + 2 |\alpha_{\text{EE},z}|^2 \Gamma_{\text{E},z} I_{\text{E},z} + \frac{\varepsilon_0}{\mu_0} \biggl[ \biggl(|\alpha_\text{HE}|^2 \Gamma_{\text{E},\perp} + \frac{\epsilon_0}{\mu_0} |\alpha_{\text{HH},\perp}|^2 \Gamma_{\text{H},\perp} \notag \\
                & \quad + 2 \sqrt{\frac{\varepsilon_0}{\mu_0}} \Re \left(\alpha_\text{HE} \alpha_{\text{HH},\perp}^{*} \Gamma_\text{c}^{*} \right) \biggr) I_{\text{H},\perp} + 2 \frac{\epsilon_0}{\mu_0} |\alpha_{\text{HH},z}|^2 \Gamma_{\text{H},z} I_{\text{H},z} \biggr] \notag \\
                & \quad + 2 \sqrt{\frac{\varepsilon_0}{\mu_0}} \Re \biggl( \biggl[ - \alpha_{\text{EE},\perp} \alpha_\text{HE}^{*} \Gamma_{\text{E},\perp} + \frac{\epsilon_0}{\mu_0} \alpha_\text{HE} \alpha_{\text{HH},\perp}^{*} \Gamma_{\text{H},\perp} \notag \\
                & \quad + \sqrt{\frac{\varepsilon_0}{\mu_0}} \alpha_{\text{EE},\perp} \alpha_{\text{HH},\perp}^{*} \Gamma_\text{c} - \sqrt{\frac{\varepsilon_0}{\mu_0}} |\alpha_\text{HE}|^2 \Gamma_\text{c}^{*} \biggr] I_\text{c} \biggr) \biggr).
\end{align}
The expressions $M_{\text{E/H},\perp/\parallel}$ and $N_\text{c}$ as well as the different integral terms $\Gamma$ and $I$ can be found in the appendix~ \ref{appC}. These three terms describing the direct emission of the nanoparticle, the amount of thermal background radiation absorbed in the nanoparticle, and the scattered contribution were already discussed in detail in Ref.~\cite{Herz3}. Nonetheless, we want to compare them with other results from literature. 

For example, when considering only the electric dipole contribution in $P_\text{scat}$ we find the expression
\begin{equation}
    P_\text{scat} = \frac{1}{8} \left(n_\text{s} - n_\text{b} \right) \hbar \omega k_0^6 \Bigl( |\alpha_{\text{EE},\perp}|^2 \Gamma_{\text{E},\perp} I_{\text{E},\perp} + 2 |\alpha_{\text{EE},z}|^2 \Gamma_{\text{E},z} I_{\text{E},z} \Bigr).
\end{equation}
for the scattered power by an electric dipole. This expression does not fully coincide with the dipole expression given in Eq.~(7) Ref.~\cite{Edalatpour1}. This deviation can be traced back to an error in the definition of the TM-mode part of the Green's function in Eq.~(4) in Ref.~\cite{Edalatpour1}. To highlight the difference and consequences, we first explicitly write the expressions for the $I$ integrals and introduce the angle $\vartheta$ between the surface normal and the wave vector by $k_\perp = k_0 \sin(\vartheta)$ and $k_z = k_0 \cos(\vartheta)$. We then obtain
\begin{equation}
\begin{split}
	P_\text{scat} &= \frac{1}{8} \left(n_\text{s} - n_\text{b} \right) \hbar \omega k_0^6 \int_0^{\pi/2}\!\! {\rm d} \vartheta\, \sin(\vartheta)\Bigl( \\
		      &\qquad|\alpha_{\text{EE},\perp}|^2 \Gamma_{\text{E},\perp} \biggl[ \cos(\vartheta)^2 |1 - r_{\rm E} \re^{2 \ri k_0 \cos(\vartheta) d}|^2 + |1 + r_{\rm H} \re^{2 \ri k_0 \cos(\vartheta) d}|^2  \biggr] \\
		      & \qquad+ 2 |\alpha_{\text{EE},z}|^2 \Gamma_{\text{E},z} \sin(\vartheta)^2 |1 + r_{\rm E} \re^{2 \ri k_0 \cos(\vartheta) d}|^2\Bigr)
\end{split}
\label{Eq:DipolScat}
\end{equation}
which is the same result as in Eq.~(38) in Ref.~\cite{Joulain2}. Note, that the TE-mode terms with $ r_{\rm E}$ have a different sign. In  Ref.~\cite{Edalatpour1} the TM-mode contribution of the scattered part of the Green's function contains the wrong combination of polarization vectors. Using this definition one would obtain the same sign in front of the $r_{\rm E}$ terms. Consequently, the scattered intensity given in Eq.~(7) in Ref.~\cite{Edalatpour1} is 
\begin{equation}
\begin{split}
	\langle |\mathbf{E}^{\rm FF}|^2 \rangle &=\frac{k_0^4}{16 \pi^2 r^2} \biggl( \bigl[ |\alpha_{\text{EE},\perp}|^2  \cos(\vartheta)^2 \langle E_x^2 \rangle  +  |\alpha_{\text{EE},z}|^2  \sin(\vartheta)^2 \langle E_z^2 \rangle \bigr] |1 + r_{\rm E} \re^{2 \ri k_0 \cos(\vartheta) d}|^2 \\
	 & \qquad + |\alpha_{\text{EE},z}|^2 \langle E_x^2 \rangle |1 + r_{\rm H} \re^{2 \ri k_0 \cos(\vartheta) d}|^2\biggr)
\end{split}
\label{Eq:DipolScatWrong}
\end{equation}
where the energy densities $\langle E_x^2 \rangle$ and $\langle E_z^2 \rangle$ of the substrate at the position of the dipole moment are connected to our $\Gamma$ integrals such that $\epsilon_0 \langle E_x^2 \rangle = \hbar \omega (n_s - n_b) \Gamma_{\text{E},\perp} k_0^2 / c$ and $\epsilon_0 \langle E_z^2 \rangle = 2 \hbar \omega (n_s - n_b) \Gamma_{\text{E},z} k_0^2 / c$ where in Ref.~\cite{Edalatpour1} $n_b = 0$. A correct use of the the polarization vectors in the Green's function would give 
\begin{equation}
\begin{split}
	\langle |\mathbf{E}^{\rm FF}|^2 \rangle &= \frac{k_0^4}{16 \pi^2 r^2} \biggl( |\alpha_{\text{EE},\perp}|^2  \cos(\vartheta)^2 \langle E_x^2 \rangle |1 - r_{\rm E} \re^{2 \ri k_0 \cos(\vartheta) d}|^2\\
	                              &\qquad +  |\alpha_{\text{EE},z}|^2  \sin(\vartheta)^2 \langle E_z^2 \rangle |1 + r_{\rm E} \re^{2 \ri k_0 \cos(\vartheta) d}|^2 \\
	 & \qquad + |\alpha_{\text{EE},z}|^2 \langle E_x^2 \rangle |1 + r_{\rm H} \re^{2 \ri k_0 \cos(\vartheta) d}|^2 \biggr).
\end{split}
\label{Eq:DipolScatcorrect}
\end{equation}
This expression coincides with the integrand of $P_\text{scat}$ in Eq.~(\ref{Eq:DipolScat}).

In the even simpler case without substrate, we are left with the expressions for the thermal emission of a nanoparticle at temperature $T_p$ in a vacuum filled with blackbody radiation at temperature $T_b = T_s$. We obtain 
\begin{equation}
  P_\text{dir} = \frac{\hbar \omega k_0^3 V_p \left( n_p - n_b \right) \Im(\chi_\text{E})}{\pi \Big|1 - k_0^2 V_p \chi_\text{E} \expval{G_{\text{EE},\text{vac}}} \Big|^2} + \frac{\epsilon_0}{\mu_0} \frac{\hbar \omega k_0^3 V_p \left( n_p - n_b \right) \Im(\chi_\text{H})}{\pi \Big|1 - k_0^2 V_p \chi_\text{H} \expval{G_{\text{HH},\text{vac}}} \Big|^2} .
\label{eq:Pdir1} 
\end{equation}
We can compare Eq. \eqref{eq:Pdir1} with the known Mie expressions of the thermal radiation of a spherical particle derived by Kattawar and Eisner~\cite{Eisner}. Taking only the contributions of the electric and magnetic dipolar terms in the limit $k_0 R \ll 1$ and $k_0 |\varepsilon| R \ll 1$ the Mie theory yields~\cite{Eisner}
\begin{equation}
\begin{split}
	P_\text{dir}^{\rm Mie} &= \hbar \omega \left(n_p - n_b \right) 12 \bigl[ \Re(a_1) - |a_1|^2  + \Re(b_1) - |b_1|^2\bigr] \\
	                       &\approx 24 k_0^3 R^3 \hbar \omega \left(n_p - n_b \right) \frac{\Im(\varepsilon)}{|\varepsilon + 2|^2} + \frac{4}{15} k_0^5 R^5 \left( n_p - n_b \right) \Im(\varepsilon). 
\end{split}
\label{Eq:PdirMie}
\end{equation}
Hence, $P_\text{dir}^{\rm Mie} = 2 P_\text{dir}$ when choosing
\begin{align}
  \chi_\text{E} & = \varepsilon - 1, \\
  \chi_\text{H} & = \frac{1}{10} k_0^2 R^2 (\varepsilon - 1)\frac{\mu_0}{\epsilon_0} 
\end{align}
together with the weak form of coupled dipole moments $\expval{G_{\text{EE},\text{vac}}} \approx \epsilon_0/\mu_0 \expval{G_{\text{HH},\text{vac}}} \approx - 1/(3 k_0^2 V)$ coinciding with S-CDM for $k_0R \ll 1$. Of course, since we consider only the thermal radiation through a single plane ``above'' the nanoparticle and neglect the contribution ``below'', we have $P_\text{dir} = P_\text{dir}^{\rm Mie}/2$. We note that the electrical part of $P_\text{dir}$ also coincides with the corresponding result for the radiation of a small sphere as derived in Ref's.~\cite{KruegerEtAlTraceFormulas2012,Herz}. 

\begin{figure}
	\centering
	\includegraphics[width=0.45\textwidth]{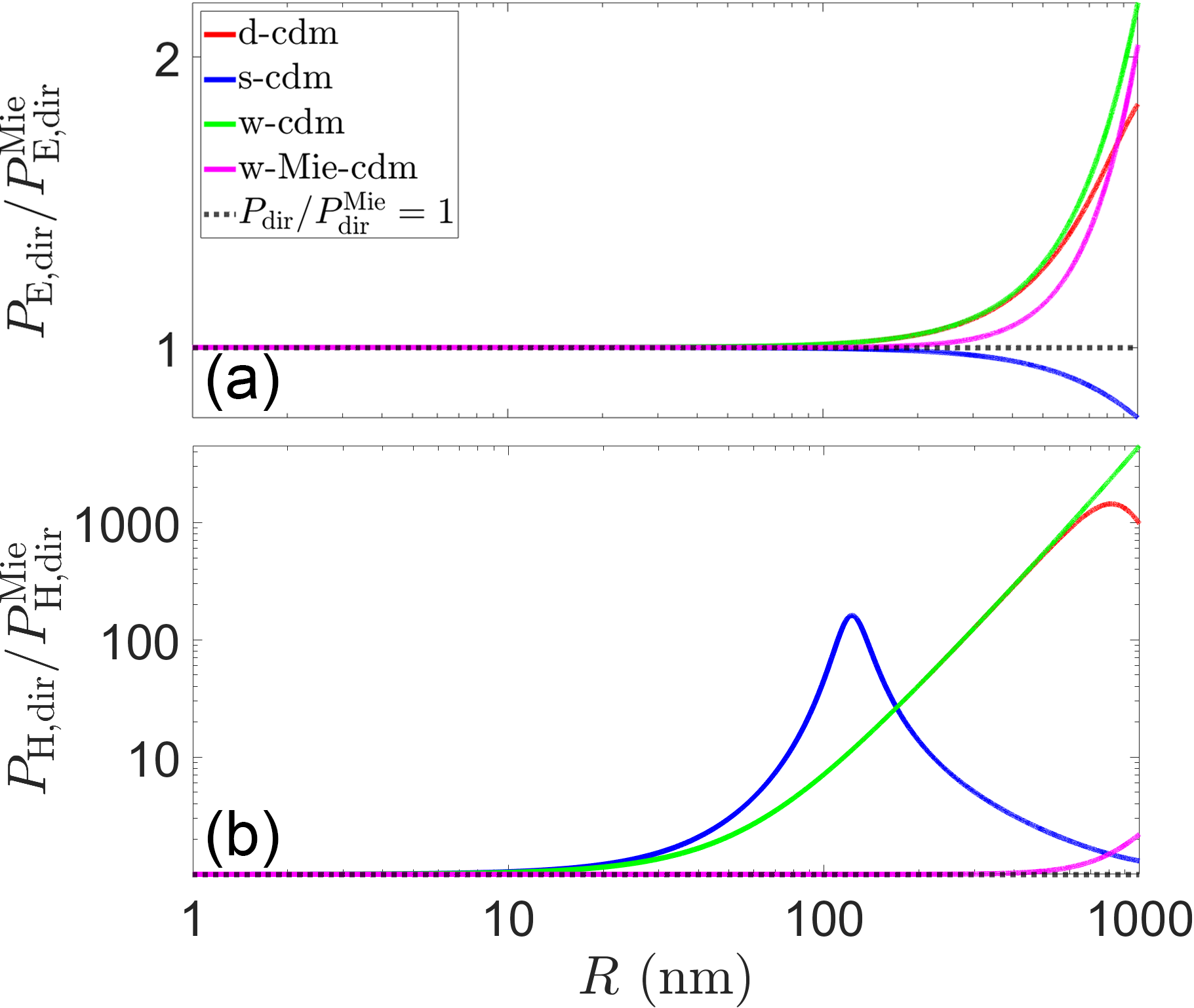}
	\caption{$P_{\rm dir}$ normalized to the exact result $P_{\rm dir}^{\rm Mie}$ in Ref.~\cite{Eisner} for a (a) SiC nanoparticle at the localized surface mode frequency $\omega = 1.787\times10^{14}\,{\rm rad}/s$ and (b) Ag nanoparticle at $\omega = 1.0\times10^{14}\,{\rm rad}/s$.}
	\label{Fig:MieComparision}
\end{figure}

As discussed before, the same result can be obtained when implementing the weak form by setting $\expval{G_{\text{EE},\text{vac}}} = \expval{G_{\text{HH},\text{vac}}} = 0$ and using the expressions for $\chi_{\rm E/H}$ from Eqs.~(\ref{ChiEappr}) and (\ref{ChiHappr}). In the same manner but by using $\chi_{\rm E/H}$ from Eqs.~(\ref{ChiE}) and (\ref{ChiH}) we obtain the weak Mie form. By comparison with Eq.~(\ref{Eq:PdirMie}) it becomes clear that the weak Mie form only covers the term $\Re(a_1) + \Re(b_1)$ and therefore is a linear approximation in the Mie coefficients~\cite{gold3}. Assuming $k_0 R \ll 1$ D-CDM behaves alike. In Fig.~\ref{Fig:MieComparision} we show the result of $P_{\rm dir}$ for the different forms of the dressed polarizability including the weak Mie form normalized to the exact Mie result $P_\text{dir}^{\rm Mie}$ in Ref.~\cite{Eisner} for the dipolar contribution for a SiC and an Ag nanoparticle. It can be seen that for SiC all different forms are equally good for $R < 100\,{\rm nm}$ which is smaller than the skin depth $d_{\rm s} = 1/2 k_0 \Im(\sqrt{\epsilon_{\rm SiC}}) \approx 830$ nm at the resonance frequency $\omega_{\rm SPhP}$ of the localized surface modes defined by $\Re\bigl(\epsilon_{\rm SiC}(\omega_{\rm SPhP})\bigr) = - 1$ and $\Im\bigl(\epsilon_{\rm SiC}(\omega_{\rm SPhP})\bigr) \ll 1$, i.e.\ by the poles of $r_{\rm E}$. For Ag the differences are much larger. The D-CDM and W-CDM are mostly the same, and they provide an equally good approximation for $R < 10\,{\rm nm}$ as the S-CDM. This could be expected because the skin depth is this time $d_{\rm s} \approx 11\,{\rm nm}$ at the chosen frequency. Interestingly, the W-Mie-CDM is very accurate even for radii over several hundreds nanometers in this case. Hence, in the numerical calculations we can safely choose nanoparticle radii up to 100 nm when using the W-Mie-CDM method, whereas for the other methods we are restricted to Ag nanoparticles with $R \leq 10\,{\rm nm}$ but for SiC nanoparticles also radii up to 100 nm can be used.

\subsection{Four nanoparticles}\label{Sec:4NP}

\begin{figure}
	\centering
	\includegraphics[width=0.8\textwidth]{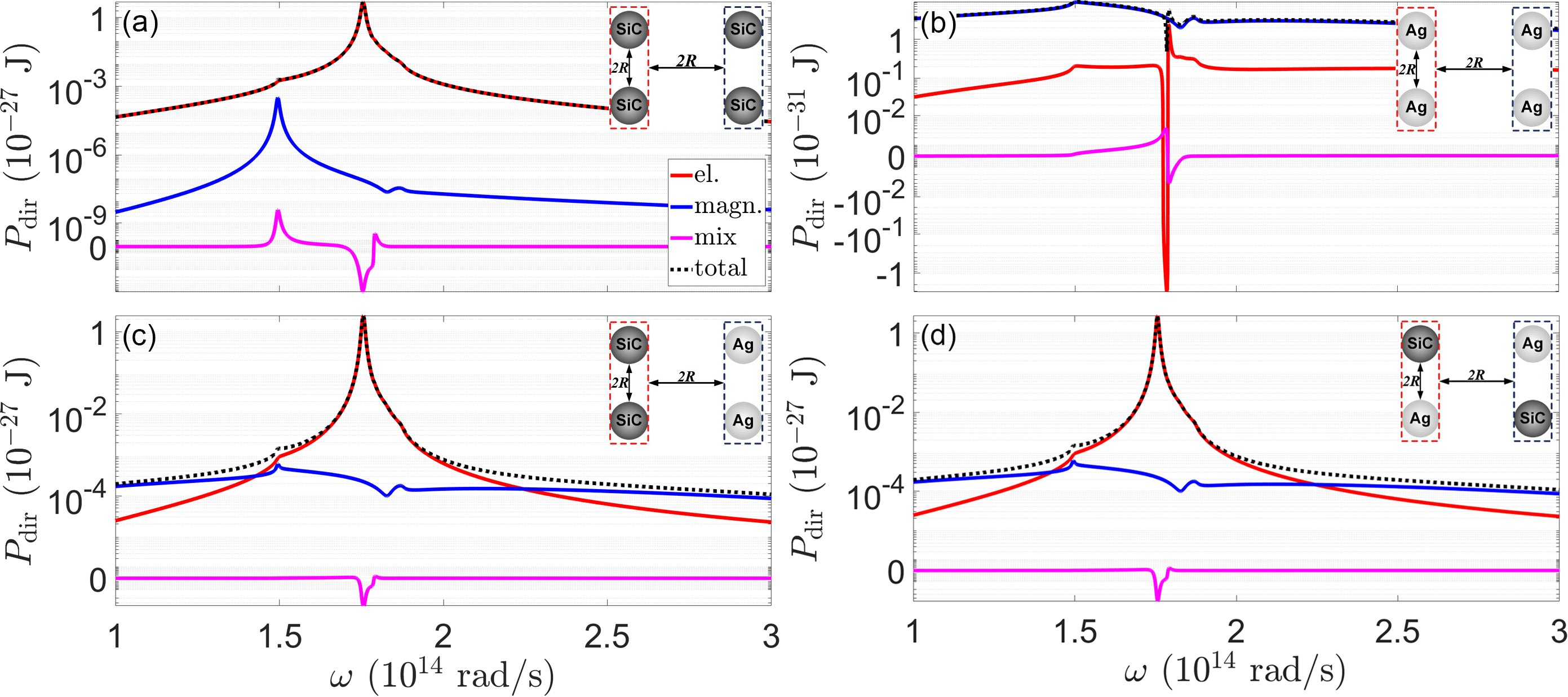}
	\caption{$P_{\rm dir}$ and its compound for four nanoparticles made of SiC or Ag of radius $R = 10\,{\rm nm}$ positioned at the corners of a square (as depicted in the inset) which is in a plane parallel to a SiC substrate at the distance $d = 52\,{\rm nm}$ (substrate - nanoparticle's center). The temperature of the two nanoparticles on the left is $350\,{\rm K}$ and those of the two particles on the right is $310\,{\rm K}$, the background and substrate have the same temperature $T_{\rm s} = T_{\rm b} = 300\,{\rm K}$.}
	\label{Fig:4NanoparticlesSiC}
\end{figure}

\begin{figure}
	\centering
	\includegraphics[width=0.8\textwidth]{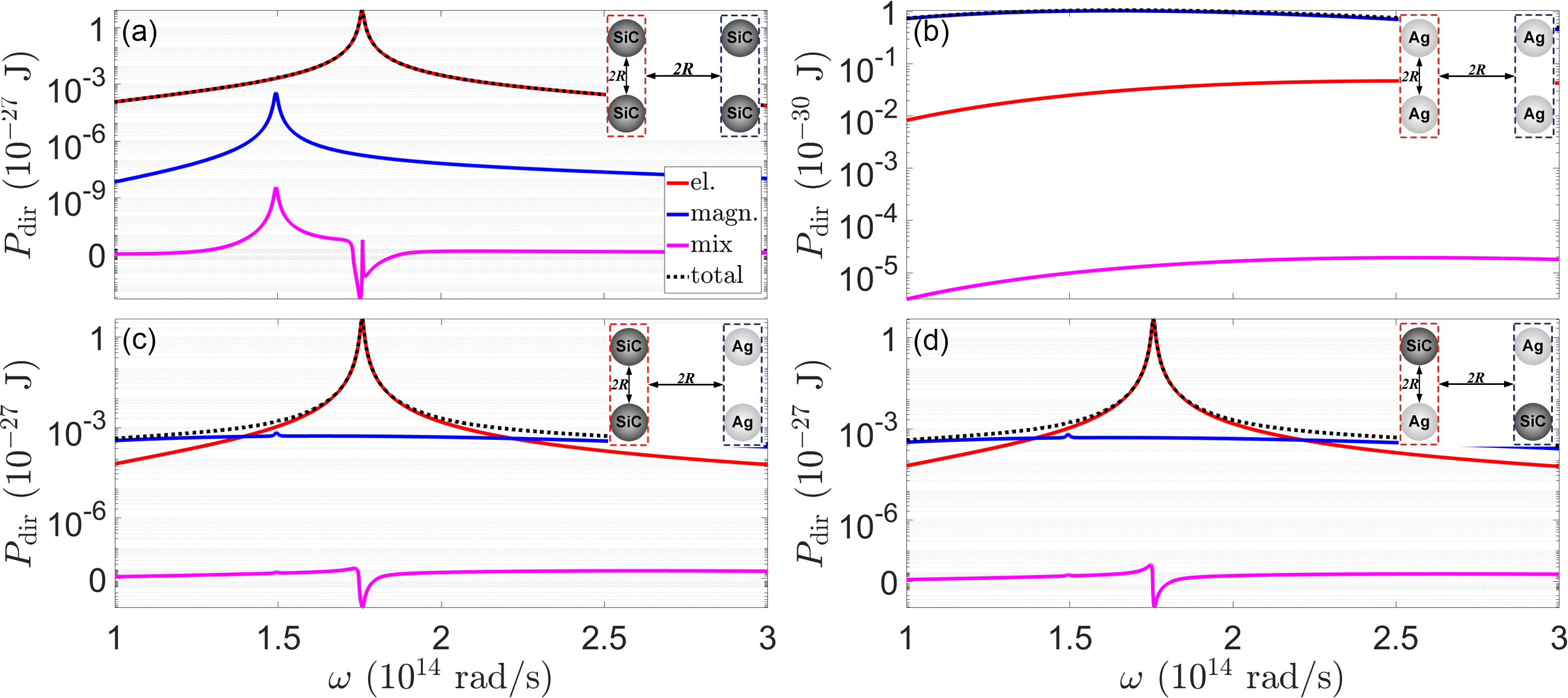}
	\caption{$P_{\rm dir}$ like in Fig.~\ref{Fig:4NanoparticlesSiC} but for a silver substrate.}
	\label{Fig:4NanoparticlesAg}
\end{figure}

Inspired by the work of Dong {\itshape et el.}~\cite{Dong} we consider four nanoparticles made of SiC or Ag in different material combinations above a SiC or Ag substrate. These nanoparticles are positioned on the corners of a square which is parallel to the substrate interface and has a side length of $4 R$. The radius of the nanoparticles is chosen to be $R = 10\,{\rm nm}$ so that we can safely use the dipole model for the metal nanoparticles using S-CDM. For convenience, we choose the same temperature for the background and the substrate which are $T_{\rm s} = T_{\rm b} = 300\,{\rm K}$ so that we only need to consider $P_{\rm dir}$ as we did for the single particle case discussed in the previous section. The two nanoparticles on the left side as depicted in the insets of Fig.~\ref{Fig:4NanoparticlesSiC} and \ref{Fig:4NanoparticlesAg} have a temperature of $350\,{\rm K}$ and the two on the right side have a temperature of $310\,{\rm K}$. Note, that, due to the different temperatures of the nanoparticles and the environment, the thermal emission does not correspond to the absorption of the whole system so that this situation cannot be treated by simply calculating the absorptivity of the nanoparticles above the substrate. To display negative values in those semi-logarithmic plots, too, we adopt the method outlined by Webber \cite{Webber}.

First of all, the numerical results in Figs.~\ref{Fig:4NanoparticlesSiC} and \ref{Fig:4NanoparticlesAg} show the known result for heat transfer between nanoparticles, namely that the electric part of $P_{\rm dir}$ dominates the thermal emission of the dielectric nanoparticles and the magnetic part of $P_{\rm dir}$ prevails in the case of metallic nanoparticles. The mixed terms, however, can be neglected in all cases. In general, we observe a strong peak at the nanoparticles localized surface phonon polariton resonance at $\omega_{\rm LPhP} = 1.78\times10^{14}\,{\rm rad/s}$ and a secondary peak at the transversal optical phonon frequency $\omega_{\rm TO} = 1.495\times10^{14}\,{\rm rad/s}$ for the SiC components. The latter frequency is mostly important for the magnetic contributions. For Ag nanoparticles above an Ag substrate no features are expected since Ag resonates in the ultraviolet as can be seen in Fig.~\ref{Fig:4NanoparticlesAg}(b) so that when involving Ag the features are due to the SiC surface or the other SiC nanoparticles. In general, the spectra resemble those of Dong {\itshape et al.}~\cite{Dong} for the heat transfer between the two particles on the left and the ones on the right. Without the substrate the resemblance would be even greater.

In the case of four Ag nanoparticles above a SiC substrate we find in Fig.~\ref{Fig:4NanoparticlesSiC}(b) reduction of thermal emission at the surface phonon polariton frequency $\omega_{\rm SPhP} = 1.787\times10^{14}\,{\rm rad/s}$ of the substrate. This shows the impact of the strong coupling of the nanoparticles to the surface phonon polaritons of the substrate which decrease the thermal emission in the far-field because the Ag nanoparticles emit their heat preferentially into the SiC substrate at the surface mode resonance. Our numerical results indicate that for smaller distances the nanoparticles and the substrate $P_{\rm dir}$ can even become negative for metallic nanoparticles above a SiC substrate at the surface phonon polariton resonance. We checked that this happens for all methods W-CDM, S-CDM, and D-CDM. Additionally, the distance at which $P_{\rm dir}$ becomes negative depends on the lateral distance between the nanoparticles. If they touch each other ($d_l = 0$) $P_{\rm dir}$ becomes negative at $d=43$ nm. For $d_l = 4R$ this distances grows up to $d=53$ and, eventually, decreases for larger lateral distances. Since $P_{\rm dir}$ should be positive for all frequencies because $T_{\rm p} > T_{\rm s} = T_{\rm b}$ we are convinced that this hints to the fact that the dipole model can be safely used for center-center distance between particles and for center-to-surface distance larger than approximately $5R$, only.

\subsection{Discrete dipole approximation - spherical nanoparticle}

Here, we want to apply our general theory to perform DDA of a macroscopic object radiating heat into the far-field. Within this approximation method the macroscopic object is replaced by a number of small polarizable subvolumes, so called voxels, which are treated with the formalism of the coupled dipole method. The idea behind this approach is that, by replacing any arbitrarily shaped macroscopic object by a sufficiently great number of subvolumes or voxels (in the ideal case an infinite number of infinitely small voxels), the electrodynamical response of the macroscopic object is described exactly. This allows for treating electrodynamic problems of, in principle, arbitrarily shaped objects~\cite{Yurkin1}. In the context of fluctuational electrodynamics this DDA method has been introduced to describe the thermal heat exchange between two macroscopic objects~\cite{EdalatpourDDA} by applying the corresponding CDM formalism first introduced in Ref.~\cite{pbaetal}. An extension to treat also the heat transfer between anisotropic and non-reciprocal objects can be found in Ref.~\cite{Ekeroth}. Furthermore, in Ref.~\cite{Edalatpour2} the DDA method for treating radiative heat exchange between an arbitrarily shaped object and a planar sample was introduced and, finally, in Ref.~\cite{Edalatpour1} the same group used the DDA method to determine the near-field of a planar sample scattered into the far-field by an arbitrarily shaped object. Thermal emission of an arbitrarily shaped object into free space, i.e. without a sample, was introduced in Ref.~\cite{Ekeroth}. When using DDA, typically all voxels of a single object are considered to have a single temperature $T_{\rm p}$. However, in principle also temperature profiles and multiple objects can be modelled.

Our here introduced method is a generalization of the two latter methods in Refs.~\cite{Ekeroth,Edalatpour1}. In Ref.~\cite{Edalatpour1}, the intensity of the scattered near-field by a single object treated within DDA is considered, only. Therefore, since no other temperatures but the temperature of the substrate $T_{\rm s} > 0\,{\rm K}$ are involved in that approach the results correspond to our expression $P_{\rm scat}$ for the scattered power assuming that  $T_{\rm b} = T_{\rm p} = 0\,{\rm K}$. But of course, even for this choice of temperature the substrate itself will directly emit heat which will be partially absorbed by the scatterer which is described by the terms $P_{\rm s}$ and $P_{\rm abs}$ in our theory. On the other hand, the approach in Ref.~\cite{Ekeroth} only considers thermal emission of a single object with temperature $T_{\rm p}$ into vacuum with temperature $T_{\rm b}$. By replacing the vacuum Green's function in the expressions in Ref.~\cite{Ekeroth} by that of a planar sample, as done in Ref.~\cite{OttSAB2020}, and integrating the mean Poynting vector over the corresponding detection plane, this method would give $P_{\rm dir}$ for the case that $T_s = T_{\rm b}$. In our approach, we do not only have access to $P_{\rm dir}$ for different temperatures of the substrate and the background but also to all other heat flux contributions which shows that our approach generalizes those previous works.

\begin{figure}
	\centering
	\includegraphics[width=0.45\textwidth]{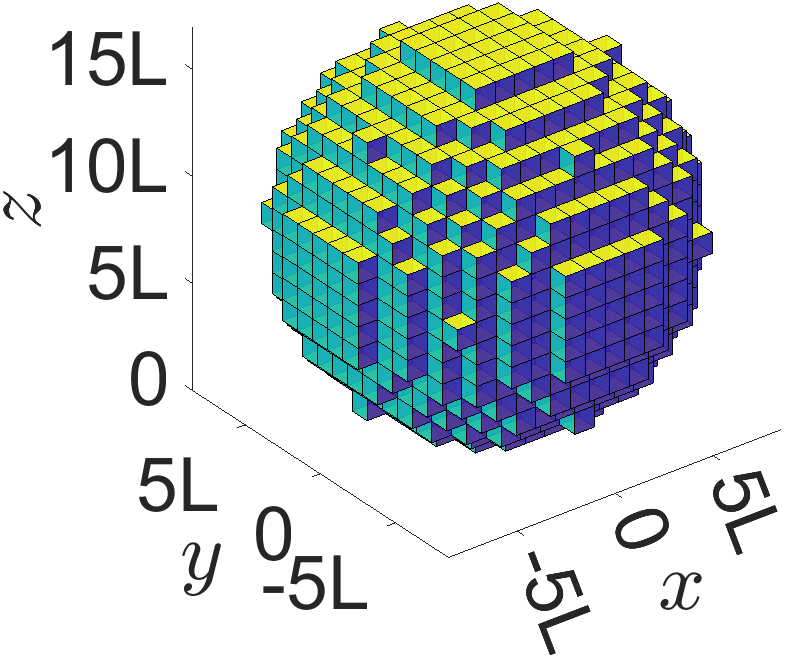}
	\caption{Sketch of the sphere used in our calculations approximated by $2533$ voxels. We scale side length $L$ in such a way that our fixed number of voxels fits the radius of the sphere.}
	\label{Fig:SphereDDA}
\end{figure}

We find that our DDA formalism works for materials where the electrical part dominates, only. That means that it cannot be applied to macroscopic metallic objects where the absorption or thermal emission is dominated by the contribution of eddy currents. Note that this does not mean that DDA is not able to describe for example metals in the optical regime. In that case, the response would, again, be dominated by the electrical fields. Mathematically, we find that the lack of resonance of the effective magnetic polarizability in the infrared regime leads to a negligible coupling between the voxels for the magnetic dipolar moments. Therefore, the overall response of the object, being replaced by a huge number of voxels, is the sum of the response of the individual voxels and it cannot be used to describe the response of the macroscopic object. That means that the contribution of eddy currents to the thermal emission of macroscopic objects cannot be treated within a DDA approximation using magnetic dipole moments. We will substantiate this claim at the end of this section. Hence, we use in the following only the electrical part of our theory to carry out the DDA calculations. As a first illustration we consider the thermal emission of a single spherical particle within the DDA as shown in Fig.~\ref{Fig:SphereDDA} in a vacuum environment and close to a substrate. 

\begin{figure}
	\centering
	\includegraphics[width=0.8\textwidth]{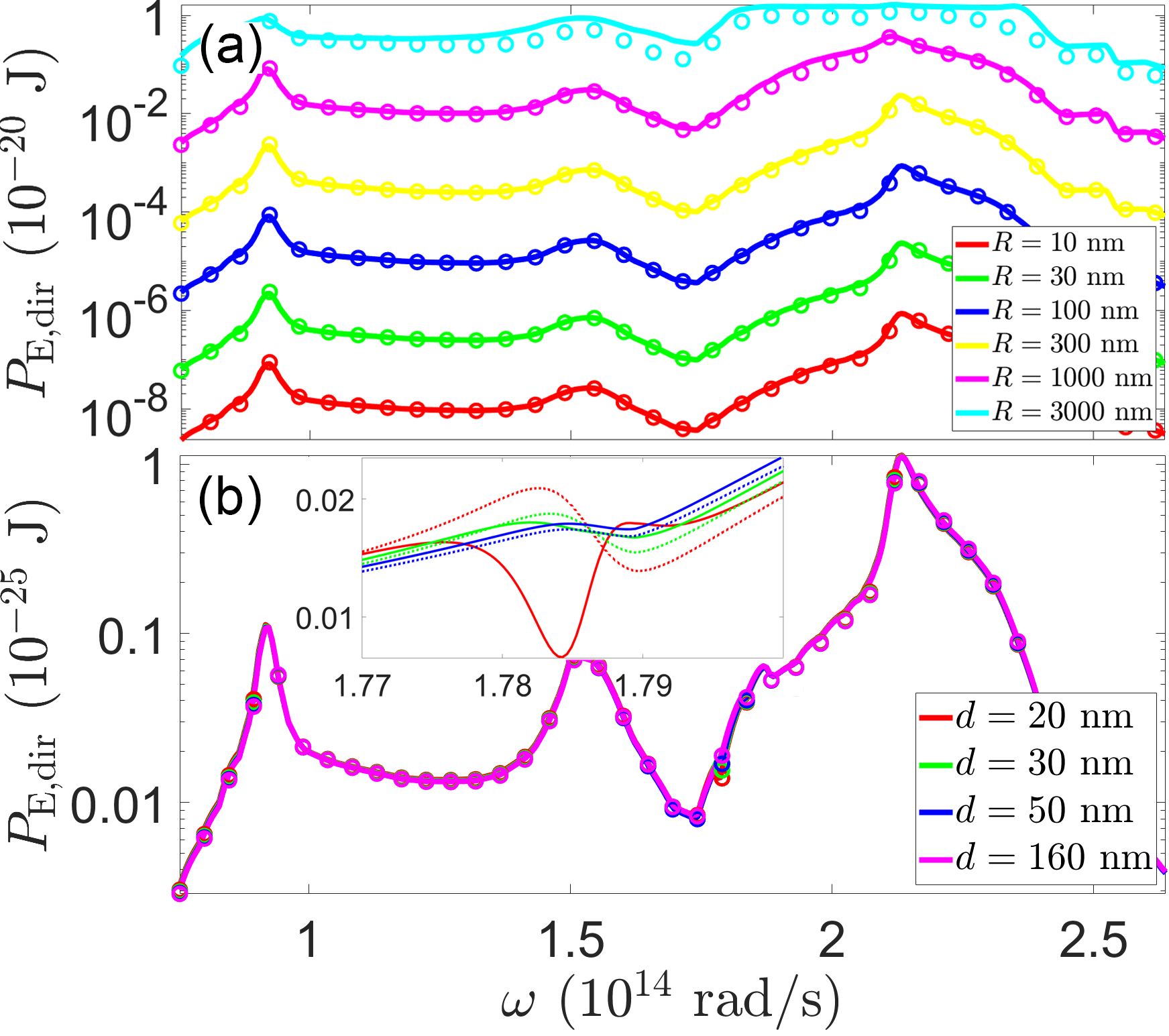}
	\caption{$P_{\rm E,dir}$ for a SiO$_2$ sphere at $T_{\rm p} = 700\,{\rm K}$ in a vacuum background with $T_{\rm b} = 300\,{\rm K}$ for S-CDM. (a) depicts the SiO$_2$ spectrum for different radii using a DDA model for the sphere with $N = 2553$ voxels (solid lines) and the full Mie result (open symbols). (b) DDA-spectra when adding a SiC substrate at $T_{\rm s} = 300K$ for different distances $d$ between substrate and sphere (solid lines) compared to the single particle solution (open symbols; dashed lines in the inset). The inset magnifies the spectra within the reststrahlen band of SiC.}
	\label{Fig:SphereMie}
\end{figure}

We have already discussed in Sec.~\ref{Sec:SingleParticle} that the dipole model provides reliable results for any choice of dressed polarizability for particle radii smaller than the skin depth of the material. Here, we will use S-CDM as in the DDA calculations in Ref.~\cite{Edalatpour1}, because we want to discuss specific results form that work in the next section. In Fig.~\ref{Fig:SphereMie}(a) we show the results of $P_{\rm E,dir}$ for spherical SiO$_2$ nanoparticles of different radii using the DDA with $N = 2553$ voxels compared with the full Mie results of Katawar and Eisner~\cite{Eisner}. It can be seen that the DDA results are in good agreement with the exact results as long as the radius is small enough -- in our case $R \leq 1$ µm. This was also found in Ref.~\cite{Ekeroth} using D-CDM. By using more voxels, also the results for larger spherical particles can be improved~\cite{Edalatpour3} but the main features of the spectrum are well captured for all radii even with our relatively small number of voxels. On the other hand, it is interesting that even for small particles with only 30nm radius DDA shows a relative deviation of about 26\% at the surface phonon-polariton resonance frequencies of SiO$_2$ which is due to the ``slow'' convergence of DDA~\cite{Edalatpour3}, which means one can only increase the accuracy by using more and more voxels which results in longer computation times to obtain a certain accuracy level.  

Since our model also includes the possibility to study the impact of a substrate to the thermal emission, we show in Fig.~\ref{Fig:SphereMie}(b) our numerical results for a SiO$_2$ nanoparticle with radius $R = 30\,{\rm nm}$ above a SiC substrate. For the nanoparticle we choose the DDA as before for a single SiO$_2$ particle with $N = 2553$ voxels and compare it to the single-dipole result from section~\ref{Sec:SingleParticle}. This allows us to check the validity of the dipole approximation when the distance $d$ is smaller than $4R$. Furthermore, we choose $T_{\rm p} = 700\,{\rm K}$ and $T_{\rm s} = T_{\rm b} = 300\,{\rm K}$. We observe a slight decline in $P_{\rm E,dir}$ for increasing distances outside of the reststrahlen band of SiC. However, the DDA approximation is still in good agreement with the single nanoparticle result (open symbols). Note that $d$ describes the distance between substrate and sphere surface so that $z_p = d + R$ holds. In the inset we can see large deviations between DDA and single dipole model for distances $d \leq 20$ nm between substrate and sphere surface due to the SPhP mode of SiC at $1.787\times10^{14}$ rad/s. If we assume that DDA is giving the correct result then this means that the deviation stems from the shortcomings of the dipole model and that higher multipole orders become important at such distances as it is expected. On the other hand, because of the ``slow'' convergence of the DDA method further detailed studies on the convergence of the DDA method are required but they are out of the scope of this work.

As a side note, we considered SiC for the sphere's material as well. In that case we obtained poor agreement between DDA and single particle model for frequencies below $1.8\times10^{14}$ rad/s. Because both results coincide for frequencies above that frequency we trace this deviation back to large values of the complex permittivity of SiC which is in agreement with fact that the complex refractive index has significant impact on the accuracy of DDA results as shown in Ref.~\cite{Edalatpour3}.

\begin{figure}
	\centering
	\includegraphics[width=0.8\textwidth]{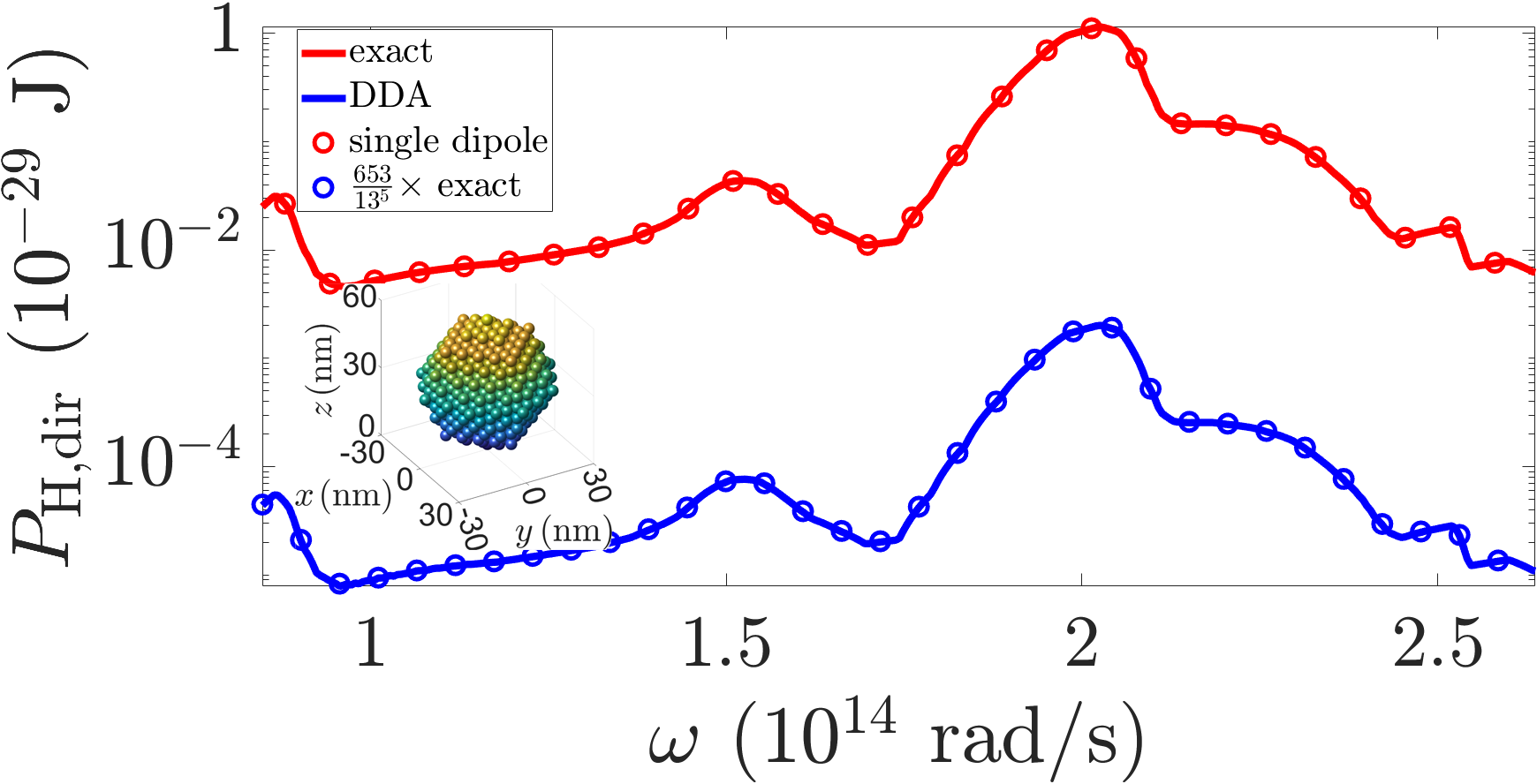}
	\caption{$P_{\rm H,dir}$ for a SiO$_2$ sphere for identical parameters like in Fig.~\ref{Fig:SphereMie} with radius $R=30$ nm. The exact Mie result (solid 
	red line) is compared with the single dipole result (open red symbols), the DDA result for 653 spherical nanoparticles (solid blue line), and the exact Mie 
	result multiplied by $653/13^5$ (open blue symbols). Inset: Sketch of the sphere used for the DDA.}
	\label{Fig:magn}
\end{figure}

We mentioned before that DDA does not work for the magnetic contribution. In Fig.~\ref{Fig:magn} we compare the purely magnetic contributions of the directly emitted power of a SiO$_2$ sphere approximated by 653 spherical nanoparticles (solid blue line, inset) for the same parameters like in Fig.~\ref{Fig:SphereMie} with the exact Mie result in Ref.~\cite{Eisner} (solid red line) and the single dipole model in Eq.~\eqref{eq:Pdir_snp} (open red symbols) at constant radius $R=30$ nm. Here, we chose spherical particles to perform DDA for a better visualization of the following point which is highlighted by the open blue symbols: As mentioned before, the magnetic contribution of our DDA calculations only provides the sum of the directly emitted power of each nanoparticle approximating the desired object. In this DDA example 13 nanoparticles fit into the whole sphere diameter. Therefore, the radius of each nanoparticle is a thirteenth of the radius of the whole sphere. Because the particle volume only appears in the magnetic polarizability (e.g. Eq.~\eqref{ChiHappr}) which is proportional to $R^5$, the sum of the directly emitted power of all nanoparticles is by factor of $653/13^5$ different from the exact result. This also means that the magnetic contribution in DDA calculations would decrease for an increasing number of nanoparticles which, obviously, leads to nonphysical results. That is also the reason why the single dipole model coincides with the exact result because, then, the number of particles and the radii are identical. Therefore, since the magnetic contribution dominates the thermally emitted heat radiation for metallic nanoparticles, the DDA method is not applicable to metals in the infrared.

\subsection{discrete dipole approximation - sharp tip}

Finally, we want to discuss the results obtained by Edalatpour {\itshape et al.} in Ref.~\cite{Edalatpour1}. In that work, the scattering of near-field thermal radiation of a sharp Si tip in 10 nm distance of a SiC substrate has been compared to the results for the scattering in the dipole model considering a dipole which has the same size as the tip. Clearly, in this scenario the dipole model is used in a regime where it should not by applicable, but Babuty {\itshape et al.} in Ref.~\cite{Joulain2} used exactly this approach to model the measured spectrum of a tip based near-field scattering microscope. Comparing the DDA result for a sharp tip with the dipole model Edalatpour {\itshape et al.} found in Ref.~\cite{Edalatpour1} that DDA predicts resonances at $944\,{\rm cm}^{-1}$ and $980\,{\rm cm}^{-1}$, whereas the dipole model predicts a resonance at $929\,{\rm cm}^{-1}$. 

\begin{figure}
	\centering
	\includegraphics[height=0.45\textwidth]{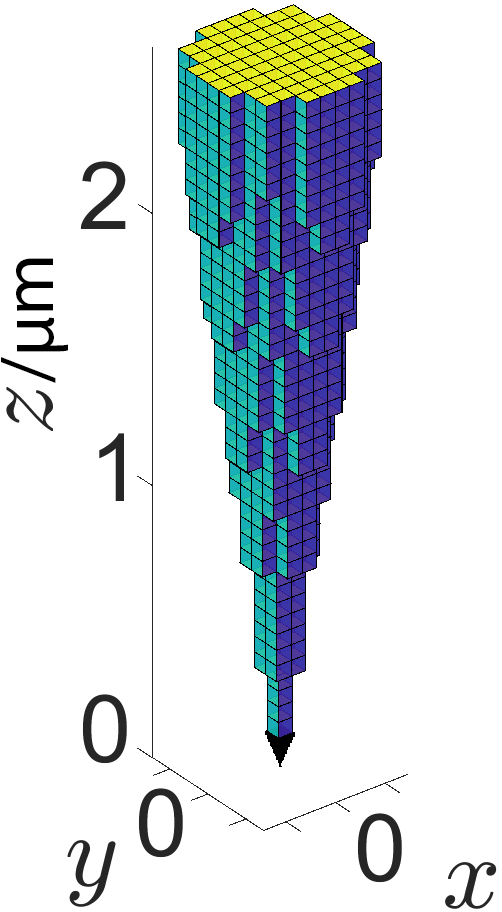}
	\caption{Sketch of the sharp tip used in our calculation. The foremost part of length 126.35 nm is approximated by $3904$ voxels with side length 3.61 nm and the remaining part of length 2485.4 nm by $1484$ voxels of side length 57.8 nm.}
	\label{Fig:TipDDA}
\end{figure}

The goal of this section is to revisit this result. To this end, we model a sharp Si tip as depicted in Fig.~\ref{Fig:TipDDA} using approximately the same size and shape as for probe B in Ref.~\cite{Joulain2}. We further use the same temperatures  $T_{\rm p} = T_{\rm b} = 0\,{\rm K}$ and $T_{\rm s} = 573\,{\rm K}$ as indicated in Ref.~\cite{Edalatpour1}, but it should be kept in mind that in this reference only the scattered intensity of the substrate's thermal near-field is calculated so that strictly speaking $T_{\rm s} = 573\,{\rm K}$ is the only temperature which can be modified in that approach. By setting  $T_{\rm p} = T_{\rm b} = 0\,{\rm K}$ the authors can argue to neglect thermal emission of the tip itself which is not treated in that work. But even when these temperatures are chosen we have still thermal emission of the SiC substrate manifesting in $P_{\rm s}$ together with shielding of that emission taken into account by $P_{\rm abs}$. To calculate $P_{\rm s}$, we use the cross section of the largest part of the tip at its backside for area $A$. The distance between substrate and foremost part of the tip is only 10 nm. We compare the DDA results as in Ref.~\cite{Edalatpour1} with those for a single dipole whose radius corresponds to the size of the foremost part ($R = 63.2\,{\rm nm}$) and with one that has the size of the whole tip ($R = 1.306$ µm) using $A = \pi R^2$ in those cases. Since they considered an emission angle of $\vartheta=\pi/4$ in Ref.~\cite{Edalatpour1}, we slightly adapt our theory by again introducing $k_\perp = k_0 \sin(\vartheta)$ and $k_z = k_0 \cos(\vartheta)$ and substituting
\begin{equation}
  \int_0^{k_0} \!\! \frac{\text{d} k_\perp}{2 \pi}  \, k_\perp f(k_\perp) = \int_0^{\pi/2} \!\! \frac{\text{d} \vartheta}{2 \pi} \, \cos(\vartheta) \sin(\vartheta) k_0^2 f(k_0 \sin(\vartheta)) 
\end{equation}
in all $\mathds{I}$ and $\mathds{R}$ integrals as well as in the one used for $P_{\rm s}$. Here, $f$ denotes the corresponding integrand and $\vartheta$ is the emission angle between wave vector $\mathbf{k}$ and surface normal. Now, we just pick the integrand value for $\vartheta=\pi/4$ instead of integrating over all emission angles.

\begin{figure}
	\centering
	\includegraphics[width=\textwidth]{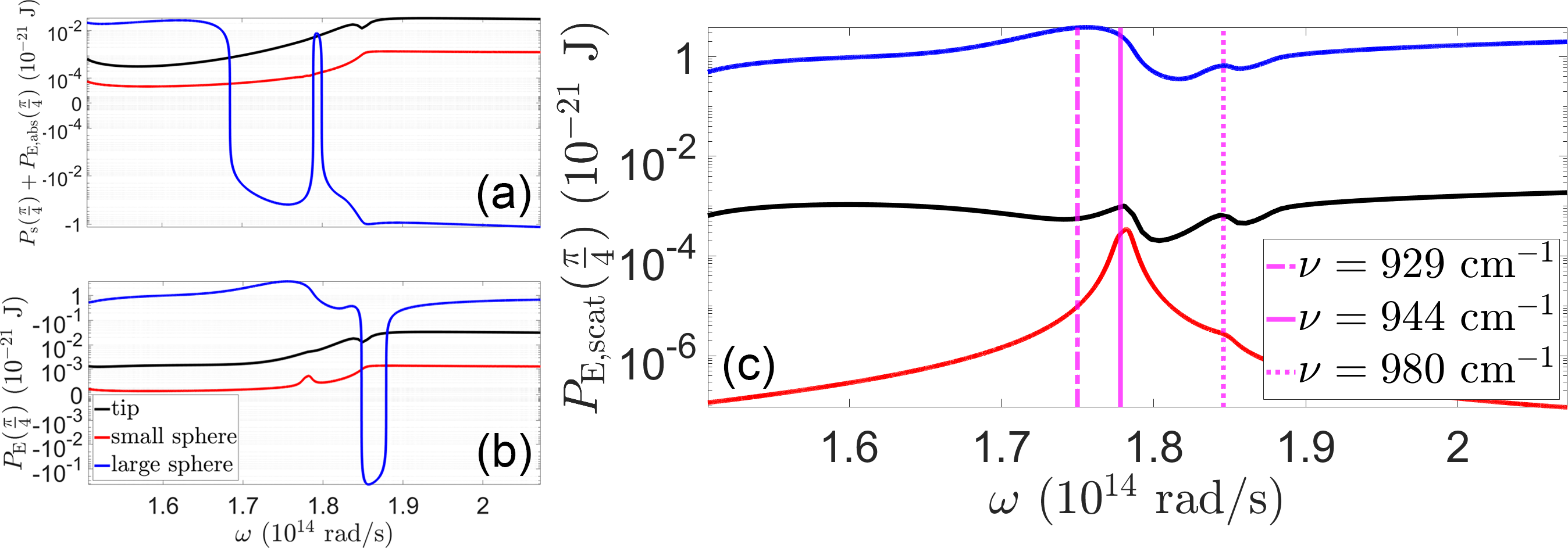}
	\caption{$P_{\rm E,abs} + P_{\rm s}$ (a), $P_{\rm E}$ (b), and $P_{\rm E,scat}$ (c) for a Si tip, small sphere with $R=63.2$ nm, and large sphere with $R=1.306$ µm at $T_{\rm p} = 0\,{\rm K}$ in a vacuum background with $T_{\rm b} = 0\,{\rm K}$ close to a SiC substrate with a gap distance of $10\,{\rm nm}$ with $T_{\rm s} = 573\,{\rm K}$ using a DDA model (S-CDM) with $N = 2553$ voxels. Additionally, in (c) we highlighted the three frequencies corresponding to the wave numbers $\nu$ mentioned in Ref.~\cite{Edalatpour1}.}
	\label{Fig:SphereTip}
\end{figure}

In Fig.~\ref{Fig:SphereTip}(a)-(b) we show the results for $P_{\rm E,abs} + P_{\rm s}$ and the full emitted power $P_{\rm E}$. It can be seen that for the large sphere within the dipole model these quantities can be negative. We are aware that this result, in general, seems to violate basic thermodynamics. Therefore, let us emphasize once more that this result does not impair the robustness of our model but is caused by the arbitrarily chosen plane $A$ contributing to $P_{\rm s}$. We could easily choose another value for $A$ to obtain purely positive results. This negative value in the emitted power can be interpreted such that the spherical dipolar object in the near-field of the substrate absorbs more heat from the substrate than predicted by the geometry of its cross section $A = \pi R^2$. This might not be astonishing but we want to emphasize that the dipole model is here used in a regime where $R \gg d$ so that the dipole approximation itself is strictly speaking not applicable and higher dipole moments need to be included. It can also be noted that the spectra for  $P_{\rm E,abs} + P_{\rm s}$ and the full emitted power $P_{\rm E}$ for the small dipolar sphere resemble those of the DDA tip in Fig.~\ref{Fig:SphereTip}(a)-(b).

However, here, we are more interested in the scattering spectra. For the scattering spectra of the DDA tip and those of the two dipoles, at first, it can be observed that the spectra of the tip and the large sphere are quite similar in Fig.~\ref{Fig:SphereTip}(c). The radiation scattered by the tip and by the small sphere possess a resonance at $\omega=1.778\times10^{14}$ rad/s ($\nu=945$ cm$^{-1}$) corresponding to the redshifted surface mode resonance of the SiC substrate at $\omega_{\rm SPhP} = 1.787\times10^{14}\,{\rm rad/s}$ as discussed in Ref.~\cite{Edalatpour1} which was found there to be at $\nu=944$ cm$^{-1}$. On the other hand, the radiation scattered by the tip and by the large sphere share a peak at $\omega=1.846\times10^{14}$ rad/s ($\nu=980$ cm$^{-1}$). There is also a small feature in the scattering spectrum of the small sphere at this frequency. At $\omega=1.75\times10^{14}$ rad/s ($\nu=929$ cm$^{-1}$) we find no peak in Fig.~\ref{Fig:SphereTip}(c). There is, however, a broad peak at $\omega\approx1.756\times10^{14}$ rad/s for radiation scattered at the large sphere. To compare our results with the ones in Ref.~\cite{Edalatpour1} we highlighted the there mentioned wave numbers in Fig.~\ref{Fig:SphereTip}(c). From our results we conclude that a large sphere is capable of approximating the scattering signal of the tip because the shapes of the two spectra resemble each other after exceeding their minimum value. Note that the first peak of the large sphere scattering spectrum is at $932$ cm$^{-1}$ rather than $929$ cm$^{-1}$ as in Fig. 3(b) in Ref~\cite{Edalatpour1}. Interestingly, the first peak of the tip scattering spectrum is very well captured by the small dipole representing the foremost part of the tip. Hence, the scattering spectrum of the tip can be understood as being a mixture of the scattering of its foremost part (represented by the small sphere contributing the first peak) and its bulk part (represented by the large sphere contributing the second peak) regarding the peak positions. This is a reasonable result which has not been pointed out in Ref.~\cite{Edalatpour1}. 

\begin{figure}
	\centering
	\includegraphics[width=0.75\textwidth]{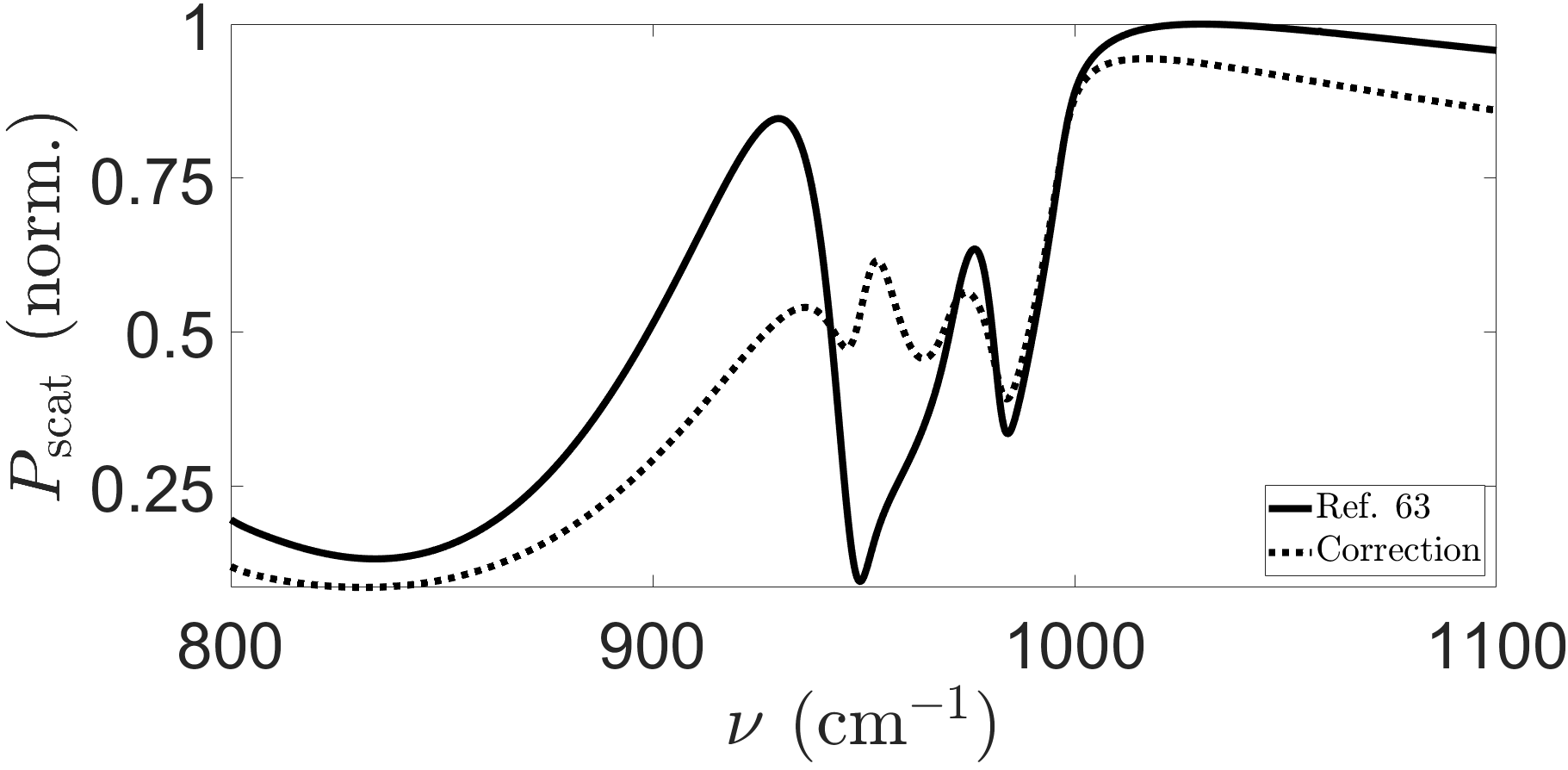}
	\caption{Comparing scattered power according to expressions from Eq.~(\ref{Eq:DipolScatWrong}) of Ref.~\cite{Edalatpour1} (solid line) 
	and to Eq.~(\ref{Eq:DipolScat}) (dashed line) for distance $d=10$ nm between substrate and nanoparticle's surface, radius $R=3.12$ µm, $T_p=T_b=0$ 
	K, $T_s=573$ K, and $\vartheta=45$°. Both graphs are normalized to the maximum value of the solid line.}
	\label{Fig:Correction}
\end{figure}

Therefore, we find that the large sphere approximation within the dipole model already gives a quite reasonable result even though the position of the first peak seems to be related to the scattering of the foremost part of the tip captured by the small sphere scattering spectrum. In Ref.~\cite{Edalatpour1} the difference between the shapes of the spectra of the tip and the large sphere is much larger than in our. It is not easy to find the source of this discrepancy to Ref.~\cite{Edalatpour1}, but we are convinced that the aforementioned error in the dipole model used in Eq.~(7) of Ref.~\cite{Edalatpour1} which are due to the wrong choice of the polarization vectors for the TM mode part in Eq.~(4) in Ref.~\cite{Edalatpour1} is a key element. Therefore, we tried to retrieve the result in Fig. 3(b) in Ref.~\cite{Edalatpour1}. To do so, we also have to use the quasi-static limit to express the dressed polarizability (compare with Eq.(8) in in Ref.~\cite{Edalatpour1}). This limit is only applicable in the near-field assuming that the information carried by the heat transfer is transferred instantaneous, e.g. $c \rightarrow \infty$ or $k_0 \rightarrow 0$. Then, it is possible to provide an analytical result for $\expval{G_{\text{EE,ref},\perp/z}}$ in the definition of the polarizabilities because the reflection coefficients inside the integral become independent of the wave vector:
\begin{equation}
	\expval{G_{\text{EE,ref},\gamma}} \approx \frac{1}{16 C_\gamma \pi k_0^2 (d + R)^3} \frac{\varepsilon_s - 1}{\varepsilon_s + 1} .
\end{equation}
with $C_\gamma = 2$ for $\gamma = \perp$ and $C_\gamma = 1$ for $\gamma = z$. For a distance $d + R$ of several microns as chosen in Ref.~\cite{Edalatpour1}, however, this approximation is not valid anymore because one leaves the quasi-static regime. Using the above quasi-static approximation for such distances means that the term $\expval{G_{\text{EE,ref},\gamma}}$ becomes, at a distance $d + R$ of several microns, so small that it can be neglected and, therefore, the corresponding polarizability is effectively the expression without the surface term as given in Eq.~(\ref{Eq:alphascdmvac}). To see that this inconsistency together with the sign error leading to Eq.~(\ref{Eq:DipolScatWrong}) have a measurable consequence, we provide a numerical example for a configuration used in Ref.~\cite{Edalatpour1} with distance $d = 10$ nm between substrate and nanoparticle's surface, radius $R=3.12$ µm, $T_p=T_b=0$ K, and $T_s=573$ K. In Fig.~\ref{Fig:Correction} we compare the spectra according to the wrong expression Eq.~(\ref{Eq:DipolScatWrong}) of Ref.~\cite{Edalatpour1} and the correct one according to Eq.~(\ref{Eq:DipolScat}) normalized to the maximal value of the graph corresponding to Eq.~(\ref{Eq:DipolScatWrong}) of Ref.~\cite{Edalatpour1}. Here, we used the angle $\vartheta=45$°.
Note that the solid line coincides with the corresponding graph in Fig.~3(b) of Ref.~\cite{Edalatpour1}. The peak mentioned in the Ref.~\cite{Edalatpour1} at $\nu = 929$ cm$^{-1}$ can be found here. Using the correct sign but the inconsistent quasi-static approximation for $\expval{G_{\text{EE,ref},\gamma}}$ this peak would shift to $\nu = 927$ cm$^{-1}$. In the corrected graph, however, the peak is located at $\nu = 936$ cm$^{-1}$; two additional peaks appear at $\nu = 953$ cm$^{-1}$ and $\nu = 974$ cm$^{-1}$. These three values are very close to the peaks at $\nu = 944$ cm$^{-1}$ and $\nu = 980$ cm$^{-1}$ observed in the DDA spectrum of the field scattered by the tip in Fig.~2(b) in Ref.~\cite{Edalatpour1}.

As a conclusion, our DDA-modelled tip reproduces the key results of the DDA simulations in Ref.~\cite{Edalatpour1}. In contrast to Ref.~\cite{Edalatpour1} we conclude that the dipole model as used in Ref.~\cite{Joulain2} to describe the spectra is not as bad as the results of Ref.~\cite{Edalatpour1} suggest, because there seems to be an error in their dipole model result. Furthermore, we also find that the position of the red-shifted SPhP resonance is well described be the results of a small spherical scatterer. It seems that the main features are captured by a mixture of the scattering of the foremost part and the bulk part of the tip which can each be described by dipole mode. Of course, in the end, the spectra depend on the choice of the tip shape and size as brought forward in Ref.~\cite{Edalatpour1}.
Finally, we want to emphasize that our model is more versatile than the pure scattering calculations in Ref.~\cite{Edalatpour1} because it also includes the direct thermal emission of the tip as well as the substrate's contribution and the part blocked by the tip. Actually, this last part can be relatively large as already pointed out by Komiyama~\cite{Komiyama,Weng} and also found in our numerical calculations so that it is important to minimize this contribution in measurements. 

\section{Conclusion}

In conclusion, we have derived the expressions for the emitted power of $N$ nanoparticles above a substrate where each particle temperature as well as the temperatures of the substrate and background radiation can be set, separately. Our expressions contain the direct power emitted by the nanoparticles, the power emitted by the substrate, the substrate emission which is absorbed by the nanoparticles, and the power due to the near-field scattering. Our model generalizes previous models and is, therefore, very versatile. It can be used to study the thermal emission of dielectric and metallic particle assemblies close to a substrate within CDM and to model thermal emission of a single or several macroscopic objects in the vicinity of a substrate within DDA. To illustrate this versatility we discussed the thermal emission of a four-particle configuration close to a dielectric and metallic substrate, the thermal emission of a sphere close to a substrate using DDA, and the thermal emission of a sharp tip as used in scattering type near-field thermal profiles, again, within DDA in the case of dielectric nanoparticles. We found that the dipole approximation can only be safely used for metallic nanoparticles close to polar substrates if the distance of the particle centers to the surface is larger than five times the particle radius. Furthermore, our study of the sharp tip showed that, in contrast to previously obtained results~\cite{Edalatpour1}, the dipole model as used in Ref.~\cite{Babuty} can give qualitatively correct spectra of a sharp scattering tip even though a DDA model for the tip shape is definitely preferable to a simple dipole. We point out that the tip spectrum seems to be a mixture of the scattering of its foremost part which can be modelled by a small spherical scatterer and its bulk part which can be modelled by a large spherical scatterer. Our results indicate that the magnetic dipole contribution can be used in CDM simulations but not DDA simulations. We are convinced that our generalized model will be very useful in further more detailed studies of thermal emission of many-particle systems including substrates and the modelling of scattering type experiments as in Ref's.~\cite{Wang, Komiyama, Kajihara2, DeWilde, Babuty,Jones,O'Callahan}.


S.-A.\ B.\ acknowledges support from Heisenberg Programme of the Deutsche Forschungsgemeinschaft (DFG, German Research Foundation) under the project No.\ 404073166 and F.\ H.\ acknowledges support from the Studienstiftung des deutschen Volkes (German Academic Scholarship Foundation).

%
%
\appendix

\section{The Green's functions \label{appA}}

All Green's functions can be decomposed into a vacuum part and a contribution due to reflections at the substrate's surface. The vacuum part of the electric Green's function for different observation and source point is well known:
\begin{align}
\mathds{G}_\text{EE,vac} (\mathbf{r}, \mathbf{r}') & = \frac{e^{\ri k_0 R}}{4 \pi |\mathbf{r} - \mathbf{r}'|} \biggl[ \frac{k_0^2 |\mathbf{r} - \mathbf{r}'|^2 + \ri k_0 |\mathbf{r} - \mathbf{r}'| - 1}{k_0^2 |\mathbf{r} - \mathbf{r}'|^2} \mathds{1} \notag \\
& \quad - \frac{k_0^2 |\mathbf{r} - \mathbf{r}'|^2 + 3 \ri k_0 |\mathbf{r} - \mathbf{r}'| - 3}{k_0^2 |\mathbf{r} - \mathbf{r}'|^2} \frac{\left( \mathbf{r} - \mathbf{r}' \right) \otimes \left( \mathbf{r} - \mathbf{r}' \right)}{|\mathbf{r} - \mathbf{r}'|^2} \biggr] .
\end{align}
All the others can be computed by exploiting relations between the Green's functions, so that we get
\begin{align}
G_{\text{HE,vac},ij} (\mathbf{r}, \mathbf{r}') & = \ri \sqrt{\frac{\varepsilon_0}{\mu_0}} \frac{e^{\ri k_0 |\mathbf{r} - \mathbf{r}'|}}{4 \pi |\mathbf{r} - \mathbf{r}'|} \frac{\ri k_0 |\mathbf{r} - \mathbf{r}'| - 1}{k_0 |\mathbf{r} - \mathbf{r}'|} \epsilon_{ijk} \frac{r_k - r_k'}{|\mathbf{r} - \mathbf{r}'|} , \\
G_{\text{EH,vac},ij} (\mathbf{r}, \mathbf{r}') & = - G_{\text{HE,vac},ij} (\mathbf{r}, \mathbf{r}') , \\
\mathds{G}_\text{HH,vac} (\mathbf{r}, \mathbf{r}') & = \frac{\varepsilon_0}{\mu_0} \mathds{G}_\text{EE,vac} (\mathbf{r}, \mathbf{r}') .
\end{align}
The volume averages of the vacuum contributions were calculated by Fikioris \cite{Fikioris}, meaning
\begin{align}
\expval{\mathds{G}_\text{EE,vac} (\mathbf{r})} & = \expval{G_\text{EE,vac}} \mathds{1}, \\
\expval{G_\text{EE,vac}} & = \frac{1}{k_0^2 V} \left[ \frac{2}{3} \left( 1 - \ri k_0 R \right) e^{\ri k_0 R} - 1 \right]
\end{align}
with radius R and volume V of a spherical nanoparticle. It is easy to show that
\begin{align}
\expval{\mathds{G}_\text{HH,vac} (\mathbf{r})} & = \frac{\varepsilon_0}{\mu_0} \expval{\mathds{G}_\text{EE,vac} (\mathbf{r})} , \\
\expval{\mathds{G}_\text{HE,vac} (\mathbf{r})} & = \expval{\mathds{G}_\text{EH,vac} (\mathbf{r})} = 0
\end{align}
holds. 

The contributions due to reflections at the substrate's surface are 
\begin{align}
\mathds{G}_\text{EE,ref} (k_\perp, z, z') & = \frac{\ri e^{\ri k_z (z + z')}}{2 k_z} \left( r_\text{H} \mathbf{a}_\perp (k_0) \otimes \mathbf{a}_\perp (k_0) + r_\text{E} \mathbf{a}_\parallel^{+} (k_0) \otimes \mathbf{a}_\parallel^{-} (k_0) \right), \\
\mathds{G}_\text{HH,ref} (k_\perp, z, z') & = \frac{\varepsilon_0}{\mu_0} \frac{\ri e^{\ri k_z (z + z')}}{2 k_z} \left( r_\text{E} \mathbf{a}_\perp (k_0) \otimes \mathbf{a}_\perp (k_0) + r_\text{H} \mathbf{a}_\parallel^{+} (k_0) \otimes \mathbf{a}_\parallel^{-} (k_0) \right), \\
\mathds{G}_\text{HE,ref} (k_\perp, z, z') & = \sqrt{\frac{\varepsilon_0}{\mu_0}} \frac{\ri e^{\ri k_z (z + z')}}{2 k_z} \left( r_\text{E} \mathbf{a}_\perp (k_0) \otimes \mathbf{a}_\parallel^{-} (k_0) - r_\text{H} \mathbf{a}_\parallel^{+} (k_0) \otimes \mathbf{a}_\perp (k_0) \right), \\
\mathds{G}_\text{EH,ref} (k_\perp, z, z') & = \sqrt{\frac{\varepsilon_0}{\mu_0}} \frac{\ri e^{\ri k_z (z + z')}}{2 k_z} \left( r_\text{E} \mathbf{a}_\parallel^{+} (k_0) \otimes \mathbf{a}_\perp (k_0) - r_\text{H} \mathbf{a}_\perp (k_0) \otimes \mathbf{a}_\parallel^{-} (k_0) \right)
\end{align}
with
\begin{align}
\mathds{G}_\text{ref} (\mathbf{r}, \mathbf{r}') & = \int \frac{\text{d}^2 k_\perp}{(2 \pi)^2} e^{\ri \mathbf{k}_\perp \cdot (\mathbf{x} - \mathbf{x}')} \mathds{G}_\text{ref} (k_\perp, z, z') .
\end{align}
Here, we introduced the polarization unit vectors
\begin{align}
\mathbf{a}_\perp (k_m) & = \frac{1}{k_\perp} (k_y, -k_x, 0)^T, \\
\mathbf{a}_\parallel^{\pm} (k_m) & = \frac{1}{k_\perp k_m} (\mp k_x k_z, \mp k_y k_z, k_\perp^2)^T, \\
\mathbf{k}_\perp & = (k_x, k_y, k_z)^T
\end{align}
and wave vector $\mathbf{k}_\perp = (k_x, k_y, 0)^T$ as well as $\mathbf{x} = (x, y, 0)^T$,  $k_z = \sqrt{k_m^2 - k_\perp^2}$, and $k_m = \sqrt{\varepsilon_m} k_0$. Additionally, we use the Fresnel amplitude reflection coefficients
\begin{align}
r_\text{H} & = \frac{k_z - k_{\text{s},z}}{k_z + k_{\text{s},z}}, \\
r_\text{E} & = \frac{\varepsilon_s k_z - k_{\text{s},z}}{\varepsilon_s k_z + k_{\text{s},z}}
\end{align}
with $k_{m,z} = \sqrt{k_m^2 - k_\perp^2}$. Since there is no singularity when $\mathbf{r}$ approaches $\mathbf{r}'$, we can safely approximate the volume average by the integrand multiplied by the volume:
\begin{align}
\expval{\mathds{G}_{kk,\text{ref}} (\mathbf{r})} & = \expval{G_{kk,\text{ref},\perp}} [\mathbf{e}_x \otimes \mathbf{e}_x + \mathbf{e}_y \otimes \mathbf{e}_y] + \expval{G_{kk,\text{ref},z}} \mathbf{e}_z \otimes \mathbf{e}_z , \\
\expval{\mathds{G}_\text{HE/EH,ref} (\mathbf{r})} & = \expval{G_\text{HE/EH,ref}} [\mathbf{e}_x \otimes \mathbf{e}_y - \mathbf{e}_y \otimes \mathbf{e}_x] 
\end{align}
with
\begin{align}
\expval{G_{\text{EE,ref},\perp}} & = \ri \int_0^\infty \frac{\text{d} k_\perp}{8 \pi} \frac{k_\perp}{k_z} e^{2 \ri k_z z} \left( r_\text{H} - r_\text{E} \frac{k_z^2}{k_0^2} \right) , \\
\expval{G_{\text{EE,ref},z}} & = \ri \int_0^\infty \frac{\text{d} k_\perp}{4 \pi} \frac{k_\perp^3}{k_z k_0^2} e^{2 \ri k_z z} r_\text{E}, \\
\expval{G_{\text{HH,ref},\perp}} & = \ri \frac{\varepsilon_0}{\mu_0} \int_0^\infty \frac{\text{d} k_\perp}{8 \pi} \frac{k_\perp}{k_z} e^{2 \ri k_z z} \left( r_\text{E} - r_\text{H} \frac{k_z^2}{k_0^2} \right) , \\
\expval{G_{\text{HH,ref},z}} & = \ri \frac{\varepsilon_0}{\mu_0} \int_0^\infty \frac{\text{d} k_\perp}{4 \pi} \frac{k_\perp^3}{k_z k_0^2} e^{2 \ri k_z z} r_\text{H} , \\
\expval{G_{\text{HE/EH,ref}}} & = - \ri \sqrt{\frac{\varepsilon_0}{\mu_0}} \int_0^\infty \frac{\text{d} k_\perp}{8 \pi} \frac{k_\perp}{k_0} e^{2 \ri k_z z} \left( r_\text{H} - r_\text{E} \right) .
\end{align}

For the LEQC we have to use different Green's functions resulting from the transmission of radiation through the substrate's surface. We employed
\begin{align}
\mathds{G}^\text{s}_\text{EE} (k_\perp, z, z') & = \frac{\ri e^{\ri (k_z z - k_{\text{s},z} z')}}{2 k_{\text{s},z}} \left( t_\text{H} \mathbf{a}_\perp (k_0) \otimes \mathbf{a}_\perp (k_\text{s}) + t_\text{E} \mathbf{a}_\parallel^{+} (k_0) \otimes \mathbf{a}_\parallel^{+} (k_\text{s} ) \right), \\
\mathds{G}^s_\text{HE} (k_\perp, z, z') & = \sqrt{\frac{\varepsilon_0}{\mu_0}} \frac{\ri e^{\ri (k_z z - k_{\text{s},z} z')}}{2 k_{\text{s},z}} \left( t_\text{E} \mathbf{a}_\perp (k_0) \otimes \mathbf{a}_\parallel^{+} (k_\text{s}) - t_\text{H} \mathbf{a}_\parallel^{+} (k_0) \otimes \mathbf{a}_\perp (k_\text{s}) \right)
\end{align}
with the Fresnel amplitude transmission coefficients
\begin{align}
t_\text{H} & = \frac{2 k_{\text{s},z}}{k_z + k_{\text{s},z}}, \\
t_\text{E} & = \frac{k_\text{s}}{k_0} \frac{2 k_{\text{s},z}}{\varepsilon_s k_z + k_{\text{s},z}}.
\end{align}

\section{List of expressions used for the power contributions \label{appB}}

For the heat transfer between all nanoparticles and background we defined the following integrals
\begin{align}
\mathds{I}_k^{\alpha \beta} & = \int_0^{k_0} \frac{\text{d} k_\perp}{2 \pi} \frac{k_\perp}{k_z k_0} \biggl[ \biggl( \frac{J_1(\xi)}{\xi} I_{\bar{k},1}^{\alpha \beta +} + \frac{k_z^2}{k_0^2} \left( J_0(\xi) - \frac{J_1(\xi)}{\xi} \right) I_{k,1}^{\alpha \beta -} \biggr) \mathbf{e}_x \otimes \mathbf{e}_x \notag \\
& \quad + \biggl( \left( J_0(\xi) - \frac{J_1(\xi)}{\xi} \right) I_{\bar{k},1}^{\alpha \beta +} + \frac{k_z^2}{k_0^2} \frac{J_1(\xi)}{\xi} I_{k,1}^{\alpha \beta -} \biggr) \mathbf{e}_y \otimes \mathbf{e}_y + \frac{k_\perp^2}{k_0^2} J_0(\xi) I_{k,1}^{\alpha \beta +} \mathbf{e}_z \otimes \mathbf{e}_z \biggl]
\end{align}
and
\begin{align}
\mathds{I}_c^{\alpha \beta} & = \int_0^{k_0} \frac{\text{d} k_\perp}{2 \pi} \frac{k_\perp}{k_0^2} \biggl[ \left( \frac{J_1(\xi)}{\xi} I_{\text{H},2}^{\alpha \beta +} + \left( J_0(\xi) - \frac{J_1(\xi)}{\xi} \right) I_{\text{E},2}^{\alpha \beta -} \right) \mathbf{e}_y \otimes \mathbf{e}_x \notag \\
& \quad - \left( \left( J_0(\xi) - \frac{J_1(\xi)}{\xi} \right) I_{\text{H},2}^{\alpha \beta +} + \frac{J_1(\xi)}{\xi} I_{\text{E},2}^{\alpha \beta -} \right) \mathbf{e}_x \otimes \mathbf{e}_y \notag \\
& \quad + \ri \frac{k_\perp}{k_z} J_1(\xi) \left( I_{\text{E},1}^{\alpha \beta +} \mathbf{e}_y \otimes \mathbf{e}_z - I_{\text{H},1}^{\alpha \beta +} \mathbf{e}_z \otimes \mathbf{e}_y \right) \biggl].
\end{align}
$J_{0,1}(\xi)$ denotes the cylindrical Bessel function of the zeroth and first kind, respectively. Additionally, we employed the abbreviations
\begin{align}
I_{k,1}^{\alpha \beta \pm} & = e^{\ri k_z (z_\alpha - z_\beta)} \pm 2 \Re \left( r_k e^{\ri k_z (z_\alpha + z_\beta)} \right) + |r_k|^2 e^{-\ri k_z (z_\alpha - z_\beta)} , \\
I_{k,2}^{\alpha \beta \pm} & = e^{\ri k_z (z_\alpha - z_\beta)} \pm 2 \ri \Im \left( r_k e^{\ri k_z (z_\alpha + z_\beta)} \right) - |r_k|^2 e^{-\ri k_z (z_\alpha - z_\beta)} , \\
\xi & = k_\perp \sqrt{(x_\alpha - x_\beta)^2 + (y_\alpha - y_\beta)^2} .
\end{align}

In the heat transfer between substrate and background we defined the integrals
\begin{align}
\mathds{R}_k^{\alpha \beta} & = \int_0^{k_0} \frac{\text{d} k_\perp}{2 \pi} \frac{k_\perp}{k_z k_0} \biggl[ \left( \frac{J_1(\xi)}{\xi} R_{\bar{k},1}^{\alpha \beta +} + \frac{k_z^2}{k_0^2} \left( J_0(\xi) - \frac{J_1(\xi)}{\xi} \right) R_{k,1}^{\alpha \beta -} \right) \mathbf{e}_x \otimes \mathbf{e}_x \notag \\
& \quad + \left( \left( J_0(\xi) - \frac{J_1(\xi)}{\xi} \right) R_{\bar{k},1}^{\alpha \beta +} + \frac{k_z^2}{k_0^2} \frac{J_1(\xi)}{\xi} R_{k,1}^{\alpha \beta -} \right) \mathbf{e}_y \otimes \mathbf{e}_y + \frac{k_\perp^2}{k_0^2} J_0(\xi_\alpha^\beta) R_{k,1}^{\alpha \beta +} \mathbf{e}_z \otimes \mathbf{e}_z \notag \\ 
& \quad + \ri \frac{k_\perp k_z}{k_0^2} J_1(\xi) \left( R_{k,1}^{\alpha \beta -} \mathbf{e}_z \otimes \mathbf{e}_x + R_{k,1}^{\alpha \beta +} \mathbf{e}_x \otimes \mathbf{e}_z \right) \biggl]
\end{align}
and 
\begin{align}
\mathds{R}_c^{\alpha \beta} & = \int_0^{k_0} \frac{\text{d} k_\perp}{2 \pi} \frac{k_\perp}{k_0^2} \biggl[ \biggl( \left( J_0(\xi) - \frac{J_1(\xi)}{\xi} \right) R_{\text{H},2}^{\beta \alpha +} + \frac{J_1(\xi)}{\xi} R_{\text{E},2}^{\beta \alpha -} \biggr) \mathbf{e}_y \otimes \mathbf{e}_x \notag \\
& \quad - \Bigl( \frac{J_1(\xi)}{\xi} R_{\text{H},2}^{\beta \alpha +} + \left( J_0(\xi) - \frac{J_1(\xi)}{\xi} \right) R_{\text{E},2}^{\beta \alpha -} \Bigr) \mathbf{e}_x \otimes \mathbf{e}_y \notag \\ 
& \quad - \frac{k_\perp}{2 k_z} J_1(\xi_\alpha^\beta) \left( [R_{\text{H},2}^{\beta \alpha +} + R_{\text{H},2}^{\beta \alpha -}] \mathbf{e}_y \otimes \mathbf{e}_z - [R_{\text{E},2}^{\beta \alpha +} + R_{\text{E},2}^{\beta \alpha -}] \mathbf{e}_z \otimes \mathbf{e}_y \right) \biggl]
\end{align}
for $P_\text{abs}$ where we used the abbreviations
\begin{align}
R_{k,1}^{\alpha \beta \pm} & = (1 - |r_k|^2) \left( 1 \pm r_k e^{\ri k_z (z_\alpha + z_\beta)} \right) , \\
R_{k,2}^{\alpha \beta \pm} & = (1 - |r_k|^2) \left( \ri \sin(k_z [z_\alpha - z_\beta]) \pm r_k e^{\ri k_z (z_\alpha + z_\beta)} \right) 
\end{align}
and four additional integrals for $P_\text{scat}$, namely
\begin{align}
\mathbf{\Gamma}_k^{\alpha \beta} & = \int_0^{k_0} \frac{\text{d} k_\perp}{2 \pi} \frac{k_\perp}{k_z k_0} \biggl[ \biggl( \frac{J_1(\xi)}{\xi} \nu_{\bar{k},1}^{\alpha \beta} + \frac{k_z^2}{k_0^2} \left( J_0(\xi) - \frac{J_1(\xi)}{\xi} \right) \nu_{k,1}^{\alpha \beta} \biggr) \mathbf{e}_x \otimes \mathbf{e}_x \notag \\
& \quad + \biggl( \left( J_0(\xi) - \frac{J_1(\xi)}{\xi} \right) \nu_{\bar{k},1}^{\alpha \beta} + \frac{k_z^2}{k_0^2} \frac{J_1(\xi)}{\xi} \nu_{k,1}^{\alpha \beta} \biggr) \mathbf{e}_y \otimes \mathbf{e}_y + \frac{k_\perp^2}{k_0^2} J_0(\xi) \nu_{k,1}^{\alpha \beta} \mathbf{e}_z \otimes \mathbf{e}_z \notag \\
& \quad + \int_{k_0}^\infty \frac{\text{d} k_\perp}{2 \pi} \frac{k_\perp}{|k_z| k_0} \biggl[ \biggl( \frac{J_1(\xi)}{\xi} \nu_{\bar{k},2}^{\alpha \beta} + \frac{k_z^2}{k_0^2} \left( J_0(\xi) - \frac{J_1(\xi)}{\xi} \right) \nu_{k,2}^{\alpha \beta} \biggr) \mathbf{e}_x \otimes \mathbf{e}_x \notag \\
& \quad + \biggl( \left( J_0(\xi) - \frac{J_1(\xi)}{\xi} \right) \nu_{\bar{k},2}^{\alpha \beta} + \frac{k_z^2}{k_0^2} \frac{J_1(\xi)}{\xi} \nu_{k,2}^{\alpha \beta} \biggr) \mathbf{e}_y \otimes \mathbf{e}_y + \frac{k_\perp^2}{k_0^2} J_0(\xi) \nu_{k,2}^{\alpha \beta} \mathbf{e}_z \otimes \mathbf{e}_z \biggr] 
\end{align}
and
\begin{align}
\mathbf{\Gamma}_c^{\alpha \beta} & = \int_0^{k_0} \frac{\text{d} k_\perp}{2 \pi} \frac{k_\perp}{k_0^2} \biggl[ \biggl( \frac{J_1(\xi)}{\xi} \nu_{\text{H},1}^{\alpha \beta} + \left( J_0(\xi) - \frac{J_1(\xi)}{\xi} \right) \nu_{\text{E},1}^{\alpha \beta} \biggr) \mathbf{e}_x \otimes \mathbf{e}_y \notag \\
& \quad - \biggl( \left( J_0(\xi) - \frac{J_1(\xi)}{\xi} \right) \nu_{\text{H},1}^{\alpha \beta} + \frac{J_1(\xi)}{\xi} \nu_{\text{E},1}^{\alpha \beta} \biggr) \mathbf{e}_y \otimes \mathbf{e}_x \notag \biggr] \\
& \quad - \ri \int_{k_0}^\infty \frac{\text{d} k_\perp}{2 \pi} \frac{k_\perp}{k_0^2} \biggl[ \biggl( \frac{J_1(\xi)}{\xi} \nu_{\text{H},2}^{\alpha \beta} - \left( J_0(\xi) - \frac{J_1(\xi)}{\xi} \right) \nu_{\text{E},2}^{\alpha \beta} \biggr) \mathbf{e}_x \otimes \mathbf{e}_y \notag \\
& \quad - \biggl( \left( J_0(\xi) - \frac{J_1(\xi)}{\xi} \right) \nu_{\text{H},2}^{\alpha \beta} - \frac{J_1(\xi)}{\xi} \nu_{\text{E},2}^{\alpha \beta} \biggr) \mathbf{e}_y \otimes \mathbf{e}_x \biggr] 
\end{align}
as well as
\begin{align}
\mathds{K}_k^{\alpha \beta} & = \ri \int_0^{k_0} \frac{\text{d} k_\perp}{2 \pi} \frac{k_\perp^2}{k_0^3} J_1(\xi) \nu_{k,1}^{\alpha \beta}\left[ \mathbf{e}_x \otimes \mathbf{e}_z + \mathbf{e}_z \otimes \mathbf{e}_x \right] \notag \\
& \quad - \int_{k_0}^\infty \frac{\text{d} k_\perp}{2 \pi} \frac{k_\perp^2}{k_0^3} J_1(\xi) \nu_{k,2}^{\alpha \beta} \left[ \mathbf{e}_x \otimes \mathbf{e}_z - \mathbf{e}_z \otimes \mathbf{e}_x \right]
\end{align}
and
\begin{align}
\mathds{K}_c^{\alpha \beta} & = - \ri \int_0^{k_0} \frac{\text{d} k_\perp}{2 \pi} \frac{k_\perp^2}{k_z k_0^2} J_1(\xi) \left( \nu_{\text{H},1}^{\alpha \beta}\mathbf{e}_y \otimes \mathbf{e}_z - \nu_{\text{E},1}^{\alpha \beta} \mathbf{e}_z \otimes \mathbf{e}_y \right) \notag \\
& \quad - \ri \int_{k_0}^\infty \frac{\text{d} k_\perp}{2 \pi} \frac{k_\perp^2}{|k_z| k_0^2} J_1(\xi) \left( \nu_{\text{H},2}^{\alpha \beta} \mathbf{e}_y \otimes \mathbf{e}_z - \nu_{\text{E},2}^{\alpha \beta} \mathbf{e}_z \otimes \mathbf{e}_y \right)
\end{align}
together with the abbreviations 
\begin{align}
\nu_{k,1}^{\alpha \beta} & = (1 - |r_k|^2) e^{\ri k_z (z_\alpha - z_\beta)} , \\
\nu_{k,2}^{\alpha \beta} & = 2 \Im(r_k) e^{- |k_z| (z_\alpha + z_\beta)} .
\end{align}

\section{Abbreviations for single nanoparticle case \label{appC}}

For the single nanoparticle case, we introduced the matrix entries 
\begin{align}
M_{k,\perp} & = \left( n_p - n_b \right) \frac{ k_0^2 V_p \Im (\chi_k) + \frac{k_0^6 V_p^3 \Im (\chi_{\bar{k}}) |\chi_k|^2 |\mathds{G_\text{HE}}|^2}{ |1 - k_0^2 V_p \chi_{\bar{k}} \expval{G_{\bar{k}\bar{k}, \perp}}|^2}}{\Big|1 - k_0^2 V_p \chi_{k} \expval{G_{kk,\perp}} + \frac{k_0^4 V_p^2 \chi_\text{E} \chi_\text{H} \expval{G_\text{HE}}^2}{1 - k_0^2 V_p \chi_{\bar{k}} \expval{G_{\bar{k}\bar{k},\perp}}}\Big|^2} , \\
M_{k,z} & = \left( n_p - n_b \right) \frac{k_0^2 V_p \Im(\chi_k)}{|1 - k_0^2 V_p \chi_k \expval{G_{kk,z}}|^2} , \\
N_\text{c} & = \frac{ k_0^4 V_p^2 \left( n_p - n_b \right) \Im (\chi_\text{H}) \chi_\text{E} \mathds{G_\text{HE}} \left( 1 - k_0^2 V_p \chi_\text{E}^{*} \expval{G_{\text{EE}, \perp}}^{*} \right)}{\Big|\left(1 - k_0^2 V_p \chi_\text{E} \expval{G_{\text{EE},\perp}} \right) \left(1 - k_0^2 V_p \chi_\text{H} \expval{G_{\text{HH}, \perp}} \right) + k_0^4 V_p^2 \chi_\text{E} \chi_\text{H} \expval{G_\text{HE}}^2 \Big|^2} \notag \\
& \quad - \frac{ k_0^4 V_p^2 \left( n_p - n_b \right) \Im (\chi_\text{E}) \chi_\text{H}^{*} \mathds{G_\text{HE}}^{*} \left( 1 - k_0^2 V_p \chi_\text{H} \expval{G_{\text{HH}, \perp}} \right)}{\Big|\left(1 - k_0^2 V_p \chi_\text{E} \expval{G_{\text{EE},\perp}} \right) \left(1 - k_0^2 V_p \chi_\text{H} \expval{G_{\text{HH}, \perp}} \right) + k_0^4 V_p^2 \chi_\text{E} \chi_\text{H} \expval{G_\text{HE}}^2 \Big|^2}.
\end{align}
The asterisk denotes the complex conjugate. We also define the integrals
\begin{align}
I_{k,\perp} & = \int_0^{k_0} \frac{\text{d} k_\perp}{2 \pi} \frac{k_\perp}{k_z k_0} \Bigl( \Big|1 + r_{\bar{k}} e^{2 \ri k_z z_p} \Big|^2 + \frac{k_z^2}{k_0^2} \Big|1 - r_k e^{2 \ri k_z z_p} \Big|^2 \Bigr) , \\
I_{k,z} & = \int_0^{k_0} \frac{\text{d} k_\perp}{2 \pi} \frac{k_\perp^3}{k_z k_0^3} \Big| 1 + r_k e^{2 \ri k_z z_p} \Big|^2 , \\
I_c & = \int_0^{k_0} \frac{\text{d} k_\perp}{2 \pi} \frac{k_\perp}{k_0^2} \left( 2 + 2 \ri \Im \left[(r_\text{H} - r_\text{E}) e^{2 \ri k_z z_p} \right] - |r_\text{H}|^2 - |r_\text{E}|^2 \right)
\end{align}
for the heat radiation transferred between nanoparticle and background. We also defined the integrals
\begin{align}
R_{k,\perp} & = \int_0^{k_0} \frac{\text{d} k_\perp}{2 \pi} \frac{k_\perp}{k_z k_0} \left( (1 - |r_{\bar{k}}|^2) \left( 1 + r_{\bar{k}} e^{2 \ri k_z z_p} \right) + \frac{k_z^2}{k_0^2} (1 - |r_k|^2) \left( 1 - r_k e^{2 \ri k_z z_p} \right) \right) , \\
R_{k,z} & = \int_0^{k_0} \frac{\text{d} k_\perp}{2 \pi} \frac{k_\perp^3}{k_z k_0^3} (1 - |r_k|^2) \left( 1 + r_k e^{2 \ri k_z z_p} \right) , \\
R_c & = \int_0^{k_0} \frac{\text{d} k_\perp}{2 \pi} \frac{k_\perp}{k_0^2} e^{2 \ri k_z z_p} \left( (1 - |r_\text{H}|^2) r_\text{H} - (1 - |r_\text{E}|^2) r_\text{E}\right) 
\end{align}
for the radiation absorbed in the nanoparticle and the integrals 
\begin{align}
\Gamma_{k,\perp} & = \int_0^{k_0} \frac{\text{d} k_\perp}{2 \pi} \frac{k_\perp}{k_z k_0} \biggl( (1 - |r_{\bar{k}}|^2) + \frac{k_z^2}{k_0^2} (1 - |r_k|^2) \biggr) + 2 \int_{k_0}^\infty \frac{\text{d} k_\perp}{2 \pi} \frac{k_\perp}{|k_z| k_0} e^{-2 |k_z| z_p} \Im \biggl( r_{\bar{k}} + \frac{|k_z|^2}{k_0^2} r_k \biggr) \\
\Gamma_{k,z} & = \int_0^{k_0} \frac{\text{d} k_\perp}{2 \pi} \frac{k_\perp^3}{k_z k_0^3} (1 - |r_k|^2) + 2 \int_{k_0}^\infty \frac{\text{d} k_\perp}{2 \pi} \frac{k_\perp^3}{|k_z| k_0^3} e^{-2 |k_z| z_p} \Im(r_k) \\
\Gamma_c & = \int_0^{k_0} \frac{\text{d} k_\perp}{2 \pi} \frac{k_\perp}{k_0^2} (2 - |r_\text{H}|^2 - |r_\text{E}|^2) - 2 \ri \int_{k_0}^\infty \frac{\text{d} k_\perp}{2 \pi} \frac{k_\perp}{k_0^2} e^{-2 |k_z| z_p} \Im(r_\text{H} - r_\text{E})
\end{align}
for the radiation scattered by the nanoparticle.

\bibliographystyle{apsrev4-2}
\bibliography{Referenzen}

\begin{thebibliography}{88}%
\makeatletter
\providecommand \@ifxundefined [1]{%
 \@ifx{#1\undefined}
}%
\providecommand \@ifnum [1]{%
 \ifnum #1\expandafter \@firstoftwo
 \else \expandafter \@secondoftwo
 \fi
}%
\providecommand \@ifx [1]{%
 \ifx #1\expandafter \@firstoftwo
 \else \expandafter \@secondoftwo
 \fi
}%
\providecommand \natexlab [1]{#1}%
\providecommand \enquote  [1]{``#1''}%
\providecommand \bibnamefont  [1]{#1}%
\providecommand \bibfnamefont [1]{#1}%
\providecommand \citenamefont [1]{#1}%
\providecommand \href@noop [0]{\@secondoftwo}%
\providecommand \href [0]{\begingroup \@sanitize@url \@href}%
\providecommand \@href[1]{\@@startlink{#1}\@@href}%
\providecommand \@@href[1]{\endgroup#1\@@endlink}%
\providecommand \@sanitize@url [0]{\catcode `\\12\catcode `\$12\catcode
  `\&12\catcode `\#12\catcode `\^12\catcode `\_12\catcode `\%12\relax}%
\providecommand \@@startlink[1]{}%
\providecommand \@@endlink[0]{}%
\providecommand \url  [0]{\begingroup\@sanitize@url \@url }%
\providecommand \@url [1]{\endgroup\@href {#1}{\urlprefix }}%
\providecommand \urlprefix  [0]{URL }%
\providecommand \Eprint [0]{\href }%
\providecommand \doibase [0]{https://doi.org/}%
\providecommand \selectlanguage [0]{\@gobble}%
\providecommand \bibinfo  [0]{\@secondoftwo}%
\providecommand \bibfield  [0]{\@secondoftwo}%
\providecommand \translation [1]{[#1]}%
\providecommand \BibitemOpen [0]{}%
\providecommand \bibitemStop [0]{}%
\providecommand \bibitemNoStop [0]{.\EOS\space}%
\providecommand \EOS [0]{\spacefactor3000\relax}%
\providecommand \BibitemShut  [1]{\csname bibitem#1\endcsname}%
\let\auto@bib@innerbib\@empty
\bibitem [{\citenamefont {Draine}(1988)}]{Draine1}%
  \BibitemOpen
  \bibfield  {author} {\bibinfo {author} {\bibfnamefont {B.}~\bibnamefont
  {Draine}},\ }\href@noop {} {\bibfield  {journal} {\bibinfo  {journal}
  {Astrophys. J.}\ }\textbf {\bibinfo {volume} {333}},\ \bibinfo {pages} {848}
  (\bibinfo {year} {1988})}\BibitemShut {NoStop}%
\bibitem [{\citenamefont {Ben-Abdallah}\ \emph {et~al.}(2011)\citenamefont
  {Ben-Abdallah}, \citenamefont {Biehs},\ and\ \citenamefont
  {K.Joulain}}]{pbaetal}%
  \BibitemOpen
  \bibfield  {author} {\bibinfo {author} {\bibfnamefont {P.}~\bibnamefont
  {Ben-Abdallah}}, \bibinfo {author} {\bibfnamefont {S.-A.}\ \bibnamefont
  {Biehs}},\ and\ \bibinfo {author} {\bibnamefont {K.Joulain}},\ }\href@noop {}
  {\bibfield  {journal} {\bibinfo  {journal} {Phys. Rev. Lett.}\ }\textbf
  {\bibinfo {volume} {107}},\ \bibinfo {pages} {114301} (\bibinfo {year}
  {2011})}\BibitemShut {NoStop}%
\bibitem [{\citenamefont {Messina}\ \emph {et~al.}(2013)\citenamefont
  {Messina}, \citenamefont {Tschikin}, \citenamefont {Biehs},\ and\
  \citenamefont {Ben-Abdallah}}]{Messina}%
  \BibitemOpen
  \bibfield  {author} {\bibinfo {author} {\bibfnamefont {R.}~\bibnamefont
  {Messina}}, \bibinfo {author} {\bibfnamefont {M.}~\bibnamefont {Tschikin}},
  \bibinfo {author} {\bibfnamefont {S.-A.}\ \bibnamefont {Biehs}},\ and\
  \bibinfo {author} {\bibfnamefont {P.}~\bibnamefont {Ben-Abdallah}},\
  }\href@noop {} {\bibfield  {journal} {\bibinfo  {journal} {Phys. Rev. B}\
  }\textbf {\bibinfo {volume} {88}},\ \bibinfo {pages} {104307} (\bibinfo
  {year} {2013})}\BibitemShut {NoStop}%
\bibitem [{\citenamefont {Dong}\ \emph {et~al.}(2017)\citenamefont {Dong},
  \citenamefont {Zhao},\ and\ \citenamefont {Liu}}]{Dong}%
  \BibitemOpen
  \bibfield  {author} {\bibinfo {author} {\bibfnamefont {J.}~\bibnamefont
  {Dong}}, \bibinfo {author} {\bibfnamefont {J.}~\bibnamefont {Zhao}},\ and\
  \bibinfo {author} {\bibfnamefont {L.}~\bibnamefont {Liu}},\ }\href@noop {}
  {\bibfield  {journal} {\bibinfo  {journal} {Phys. Rev. B}\ }\textbf {\bibinfo
  {volume} {95}},\ \bibinfo {pages} {125411} (\bibinfo {year}
  {2017})}\BibitemShut {NoStop}%
\bibitem [{\citenamefont {Tomchuk}\ and\ \citenamefont
  {Grigorchuk}(2006)}]{wirbelM}%
  \BibitemOpen
  \bibfield  {author} {\bibinfo {author} {\bibfnamefont {P.~M.}\ \bibnamefont
  {Tomchuk}}\ and\ \bibinfo {author} {\bibfnamefont {N.~I.}\ \bibnamefont
  {Grigorchuk}},\ }\href@noop {} {\bibfield  {journal} {\bibinfo  {journal}
  {Phys. Rev. B}\ }\textbf {\bibinfo {volume} {73}},\ \bibinfo {pages} {155423}
  (\bibinfo {year} {2006})}\BibitemShut {NoStop}%
\bibitem [{\citenamefont {Dedkov}\ and\ \citenamefont
  {Kyasov}(2007)}]{Dedkov2}%
  \BibitemOpen
  \bibfield  {author} {\bibinfo {author} {\bibfnamefont {G.}~\bibnamefont
  {Dedkov}}\ and\ \bibinfo {author} {\bibfnamefont {A.~A.}\ \bibnamefont
  {Kyasov}},\ }\href@noop {} {\bibfield  {journal} {\bibinfo  {journal} {Tech.
  Phys. Lett.}\ }\textbf {\bibinfo {volume} {33}},\ \bibinfo {pages} {305}
  (\bibinfo {year} {2007})}\BibitemShut {NoStop}%
\bibitem [{\citenamefont {Chapuis}\ \emph
  {et~al.}(2008{\natexlab{a}})\citenamefont {Chapuis}, \citenamefont {Laroche},
  \citenamefont {Volz},\ and\ \citenamefont {Greffet}}]{Chapuis}%
  \BibitemOpen
  \bibfield  {author} {\bibinfo {author} {\bibfnamefont {P.-O.}\ \bibnamefont
  {Chapuis}}, \bibinfo {author} {\bibfnamefont {M.}~\bibnamefont {Laroche}},
  \bibinfo {author} {\bibfnamefont {S.}~\bibnamefont {Volz}},\ and\ \bibinfo
  {author} {\bibfnamefont {J.-J.}\ \bibnamefont {Greffet}},\ }\href@noop {}
  {\bibfield  {journal} {\bibinfo  {journal} {Phys. Rev. B}\ }\textbf {\bibinfo
  {volume} {77}},\ \bibinfo {pages} {125402} (\bibinfo {year}
  {2008}{\natexlab{a}})}\BibitemShut {NoStop}%
\bibitem [{\citenamefont {Chapuis}\ \emph
  {et~al.}(2008{\natexlab{b}})\citenamefont {Chapuis}, \citenamefont {Laroche},
  \citenamefont {Volz},\ and\ \citenamefont {Greffet}}]{gold3}%
  \BibitemOpen
  \bibfield  {author} {\bibinfo {author} {\bibfnamefont {P.-O.}\ \bibnamefont
  {Chapuis}}, \bibinfo {author} {\bibfnamefont {M.}~\bibnamefont {Laroche}},
  \bibinfo {author} {\bibfnamefont {S.}~\bibnamefont {Volz}},\ and\ \bibinfo
  {author} {\bibfnamefont {J.-J.}\ \bibnamefont {Greffet}},\ }\href@noop {}
  {\bibfield  {journal} {\bibinfo  {journal} {Appl. Phys. Lett.}\ }\textbf
  {\bibinfo {volume} {92}},\ \bibinfo {pages} {201906} (\bibinfo {year}
  {2008}{\natexlab{b}})}\BibitemShut {NoStop}%
\bibitem [{\citenamefont {Kr\"{u}ger}\ \emph {et~al.}(2012)\citenamefont
  {Kr\"{u}ger}, \citenamefont {Bimonte}, \citenamefont {Emig},\ and\
  \citenamefont {Kardar}}]{KruegerEtAlTraceFormulas2012}%
  \BibitemOpen
  \bibfield  {author} {\bibinfo {author} {\bibfnamefont {M.}~\bibnamefont
  {Kr\"{u}ger}}, \bibinfo {author} {\bibfnamefont {G.}~\bibnamefont {Bimonte}},
  \bibinfo {author} {\bibfnamefont {T.}~\bibnamefont {Emig}},\ and\ \bibinfo
  {author} {\bibfnamefont {M.}~\bibnamefont {Kardar}},\ }\href@noop {}
  {\bibfield  {journal} {\bibinfo  {journal} {Phys. Rev. B}\ }\textbf {\bibinfo
  {volume} {86}},\ \bibinfo {pages} {115423} (\bibinfo {year}
  {2012})}\BibitemShut {NoStop}%
\bibitem [{\citenamefont {Messina}\ and\ \citenamefont
  {Antezza}(2011{\natexlab{a}})}]{scatt}%
  \BibitemOpen
  \bibfield  {author} {\bibinfo {author} {\bibfnamefont {R.}~\bibnamefont
  {Messina}}\ and\ \bibinfo {author} {\bibfnamefont {M.}~\bibnamefont
  {Antezza}},\ }\href@noop {} {\bibfield  {journal} {\bibinfo  {journal}
  {Europhys. Lett.}\ }\textbf {\bibinfo {volume} {95}},\ \bibinfo {pages}
  {61002} (\bibinfo {year} {2011}{\natexlab{a}})}\BibitemShut {NoStop}%
\bibitem [{\citenamefont {Messina}\ and\ \citenamefont
  {Antezza}(2011{\natexlab{b}})}]{scatt2}%
  \BibitemOpen
  \bibfield  {author} {\bibinfo {author} {\bibfnamefont {R.}~\bibnamefont
  {Messina}}\ and\ \bibinfo {author} {\bibfnamefont {M.}~\bibnamefont
  {Antezza}},\ }\href@noop {} {\bibfield  {journal} {\bibinfo  {journal} {Phys.
  Rev. A}\ }\textbf {\bibinfo {volume} {84}},\ \bibinfo {pages} {042102}
  (\bibinfo {year} {2011}{\natexlab{b}})}\BibitemShut {NoStop}%
\bibitem [{\citenamefont {Zhu}\ \emph {et~al.}(2018)\citenamefont {Zhu},
  \citenamefont {Guo},\ and\ \citenamefont {Fan}}]{ZhuEtAl2018}%
  \BibitemOpen
  \bibfield  {author} {\bibinfo {author} {\bibfnamefont {L.}~\bibnamefont
  {Zhu}}, \bibinfo {author} {\bibfnamefont {Y.}~\bibnamefont {Guo}},\ and\
  \bibinfo {author} {\bibfnamefont {S.}~\bibnamefont {Fan}},\ }\href@noop {}
  {\bibfield  {journal} {\bibinfo  {journal} {Phys. Rev. B}\ }\textbf {\bibinfo
  {volume} {97}},\ \bibinfo {pages} {094302} (\bibinfo {year}
  {2018})}\BibitemShut {NoStop}%
\bibitem [{\citenamefont {Rodriguez}\ \emph {et~al.}(2013)\citenamefont
  {Rodriguez}, \citenamefont {Reid},\ and\ \citenamefont
  {Johnson}}]{RodriguezEtAl2013}%
  \BibitemOpen
  \bibfield  {author} {\bibinfo {author} {\bibfnamefont {A.~W.}\ \bibnamefont
  {Rodriguez}}, \bibinfo {author} {\bibfnamefont {M.~T.~H.}\ \bibnamefont
  {Reid}},\ and\ \bibinfo {author} {\bibfnamefont {S.~G.}\ \bibnamefont
  {Johnson}},\ }\href@noop {} {\bibfield  {journal} {\bibinfo  {journal} {Phys.
  Rev. B}\ }\textbf {\bibinfo {volume} {88}},\ \bibinfo {pages} {054305}
  (\bibinfo {year} {2013})}\BibitemShut {NoStop}%
\bibitem [{\citenamefont {Polimeridis}\ \emph {et~al.}(2015)\citenamefont
  {Polimeridis}, \citenamefont {Reid}, \citenamefont {Jin}, \citenamefont
  {Johnson}, \citenamefont {White},\ and\ \citenamefont {Rodriguez}}]{Pol2015}%
  \BibitemOpen
  \bibfield  {author} {\bibinfo {author} {\bibfnamefont {A.~G.}\ \bibnamefont
  {Polimeridis}}, \bibinfo {author} {\bibfnamefont {M.~T.~H.}\ \bibnamefont
  {Reid}}, \bibinfo {author} {\bibfnamefont {W.}~\bibnamefont {Jin}}, \bibinfo
  {author} {\bibfnamefont {S.~G.}\ \bibnamefont {Johnson}}, \bibinfo {author}
  {\bibfnamefont {J.~K.}\ \bibnamefont {White}},\ and\ \bibinfo {author}
  {\bibfnamefont {A.~W.}\ \bibnamefont {Rodriguez}},\ }\href@noop {} {\bibfield
   {journal} {\bibinfo  {journal} {Phys. Rev. B}\ }\textbf {\bibinfo {volume}
  {92}},\ \bibinfo {pages} {134202} (\bibinfo {year} {2015})}\BibitemShut
  {NoStop}%
\bibitem [{\citenamefont {Jin}\ \emph {et~al.}(2016)\citenamefont {Jin},
  \citenamefont {Polimeridis},\ and\ \citenamefont {Rodriguez}}]{JinWEtAl2016}%
  \BibitemOpen
  \bibfield  {author} {\bibinfo {author} {\bibfnamefont {W.}~\bibnamefont
  {Jin}}, \bibinfo {author} {\bibfnamefont {A.~G.}\ \bibnamefont
  {Polimeridis}},\ and\ \bibinfo {author} {\bibfnamefont {A.~W.}\ \bibnamefont
  {Rodriguez}},\ }\href@noop {} {\bibfield  {journal} {\bibinfo  {journal}
  {Phys. Rev. B}\ }\textbf {\bibinfo {volume} {93}},\ \bibinfo {pages}
  {121403(R)} (\bibinfo {year} {2016})}\BibitemShut {NoStop}%
\bibitem [{\citenamefont {S\"{a}\"{a}skilahti}\ \emph
  {et~al.}(2014)\citenamefont {S\"{a}\"{a}skilahti}, \citenamefont {Oksanen},\
  and\ \citenamefont {Tulkki}}]{SaaskilahtiEtAl2014}%
  \BibitemOpen
  \bibfield  {author} {\bibinfo {author} {\bibfnamefont {K.}~\bibnamefont
  {S\"{a}\"{a}skilahti}}, \bibinfo {author} {\bibfnamefont {J.}~\bibnamefont
  {Oksanen}},\ and\ \bibinfo {author} {\bibfnamefont {J.}~\bibnamefont
  {Tulkki}},\ }\href@noop {} {\bibfield  {journal} {\bibinfo  {journal} {Phys.
  Rev. B}\ }\textbf {\bibinfo {volume} {89}},\ \bibinfo {pages} {134301}
  (\bibinfo {year} {2014})}\BibitemShut {NoStop}%
\bibitem [{\citenamefont {Dong}\ \emph {et~al.}(2018)\citenamefont {Dong},
  \citenamefont {Zhao},\ and\ \citenamefont {Liu}}]{DongEtAl2018}%
  \BibitemOpen
  \bibfield  {author} {\bibinfo {author} {\bibfnamefont {J.}~\bibnamefont
  {Dong}}, \bibinfo {author} {\bibfnamefont {J.}~\bibnamefont {Zhao}},\ and\
  \bibinfo {author} {\bibfnamefont {L.}~\bibnamefont {Liu}},\ }\href@noop {}
  {\bibfield  {journal} {\bibinfo  {journal} {Phys. Rev. B}\ }\textbf {\bibinfo
  {volume} {97}},\ \bibinfo {pages} {075422} (\bibinfo {year}
  {2018})}\BibitemShut {NoStop}%
\bibitem [{\citenamefont {Messina}\ \emph {et~al.}(2018)\citenamefont
  {Messina}, \citenamefont {Biehs},\ and\ \citenamefont
  {Ben-Abdallah}}]{gold1}%
  \BibitemOpen
  \bibfield  {author} {\bibinfo {author} {\bibfnamefont {R.}~\bibnamefont
  {Messina}}, \bibinfo {author} {\bibfnamefont {S.-A.}\ \bibnamefont {Biehs}},\
  and\ \bibinfo {author} {\bibfnamefont {P.}~\bibnamefont {Ben-Abdallah}},\
  }\href@noop {} {\bibfield  {journal} {\bibinfo  {journal} {Phys. Rev. B}\
  }\textbf {\bibinfo {volume} {97}},\ \bibinfo {pages} {165437} (\bibinfo
  {year} {2018})}\BibitemShut {NoStop}%
\bibitem [{\citenamefont {Asheichyk}\ \emph {et~al.}(2017)\citenamefont
  {Asheichyk}, \citenamefont {M\"{u}ller},\ and\ \citenamefont
  {Kr\"{u}ger}}]{AsheichykEtAl2017}%
  \BibitemOpen
  \bibfield  {author} {\bibinfo {author} {\bibfnamefont {K.}~\bibnamefont
  {Asheichyk}}, \bibinfo {author} {\bibfnamefont {B.}~\bibnamefont
  {M\"{u}ller}},\ and\ \bibinfo {author} {\bibfnamefont {M.}~\bibnamefont
  {Kr\"{u}ger}},\ }\href@noop {} {\bibfield  {journal} {\bibinfo  {journal}
  {Phys. Rev. B}\ }\textbf {\bibinfo {volume} {96}},\ \bibinfo {pages} {155402}
  (\bibinfo {year} {2017})}\BibitemShut {NoStop}%
\bibitem [{\citenamefont {Asheichyk}\ and\ \citenamefont
  {Kr\"{u}ger}(2018)}]{Asheichyk2018}%
  \BibitemOpen
  \bibfield  {author} {\bibinfo {author} {\bibfnamefont {K.}~\bibnamefont
  {Asheichyk}}\ and\ \bibinfo {author} {\bibfnamefont {M.}~\bibnamefont
  {Kr\"{u}ger}},\ }\href@noop {} {\bibfield  {journal} {\bibinfo  {journal}
  {Phys. Rev. B}\ }\textbf {\bibinfo {volume} {98}},\ \bibinfo {pages} {195401}
  (\bibinfo {year} {2018})}\BibitemShut {NoStop}%
\bibitem [{\citenamefont {Yang}\ \emph {et~al.}(2021)\citenamefont {Yang},
  \citenamefont {Zhang}, \citenamefont {Yuan}, \citenamefont {Zhou},\ and\
  \citenamefont {Yi}}]{YangEtAl2021}%
  \BibitemOpen
  \bibfield  {author} {\bibinfo {author} {\bibfnamefont {S.-H.}\ \bibnamefont
  {Yang}}, \bibinfo {author} {\bibfnamefont {Y.}~\bibnamefont {Zhang}},
  \bibinfo {author} {\bibfnamefont {M.-Q.}\ \bibnamefont {Yuan}}, \bibinfo
  {author} {\bibfnamefont {C.-L.}\ \bibnamefont {Zhou}},\ and\ \bibinfo
  {author} {\bibfnamefont {H.-L.}\ \bibnamefont {Yi}},\ }\href@noop {}
  {\bibfield  {journal} {\bibinfo  {journal} {Phys. Rev. B}\ }\textbf {\bibinfo
  {volume} {104}},\ \bibinfo {pages} {125417} (\bibinfo {year}
  {2021})}\BibitemShut {NoStop}%
\bibitem [{\citenamefont {Zhang}\ \emph {et~al.}(2019)\citenamefont {Zhang},
  \citenamefont {Antezza}, \citenamefont {Yi},\ and\ \citenamefont
  {Tan}}]{ZhangEtAl2019b}%
  \BibitemOpen
  \bibfield  {author} {\bibinfo {author} {\bibfnamefont {Y.}~\bibnamefont
  {Zhang}}, \bibinfo {author} {\bibfnamefont {M.}~\bibnamefont {Antezza}},
  \bibinfo {author} {\bibfnamefont {H.-L.}\ \bibnamefont {Yi}},\ and\ \bibinfo
  {author} {\bibfnamefont {H.-P.}\ \bibnamefont {Tan}},\ }\href@noop {}
  {\bibfield  {journal} {\bibinfo  {journal} {Phys. Rev. B}\ }\textbf {\bibinfo
  {volume} {100}},\ \bibinfo {pages} {085426} (\bibinfo {year}
  {2019})}\BibitemShut {NoStop}%
\bibitem [{\citenamefont {Ott}\ \emph {et~al.}(2021{\natexlab{a}})\citenamefont
  {Ott}, \citenamefont {Hu}, \citenamefont {Wu},\ and\ \citenamefont
  {Biehs}}]{OttEtAl2021}%
  \BibitemOpen
  \bibfield  {author} {\bibinfo {author} {\bibfnamefont {A.}~\bibnamefont
  {Ott}}, \bibinfo {author} {\bibfnamefont {Y.}~\bibnamefont {Hu}}, \bibinfo
  {author} {\bibfnamefont {X.-H.}\ \bibnamefont {Wu}},\ and\ \bibinfo {author}
  {\bibfnamefont {S.-A.}\ \bibnamefont {Biehs}},\ }\href@noop {} {\bibfield
  {journal} {\bibinfo  {journal} {Phys. Rev. Appl.}\ }\textbf {\bibinfo
  {volume} {15}},\ \bibinfo {pages} {064073} (\bibinfo {year}
  {2021}{\natexlab{a}})}\BibitemShut {NoStop}%
\bibitem [{\citenamefont {Zhu}\ and\ \citenamefont {Fan}(2016)}]{Linxiao2016}%
  \BibitemOpen
  \bibfield  {author} {\bibinfo {author} {\bibfnamefont {L.}~\bibnamefont
  {Zhu}}\ and\ \bibinfo {author} {\bibfnamefont {S.}~\bibnamefont {Fan}},\
  }\href@noop {} {\bibfield  {journal} {\bibinfo  {journal} {Phys. Rev. Lett.}\
  }\textbf {\bibinfo {volume} {117}},\ \bibinfo {pages} {134303} (\bibinfo
  {year} {2016})}\BibitemShut {NoStop}%
\bibitem [{\citenamefont {Silveirinha}(2017)}]{Silveirinha2017}%
  \BibitemOpen
  \bibfield  {author} {\bibinfo {author} {\bibfnamefont {M.~G.}\ \bibnamefont
  {Silveirinha}},\ }\href@noop {} {\bibfield  {journal} {\bibinfo  {journal}
  {Phys. Rev. B}\ }\textbf {\bibinfo {volume} {95}},\ \bibinfo {pages} {115103}
  (\bibinfo {year} {2017})}\BibitemShut {NoStop}%
\bibitem [{\citenamefont {Guo}\ \emph {et~al.}(2019)\citenamefont {Guo},
  \citenamefont {Guo},\ and\ \citenamefont {Fan}}]{GueEtAl2019}%
  \BibitemOpen
  \bibfield  {author} {\bibinfo {author} {\bibfnamefont {C.}~\bibnamefont
  {Guo}}, \bibinfo {author} {\bibfnamefont {Y.}~\bibnamefont {Guo}},\ and\
  \bibinfo {author} {\bibfnamefont {S.}~\bibnamefont {Fan}},\ }\href@noop {}
  {\bibfield  {journal} {\bibinfo  {journal} {Phys. Rev. B}\ }\textbf {\bibinfo
  {volume} {100}},\ \bibinfo {pages} {205416} (\bibinfo {year}
  {2019})}\BibitemShut {NoStop}%
\bibitem [{\citenamefont {Ott}\ \emph {et~al.}(2019{\natexlab{a}})\citenamefont
  {Ott}, \citenamefont {Messina}, \citenamefont {Ben-Abdallah},\ and\
  \citenamefont {Biehs}}]{OttEtAl2019}%
  \BibitemOpen
  \bibfield  {author} {\bibinfo {author} {\bibfnamefont {A.}~\bibnamefont
  {Ott}}, \bibinfo {author} {\bibfnamefont {R.}~\bibnamefont {Messina}},
  \bibinfo {author} {\bibfnamefont {P.}~\bibnamefont {Ben-Abdallah}},\ and\
  \bibinfo {author} {\bibfnamefont {S.-A.}\ \bibnamefont {Biehs}},\ }\href@noop
  {} {\bibfield  {journal} {\bibinfo  {journal} {J. Photonics Energy}\ }\textbf
  {\bibinfo {volume} {9}},\ \bibinfo {pages} {032711} (\bibinfo {year}
  {2019}{\natexlab{a}})}\BibitemShut {NoStop}%
\bibitem [{\citenamefont {Latella}\ and\ \citenamefont
  {Ben-Abdallah}(2017)}]{LatellaPBA2018}%
  \BibitemOpen
  \bibfield  {author} {\bibinfo {author} {\bibfnamefont {I.}~\bibnamefont
  {Latella}}\ and\ \bibinfo {author} {\bibfnamefont {P.}~\bibnamefont
  {Ben-Abdallah}},\ }\href@noop {} {\bibfield  {journal} {\bibinfo  {journal}
  {Phys. Rev. Lett.}\ }\textbf {\bibinfo {volume} {118}},\ \bibinfo {pages}
  {173902} (\bibinfo {year} {2017})}\BibitemShut {NoStop}%
\bibitem [{\citenamefont {Moncada-Villa}\ \emph {et~al.}(2015)\citenamefont
  {Moncada-Villa}, \citenamefont {Fern\'{a}ndez-Hurtado}, \citenamefont
  {Garc\'{i}a-Vidal}, \citenamefont {Garc\'{i}a-Mart\'{i}n},\ and\
  \citenamefont {Cuevas}}]{VillaEtAL2015}%
  \BibitemOpen
  \bibfield  {author} {\bibinfo {author} {\bibfnamefont {E.}~\bibnamefont
  {Moncada-Villa}}, \bibinfo {author} {\bibfnamefont {V.}~\bibnamefont
  {Fern\'{a}ndez-Hurtado}}, \bibinfo {author} {\bibfnamefont {F.~J.}\
  \bibnamefont {Garc\'{i}a-Vidal}}, \bibinfo {author} {\bibfnamefont
  {A.}~\bibnamefont {Garc\'{i}a-Mart\'{i}n}},\ and\ \bibinfo {author}
  {\bibfnamefont {J.~C.}\ \bibnamefont {Cuevas}},\ }\href@noop {} {\bibfield
  {journal} {\bibinfo  {journal} {Phys. Rev. B}\ }\textbf {\bibinfo {volume}
  {92}},\ \bibinfo {pages} {125418} (\bibinfo {year} {2015})}\BibitemShut
  {NoStop}%
\bibitem [{\citenamefont {Ben-Abdallah}(2016)}]{PBA2017}%
  \BibitemOpen
  \bibfield  {author} {\bibinfo {author} {\bibfnamefont {P.}~\bibnamefont
  {Ben-Abdallah}},\ }\href@noop {} {\bibfield  {journal} {\bibinfo  {journal}
  {Phys. Rev. Lett.}\ }\textbf {\bibinfo {volume} {116}},\ \bibinfo {pages}
  {084301} (\bibinfo {year} {2016})}\BibitemShut {NoStop}%
\bibitem [{\citenamefont {Ott}\ \emph {et~al.}(2020)\citenamefont {Ott},
  \citenamefont {Biehs},\ and\ \citenamefont {Ben-Abdallah}}]{OttEtAlanom2020}%
  \BibitemOpen
  \bibfield  {author} {\bibinfo {author} {\bibfnamefont {A.}~\bibnamefont
  {Ott}}, \bibinfo {author} {\bibfnamefont {S.-A.}\ \bibnamefont {Biehs}},\
  and\ \bibinfo {author} {\bibfnamefont {P.}~\bibnamefont {Ben-Abdallah}},\
  }\href@noop {} {\bibfield  {journal} {\bibinfo  {journal} {Phys. Rev. B}\
  }\textbf {\bibinfo {volume} {101}},\ \bibinfo {pages} {241411(R)} (\bibinfo
  {year} {2020})}\BibitemShut {NoStop}%
\bibitem [{\citenamefont {Ott}\ \emph {et~al.}(2018)\citenamefont {Ott},
  \citenamefont {Ben-Abdallah},\ and\ \citenamefont {Biehs}}]{OttEtAl2018}%
  \BibitemOpen
  \bibfield  {author} {\bibinfo {author} {\bibfnamefont {A.}~\bibnamefont
  {Ott}}, \bibinfo {author} {\bibfnamefont {P.}~\bibnamefont {Ben-Abdallah}},\
  and\ \bibinfo {author} {\bibfnamefont {S.-A.}\ \bibnamefont {Biehs}},\
  }\href@noop {} {\bibfield  {journal} {\bibinfo  {journal} {Phys. Rev. B}\
  }\textbf {\bibinfo {volume} {97}},\ \bibinfo {pages} {205414} (\bibinfo
  {year} {2018})}\BibitemShut {NoStop}%
\bibitem [{\citenamefont {Khandekar}\ and\ \citenamefont
  {Jacob}(2019)}]{Khandekar2019b}%
  \BibitemOpen
  \bibfield  {author} {\bibinfo {author} {\bibfnamefont {C.}~\bibnamefont
  {Khandekar}}\ and\ \bibinfo {author} {\bibfnamefont {Z.}~\bibnamefont
  {Jacob}},\ }\href@noop {} {\bibfield  {journal} {\bibinfo  {journal} {New J.
  Phys.}\ }\textbf {\bibinfo {volume} {21}},\ \bibinfo {pages} {103030}
  (\bibinfo {year} {2019})}\BibitemShut {NoStop}%
\bibitem [{\citenamefont {Ott}\ \emph {et~al.}(2019{\natexlab{b}})\citenamefont
  {Ott}, \citenamefont {Messina}, \citenamefont {Ben-Abdallah},\ and\
  \citenamefont {Biehs}}]{Konzeptdiode10}%
  \BibitemOpen
  \bibfield  {author} {\bibinfo {author} {\bibfnamefont {A.}~\bibnamefont
  {Ott}}, \bibinfo {author} {\bibfnamefont {R.}~\bibnamefont {Messina}},
  \bibinfo {author} {\bibfnamefont {P.}~\bibnamefont {Ben-Abdallah}},\ and\
  \bibinfo {author} {\bibfnamefont {S.-A.}\ \bibnamefont {Biehs}},\ }\href@noop
  {} {\bibfield  {journal} {\bibinfo  {journal} {Appl. Phys. Lett.}\ }\textbf
  {\bibinfo {volume} {114}},\ \bibinfo {pages} {163105} (\bibinfo {year}
  {2019}{\natexlab{b}})}\BibitemShut {NoStop}%
\bibitem [{\citenamefont {.Ott}\ and\ \citenamefont
  {Biehs}(2020)}]{OttSAB2020}%
  \BibitemOpen
  \bibfield  {author} {\bibinfo {author} {\bibfnamefont {A.}~\bibnamefont
  {.Ott}}\ and\ \bibinfo {author} {\bibfnamefont {S.-A.}\ \bibnamefont
  {Biehs}},\ }\href@noop {} {\bibfield  {journal} {\bibinfo  {journal} {Phys.
  Rev. B}\ }\textbf {\bibinfo {volume} {101}},\ \bibinfo {pages} {155428}
  (\bibinfo {year} {2020})}\BibitemShut {NoStop}%
\bibitem [{\citenamefont {Dong}\ \emph {et~al.}(2021)\citenamefont {Dong},
  \citenamefont {Zhang},\ and\ \citenamefont {Liu}}]{DongEtAl2021}%
  \BibitemOpen
  \bibfield  {author} {\bibinfo {author} {\bibfnamefont {J.}~\bibnamefont
  {Dong}}, \bibinfo {author} {\bibfnamefont {W.}~\bibnamefont {Zhang}},\ and\
  \bibinfo {author} {\bibfnamefont {L.}~\bibnamefont {Liu}},\ }\href@noop {}
  {\bibfield  {journal} {\bibinfo  {journal} {Appl. Phys. Lett.}\ }\textbf
  {\bibinfo {volume} {119}},\ \bibinfo {pages} {021104} (\bibinfo {year}
  {2021})}\BibitemShut {NoStop}%
\bibitem [{\citenamefont {Nikbakht}(2014)}]{Nikbakht2014a}%
  \BibitemOpen
  \bibfield  {author} {\bibinfo {author} {\bibfnamefont {M.}~\bibnamefont
  {Nikbakht}},\ }\href@noop {} {\bibfield  {journal} {\bibinfo  {journal} {J.
  Appl. Phys.}\ }\textbf {\bibinfo {volume} {116}},\ \bibinfo {pages} {094307}
  (\bibinfo {year} {2014})}\BibitemShut {NoStop}%
\bibitem [{\citenamefont {Incardone}\ \emph {et~al.}(2014)\citenamefont
  {Incardone}, \citenamefont {Emig},\ and\ \citenamefont
  {Kr\"{u}ger}}]{IncardoneEtAl2014}%
  \BibitemOpen
  \bibfield  {author} {\bibinfo {author} {\bibfnamefont {R.}~\bibnamefont
  {Incardone}}, \bibinfo {author} {\bibfnamefont {T.}~\bibnamefont {Emig}},\
  and\ \bibinfo {author} {\bibfnamefont {M.}~\bibnamefont {Kr\"{u}ger}},\
  }\href@noop {} {\bibfield  {journal} {\bibinfo  {journal} {Europhys. Lett.}\
  }\textbf {\bibinfo {volume} {106}},\ \bibinfo {pages} {41001} (\bibinfo
  {year} {2014})}\BibitemShut {NoStop}%
\bibitem [{\citenamefont {Nikbakht}(2015)}]{Nikbakht2015}%
  \BibitemOpen
  \bibfield  {author} {\bibinfo {author} {\bibfnamefont {M.}~\bibnamefont
  {Nikbakht}},\ }\href@noop {} {\bibfield  {journal} {\bibinfo  {journal}
  {Europhys. Lett.}\ }\textbf {\bibinfo {volume} {110}},\ \bibinfo {pages}
  {14004} (\bibinfo {year} {2015})}\BibitemShut {NoStop}%
\bibitem [{\citenamefont {Nikbakht}(2017)}]{Nikbakht2017}%
  \BibitemOpen
  \bibfield  {author} {\bibinfo {author} {\bibfnamefont {M.}~\bibnamefont
  {Nikbakht}},\ }\href@noop {} {\bibfield  {journal} {\bibinfo  {journal}
  {Phys. Rev. B}\ }\textbf {\bibinfo {volume} {96}},\ \bibinfo {pages} {125436}
  (\bibinfo {year} {2017})}\BibitemShut {NoStop}%
\bibitem [{\citenamefont {Luo}\ \emph {et~al.}(2019)\citenamefont {Luo},
  \citenamefont {Dong}, \citenamefont {Zhao}, \citenamefont {Liu},\ and\
  \citenamefont {Antezza}}]{Luo}%
  \BibitemOpen
  \bibfield  {author} {\bibinfo {author} {\bibfnamefont {M.}~\bibnamefont
  {Luo}}, \bibinfo {author} {\bibfnamefont {J.}~\bibnamefont {Dong}}, \bibinfo
  {author} {\bibfnamefont {J.}~\bibnamefont {Zhao}}, \bibinfo {author}
  {\bibfnamefont {L.}~\bibnamefont {Liu}},\ and\ \bibinfo {author}
  {\bibfnamefont {M.}~\bibnamefont {Antezza}},\ }\href@noop {} {\bibfield
  {journal} {\bibinfo  {journal} {Phys. Rev. B}\ }\textbf {\bibinfo {volume}
  {99}},\ \bibinfo {pages} {134207} (\bibinfo {year} {2019})}\BibitemShut
  {NoStop}%
\bibitem [{\citenamefont {Phan}\ \emph {et~al.}(2013)\citenamefont {Phan},
  \citenamefont {Phan},\ and\ \citenamefont {Woods}}]{PhanEtAl2013}%
  \BibitemOpen
  \bibfield  {author} {\bibinfo {author} {\bibfnamefont {A.~D.}\ \bibnamefont
  {Phan}}, \bibinfo {author} {\bibfnamefont {T.-L.}\ \bibnamefont {Phan}},\
  and\ \bibinfo {author} {\bibfnamefont {L.~M.}\ \bibnamefont {Woods}},\
  }\href@noop {} {\bibfield  {journal} {\bibinfo  {journal} {J. Appl. Phys.}\
  }\textbf {\bibinfo {volume} {114}},\ \bibinfo {pages} {214306} (\bibinfo
  {year} {2013})}\BibitemShut {NoStop}%
\bibitem [{\citenamefont {Luo}\ \emph {et~al.}(2020)\citenamefont {Luo},
  \citenamefont {Zhao}, \citenamefont {Liu},\ and\ \citenamefont
  {Antezza}}]{LuoEtAl2020}%
  \BibitemOpen
  \bibfield  {author} {\bibinfo {author} {\bibfnamefont {M.}~\bibnamefont
  {Luo}}, \bibinfo {author} {\bibfnamefont {J.}~\bibnamefont {Zhao}}, \bibinfo
  {author} {\bibfnamefont {L.}~\bibnamefont {Liu}},\ and\ \bibinfo {author}
  {\bibfnamefont {M.}~\bibnamefont {Antezza}},\ }\href@noop {} {\bibfield
  {journal} {\bibinfo  {journal} {Phys. Rev. B}\ }\textbf {\bibinfo {volume}
  {102}},\ \bibinfo {pages} {024203} (\bibinfo {year} {2020})}\BibitemShut
  {NoStop}%
\bibitem [{\citenamefont {Ben-Abdallah}\ \emph {et~al.}(2013)\citenamefont
  {Ben-Abdallah}, \citenamefont {Messina}, \citenamefont {Biehs}, \citenamefont
  {Tschikin}, \citenamefont {Joulain},\ and\ \citenamefont
  {Henkel}}]{PBAEtAlSuperdiff2013}%
  \BibitemOpen
  \bibfield  {author} {\bibinfo {author} {\bibfnamefont {P.}~\bibnamefont
  {Ben-Abdallah}}, \bibinfo {author} {\bibfnamefont {R.}~\bibnamefont
  {Messina}}, \bibinfo {author} {\bibfnamefont {S.-A.}\ \bibnamefont {Biehs}},
  \bibinfo {author} {\bibfnamefont {M.}~\bibnamefont {Tschikin}}, \bibinfo
  {author} {\bibfnamefont {K.}~\bibnamefont {Joulain}},\ and\ \bibinfo {author}
  {\bibfnamefont {C.}~\bibnamefont {Henkel}},\ }\href@noop {} {\bibfield
  {journal} {\bibinfo  {journal} {Phys. Rev. Lett.}\ }\textbf {\bibinfo
  {volume} {111}},\ \bibinfo {pages} {174301} (\bibinfo {year}
  {2013})}\BibitemShut {NoStop}%
\bibitem [{\citenamefont {Kathmann}\ \emph {et~al.}(2018)\citenamefont
  {Kathmann}, \citenamefont {Messina}, \citenamefont {Ben-Abdallah},\ and\
  \citenamefont {Biehs}}]{KathmannEtAl2018}%
  \BibitemOpen
  \bibfield  {author} {\bibinfo {author} {\bibfnamefont {C.}~\bibnamefont
  {Kathmann}}, \bibinfo {author} {\bibfnamefont {R.}~\bibnamefont {Messina}},
  \bibinfo {author} {\bibfnamefont {P.}~\bibnamefont {Ben-Abdallah}},\ and\
  \bibinfo {author} {\bibfnamefont {S.-A.}\ \bibnamefont {Biehs}},\ }\href@noop
  {} {\bibfield  {journal} {\bibinfo  {journal} {Phys. Rev. B}\ }\textbf
  {\bibinfo {volume} {98}},\ \bibinfo {pages} {115434} (\bibinfo {year}
  {2018})}\BibitemShut {NoStop}%
\bibitem [{\citenamefont {Tervo}\ \emph {et~al.}(2019)\citenamefont {Tervo},
  \citenamefont {Francoeur}, \citenamefont {Cola},\ and\ \citenamefont
  {Zhang}}]{TervoEtAl2019}%
  \BibitemOpen
  \bibfield  {author} {\bibinfo {author} {\bibfnamefont {E.}~\bibnamefont
  {Tervo}}, \bibinfo {author} {\bibfnamefont {M.}~\bibnamefont {Francoeur}},
  \bibinfo {author} {\bibfnamefont {B.}~\bibnamefont {Cola}},\ and\ \bibinfo
  {author} {\bibfnamefont {Z.}~\bibnamefont {Zhang}},\ }\href@noop {}
  {\bibfield  {journal} {\bibinfo  {journal} {Phys. Rev. B}\ }\textbf {\bibinfo
  {volume} {100}},\ \bibinfo {pages} {205422} (\bibinfo {year}
  {2019})}\BibitemShut {NoStop}%
\bibitem [{\citenamefont {Tervo}\ \emph {et~al.}(2020)\citenamefont {Tervo},
  \citenamefont {Cola},\ and\ \citenamefont {Zhang}}]{tervo}%
  \BibitemOpen
  \bibfield  {author} {\bibinfo {author} {\bibfnamefont {E.}~\bibnamefont
  {Tervo}}, \bibinfo {author} {\bibfnamefont {B.}~\bibnamefont {Cola}},\ and\
  \bibinfo {author} {\bibfnamefont {Z.}~\bibnamefont {Zhang}},\ }\href@noop {}
  {\bibfield  {journal} {\bibinfo  {journal} {J. Quant. Spectrosc. Radiat.
  Transf.}\ }\textbf {\bibinfo {volume} {246}},\ \bibinfo {pages} {106947}
  (\bibinfo {year} {2020})}\BibitemShut {NoStop}%
\bibitem [{\citenamefont {Ott}\ and\ \citenamefont
  {Biehs}(2020)}]{OttSAB2020b}%
  \BibitemOpen
  \bibfield  {author} {\bibinfo {author} {\bibfnamefont {A.}~\bibnamefont
  {Ott}}\ and\ \bibinfo {author} {\bibfnamefont {S.-A.}\ \bibnamefont
  {Biehs}},\ }\href@noop {} {\bibfield  {journal} {\bibinfo  {journal} {Phys.
  Rev. B}\ }\textbf {\bibinfo {volume} {102}},\ \bibinfo {pages} {115417}
  (\bibinfo {year} {2020})}\BibitemShut {NoStop}%
\bibitem [{\citenamefont {Neuh\"{o}fer}\ \emph {et~al.}(2021)\citenamefont
  {Neuh\"{o}fer}, \citenamefont {Herrmann}, \citenamefont {Lebeda},
  \citenamefont {Lauster}, \citenamefont {Kathmann}, \citenamefont {Biehs},\
  and\ \citenamefont {Retsch}}]{Retsch2021}%
  \BibitemOpen
  \bibfield  {author} {\bibinfo {author} {\bibfnamefont {A.~M.}\ \bibnamefont
  {Neuh\"{o}fer}}, \bibinfo {author} {\bibfnamefont {K.}~\bibnamefont
  {Herrmann}}, \bibinfo {author} {\bibfnamefont {F.}~\bibnamefont {Lebeda}},
  \bibinfo {author} {\bibfnamefont {T.}~\bibnamefont {Lauster}}, \bibinfo
  {author} {\bibfnamefont {C.}~\bibnamefont {Kathmann}}, \bibinfo {author}
  {\bibfnamefont {S.-A.}\ \bibnamefont {Biehs}},\ and\ \bibinfo {author}
  {\bibfnamefont {M.}~\bibnamefont {Retsch}},\ }\href@noop {} {\bibfield
  {journal} {\bibinfo  {journal} {Adv. Funct. Mater.}\ }\textbf {\bibinfo
  {volume} {32}},\ \bibinfo {pages} {2108370} (\bibinfo {year}
  {2021})}\BibitemShut {NoStop}%
\bibitem [{\citenamefont {Biehs}\ \emph {et~al.}(2021)\citenamefont {Biehs},
  \citenamefont {Messina}, \citenamefont {Venkataram}, \citenamefont
  {Rodriguez}, \citenamefont {Cuevas},\ and\ \citenamefont
  {Ben-Abdallah}}]{SABRMB2021}%
  \BibitemOpen
  \bibfield  {author} {\bibinfo {author} {\bibfnamefont {S.-A.}\ \bibnamefont
  {Biehs}}, \bibinfo {author} {\bibfnamefont {R.}~\bibnamefont {Messina}},
  \bibinfo {author} {\bibfnamefont {P.~S.}\ \bibnamefont {Venkataram}},
  \bibinfo {author} {\bibfnamefont {A.~W.}\ \bibnamefont {Rodriguez}}, \bibinfo
  {author} {\bibfnamefont {J.~C.}\ \bibnamefont {Cuevas}},\ and\ \bibinfo
  {author} {\bibfnamefont {P.}~\bibnamefont {Ben-Abdallah}},\ }\href@noop {}
  {\bibfield  {journal} {\bibinfo  {journal} {Rev. Mod. Phys.}\ }\textbf
  {\bibinfo {volume} {93}},\ \bibinfo {pages} {025009} (\bibinfo {year}
  {2021})}\BibitemShut {NoStop}%
\bibitem [{\citenamefont {Song}\ \emph {et~al.}(2021)\citenamefont {Song},
  \citenamefont {Cheng}, \citenamefont {Zhang}, \citenamefont {L.~Lu},
  \citenamefont {Luo},\ and\ \citenamefont {Hu}}]{SongPlagiat2021}%
  \BibitemOpen
  \bibfield  {author} {\bibinfo {author} {\bibfnamefont {J.}~\bibnamefont
  {Song}}, \bibinfo {author} {\bibfnamefont {Q.}~\bibnamefont {Cheng}},
  \bibinfo {author} {\bibfnamefont {B.}~\bibnamefont {Zhang}}, \bibinfo
  {author} {\bibfnamefont {X.~Z.}\ \bibnamefont {L.~Lu}}, \bibinfo {author}
  {\bibfnamefont {Z.}~\bibnamefont {Luo}},\ and\ \bibinfo {author}
  {\bibfnamefont {R.}~\bibnamefont {Hu}},\ }\href@noop {} {\bibfield  {journal}
  {\bibinfo  {journal} {Rep. Prog. Phys.}\ }\textbf {\bibinfo {volume} {84}},\
  \bibinfo {pages} {036501} (\bibinfo {year} {2021})}\BibitemShut {NoStop}%
\bibitem [{\citenamefont {Latella}\ \emph {et~al.}(2021)\citenamefont
  {Latella}, \citenamefont {Biehs},\ and\ \citenamefont
  {Ben-Abdallah}}]{LatellaEtAl2021}%
  \BibitemOpen
  \bibfield  {author} {\bibinfo {author} {\bibfnamefont {I.}~\bibnamefont
  {Latella}}, \bibinfo {author} {\bibfnamefont {S.-A.}\ \bibnamefont {Biehs}},\
  and\ \bibinfo {author} {\bibfnamefont {P.}~\bibnamefont {Ben-Abdallah}},\
  }\href@noop {} {\bibfield  {journal} {\bibinfo  {journal} {Opt. Express}\
  }\textbf {\bibinfo {volume} {29}},\ \bibinfo {pages} {24816} (\bibinfo {year}
  {2021})}\BibitemShut {NoStop}%
\bibitem [{\citenamefont {Edalatpour}\ and\ \citenamefont
  {Francoeur}(2014)}]{EdalatpourDDA}%
  \BibitemOpen
  \bibfield  {author} {\bibinfo {author} {\bibfnamefont {S.}~\bibnamefont
  {Edalatpour}}\ and\ \bibinfo {author} {\bibfnamefont {M.}~\bibnamefont
  {Francoeur}},\ }\href@noop {} {\bibfield  {journal} {\bibinfo  {journal} {J.
  Quant. Spectrosc. Radiat. Transf.}\ }\textbf {\bibinfo {volume} {133}},\
  \bibinfo {pages} {364} (\bibinfo {year} {2014})}\BibitemShut {NoStop}%
\bibitem [{\citenamefont {Ekeroth}\ \emph {et~al.}(2017)\citenamefont
  {Ekeroth}, \citenamefont {Garc\'{\i}a-Mart\'{\i}n},\ and\ \citenamefont
  {Cuevas}}]{Ekeroth}%
  \BibitemOpen
  \bibfield  {author} {\bibinfo {author} {\bibfnamefont {R.~A.}\ \bibnamefont
  {Ekeroth}}, \bibinfo {author} {\bibfnamefont {A.}~\bibnamefont
  {Garc\'{\i}a-Mart\'{\i}n}},\ and\ \bibinfo {author} {\bibfnamefont
  {J.}~\bibnamefont {Cuevas}},\ }\href@noop {} {\bibfield  {journal} {\bibinfo
  {journal} {Phys. Rev. B}\ }\textbf {\bibinfo {volume} {95}},\ \bibinfo
  {pages} {235428} (\bibinfo {year} {2017})}\BibitemShut {NoStop}%
\bibitem [{\citenamefont {Joulain}\ \emph {et~al.}(2014)\citenamefont
  {Joulain}, \citenamefont {Ben-Abdallah}, \citenamefont {Chapuis},
  \citenamefont {Wilde}, \citenamefont {Babuty},\ and\ \citenamefont
  {Henkel}}]{Joulain2}%
  \BibitemOpen
  \bibfield  {author} {\bibinfo {author} {\bibfnamefont {K.}~\bibnamefont
  {Joulain}}, \bibinfo {author} {\bibfnamefont {P.}~\bibnamefont
  {Ben-Abdallah}}, \bibinfo {author} {\bibfnamefont {P.-O.}\ \bibnamefont
  {Chapuis}}, \bibinfo {author} {\bibfnamefont {Y.~D.}\ \bibnamefont {Wilde}},
  \bibinfo {author} {\bibfnamefont {A.}~\bibnamefont {Babuty}},\ and\ \bibinfo
  {author} {\bibfnamefont {C.}~\bibnamefont {Henkel}},\ }\href@noop {}
  {\bibfield  {journal} {\bibinfo  {journal} {J. Quant. Spectrosc. Radiat.
  Transf.}\ }\textbf {\bibinfo {volume} {136}},\ \bibinfo {pages} {1} (\bibinfo
  {year} {2014})}\BibitemShut {NoStop}%
\bibitem [{\citenamefont {Jarzembski}\ and\ \citenamefont
  {Park}(2017)}]{Jarzembski}%
  \BibitemOpen
  \bibfield  {author} {\bibinfo {author} {\bibfnamefont {A.}~\bibnamefont
  {Jarzembski}}\ and\ \bibinfo {author} {\bibfnamefont {K.}~\bibnamefont
  {Park}},\ }\href@noop {} {\bibfield  {journal} {\bibinfo  {journal} {J.
  Quant. Spectrosc. Radiat. Transf.}\ }\textbf {\bibinfo {volume} {191}},\
  \bibinfo {pages} {67} (\bibinfo {year} {2017})}\BibitemShut {NoStop}%
\bibitem [{\citenamefont {Herz}\ \emph {et~al.}(2018)\citenamefont {Herz},
  \citenamefont {An}, \citenamefont {Komiyama},\ and\ \citenamefont
  {Biehs}}]{Herz}%
  \BibitemOpen
  \bibfield  {author} {\bibinfo {author} {\bibfnamefont {F.}~\bibnamefont
  {Herz}}, \bibinfo {author} {\bibfnamefont {Z.}~\bibnamefont {An}}, \bibinfo
  {author} {\bibfnamefont {S.}~\bibnamefont {Komiyama}},\ and\ \bibinfo
  {author} {\bibfnamefont {S.-A.}\ \bibnamefont {Biehs}},\ }\href@noop {}
  {\bibfield  {journal} {\bibinfo  {journal} {Phys. Rev. Appl.}\ }\textbf
  {\bibinfo {volume} {10}},\ \bibinfo {pages} {044051} (\bibinfo {year}
  {2018})}\BibitemShut {NoStop}%
\bibitem [{\citenamefont {Herz}\ and\ \citenamefont {Biehs}(2021)}]{Herz3}%
  \BibitemOpen
  \bibfield  {author} {\bibinfo {author} {\bibfnamefont {F.}~\bibnamefont
  {Herz}}\ and\ \bibinfo {author} {\bibfnamefont {S.-A.}\ \bibnamefont
  {Biehs}},\ }\href@noop {} {\bibfield  {journal} {\bibinfo  {journal} {J.
  Quant. Spectrosc. Radiat. Transf.}\ }\textbf {\bibinfo {volume} {266}},\
  \bibinfo {pages} {107572} (\bibinfo {year} {2021})}\BibitemShut {NoStop}%
\bibitem [{\citenamefont {Edalatpour}\ and\ \citenamefont
  {Francoeur}(2016)}]{Edalatpour2}%
  \BibitemOpen
  \bibfield  {author} {\bibinfo {author} {\bibfnamefont {S.}~\bibnamefont
  {Edalatpour}}\ and\ \bibinfo {author} {\bibfnamefont {M.}~\bibnamefont
  {Francoeur}},\ }\href@noop {} {\bibfield  {journal} {\bibinfo  {journal}
  {Phys. Rev. B}\ }\textbf {\bibinfo {volume} {94}},\ \bibinfo {pages} {045406}
  (\bibinfo {year} {2016})}\BibitemShut {NoStop}%
\bibitem [{\citenamefont {Lakhtakia}(1992)}]{Lakhtakia}%
  \BibitemOpen
  \bibfield  {author} {\bibinfo {author} {\bibfnamefont {A.}~\bibnamefont
  {Lakhtakia}},\ }\href@noop {} {\bibfield  {journal} {\bibinfo  {journal}
  {Int. J. Mod. Phys. C}\ }\textbf {\bibinfo {volume} {3}},\ \bibinfo {pages}
  {583} (\bibinfo {year} {1992})}\BibitemShut {NoStop}%
\bibitem [{\citenamefont {Fikioris}(1965)}]{Fikioris}%
  \BibitemOpen
  \bibfield  {author} {\bibinfo {author} {\bibfnamefont {J.}~\bibnamefont
  {Fikioris}},\ }\href@noop {} {\bibfield  {journal} {\bibinfo  {journal} {J.
  Math. Phys.}\ }\textbf {\bibinfo {volume} {6}},\ \bibinfo {pages} {1617}
  (\bibinfo {year} {1965})}\BibitemShut {NoStop}%
\bibitem [{\citenamefont {Yaghjian}(1980)}]{Yaghjian}%
  \BibitemOpen
  \bibfield  {author} {\bibinfo {author} {\bibfnamefont {A.}~\bibnamefont
  {Yaghjian}},\ }\href@noop {} {\bibfield  {journal} {\bibinfo  {journal}
  {Proc. IEEE}\ }\textbf {\bibinfo {volume} {68}},\ \bibinfo {pages} {246}
  (\bibinfo {year} {1980})}\BibitemShut {NoStop}%
\bibitem [{\citenamefont {Albaladejo}\ \emph {et~al.}(2010)\citenamefont
  {Albaladejo}, \citenamefont {G\'{o}mez-Medina}, \citenamefont
  {Froufe-P\'{e}rez}, \citenamefont {Marinchio}, \citenamefont {Carminati},
  \citenamefont {Torrado}, \citenamefont {Armelles}, \citenamefont
  {Garc\'{\i}a-Mart\'{i}n},\ and\ \citenamefont {S\'{a}enz}}]{Albaladejo}%
  \BibitemOpen
  \bibfield  {author} {\bibinfo {author} {\bibfnamefont {S.}~\bibnamefont
  {Albaladejo}}, \bibinfo {author} {\bibfnamefont {R.}~\bibnamefont
  {G\'{o}mez-Medina}}, \bibinfo {author} {\bibfnamefont {L.}~\bibnamefont
  {Froufe-P\'{e}rez}}, \bibinfo {author} {\bibfnamefont {H.}~\bibnamefont
  {Marinchio}}, \bibinfo {author} {\bibfnamefont {R.}~\bibnamefont
  {Carminati}}, \bibinfo {author} {\bibfnamefont {J.}~\bibnamefont {Torrado}},
  \bibinfo {author} {\bibfnamefont {G.}~\bibnamefont {Armelles}}, \bibinfo
  {author} {\bibfnamefont {A.}~\bibnamefont {Garc\'{\i}a-Mart\'{i}n}},\ and\
  \bibinfo {author} {\bibfnamefont {J.}~\bibnamefont {S\'{a}enz}},\ }\href@noop
  {} {\bibfield  {journal} {\bibinfo  {journal} {Opt. Express}\ }\textbf
  {\bibinfo {volume} {18}},\ \bibinfo {pages} {3556} (\bibinfo {year}
  {2010})}\BibitemShut {NoStop}%
\bibitem [{\citenamefont {Edalatpour}\ \emph {et~al.}(2019)\citenamefont
  {Edalatpour}, \citenamefont {Hatamipour},\ and\ \citenamefont
  {Francoeur}}]{Edalatpour1}%
  \BibitemOpen
  \bibfield  {author} {\bibinfo {author} {\bibfnamefont {S.}~\bibnamefont
  {Edalatpour}}, \bibinfo {author} {\bibfnamefont {V.}~\bibnamefont
  {Hatamipour}},\ and\ \bibinfo {author} {\bibfnamefont {M.}~\bibnamefont
  {Francoeur}},\ }\href@noop {} {\bibfield  {journal} {\bibinfo  {journal}
  {Phys. Rev. B}\ }\textbf {\bibinfo {volume} {99}},\ \bibinfo {pages} {165401}
  (\bibinfo {year} {2019})}\BibitemShut {NoStop}%
\bibitem [{\citenamefont {Goedecke}\ and\ \citenamefont
  {O'Brien}(1988)}]{Goedecke}%
  \BibitemOpen
  \bibfield  {author} {\bibinfo {author} {\bibfnamefont {G.}~\bibnamefont
  {Goedecke}}\ and\ \bibinfo {author} {\bibfnamefont {S.}~\bibnamefont
  {O'Brien}},\ }\href@noop {} {\bibfield  {journal} {\bibinfo  {journal} {Appl.
  Opt.}\ }\textbf {\bibinfo {volume} {27}},\ \bibinfo {pages} {2431} (\bibinfo
  {year} {1988})}\BibitemShut {NoStop}%
\bibitem [{\citenamefont {Draine}\ and\ \citenamefont
  {Goodman}(1993)}]{Draine2}%
  \BibitemOpen
  \bibfield  {author} {\bibinfo {author} {\bibfnamefont {B.}~\bibnamefont
  {Draine}}\ and\ \bibinfo {author} {\bibfnamefont {J.}~\bibnamefont
  {Goodman}},\ }\href@noop {} {\bibfield  {journal} {\bibinfo  {journal}
  {Astrophys. J.}\ }\textbf {\bibinfo {volume} {405}},\ \bibinfo {pages} {685}
  (\bibinfo {year} {1993})}\BibitemShut {NoStop}%
\bibitem [{\citenamefont {Carminati}\ \emph {et~al.}(2006)\citenamefont
  {Carminati}, \citenamefont {Greffet}, \citenamefont {Henkel},\ and\
  \citenamefont {Vigoureux}}]{Carminati}%
  \BibitemOpen
  \bibfield  {author} {\bibinfo {author} {\bibfnamefont {R.}~\bibnamefont
  {Carminati}}, \bibinfo {author} {\bibfnamefont {J.-J.}\ \bibnamefont
  {Greffet}}, \bibinfo {author} {\bibfnamefont {C.}~\bibnamefont {Henkel}},\
  and\ \bibinfo {author} {\bibfnamefont {J.}~\bibnamefont {Vigoureux}},\
  }\href@noop {} {\bibfield  {journal} {\bibinfo  {journal} {Opt. Commun.}\
  }\textbf {\bibinfo {volume} {261}},\ \bibinfo {pages} {368} (\bibinfo {year}
  {2006})}\BibitemShut {NoStop}%
\bibitem [{\citenamefont {Bohren}\ and\ \citenamefont
  {Huffman}(1983)}]{bohren}%
  \BibitemOpen
  \bibfield  {author} {\bibinfo {author} {\bibfnamefont {C.~F.}\ \bibnamefont
  {Bohren}}\ and\ \bibinfo {author} {\bibfnamefont {D.~R.}\ \bibnamefont
  {Huffman}},\ }\href@noop {} {\emph {\bibinfo {title} {Absorption and
  Scattering of Light by Small Particles}}}\ (\bibinfo {year}
  {1983})\BibitemShut {NoStop}%
\bibitem [{\citenamefont {Smunev}\ \emph {et~al.}(2015)\citenamefont {Smunev},
  \citenamefont {Chaumet},\ and\ \citenamefont {Yurkin}}]{Yurkin2}%
  \BibitemOpen
  \bibfield  {author} {\bibinfo {author} {\bibfnamefont {D.~A.}\ \bibnamefont
  {Smunev}}, \bibinfo {author} {\bibfnamefont {P.~C.}\ \bibnamefont
  {Chaumet}},\ and\ \bibinfo {author} {\bibfnamefont {M.~A.}\ \bibnamefont
  {Yurkin}},\ }\href@noop {} {\bibfield  {journal} {\bibinfo  {journal} {J.
  Quant. Spectrosc. Radiat. Transf.}\ }\textbf {\bibinfo {volume} {156}},\
  \bibinfo {pages} {67} (\bibinfo {year} {2015})}\BibitemShut {NoStop}%
\bibitem [{\citenamefont {Edalatpour}\ \emph {et~al.}(2015)\citenamefont
  {Edalatpour}, \citenamefont {\u{C}uma}, \citenamefont {Trueax}, \citenamefont
  {Backman},\ and\ \citenamefont {Francoeur}}]{Edalatpour3}%
  \BibitemOpen
  \bibfield  {author} {\bibinfo {author} {\bibfnamefont {S.}~\bibnamefont
  {Edalatpour}}, \bibinfo {author} {\bibfnamefont {M.}~\bibnamefont
  {\u{C}uma}}, \bibinfo {author} {\bibfnamefont {R.}~\bibnamefont {Trueax}},
  \bibinfo {author} {\bibfnamefont {R.}~\bibnamefont {Backman}},\ and\ \bibinfo
  {author} {\bibfnamefont {M.}~\bibnamefont {Francoeur}},\ }\href@noop {}
  {\bibfield  {journal} {\bibinfo  {journal} {Phys. Rev. E}\ }\textbf {\bibinfo
  {volume} {91}},\ \bibinfo {pages} {063307} (\bibinfo {year}
  {2015})}\BibitemShut {NoStop}%
\bibitem [{\citenamefont {Yurkin}\ and\ \citenamefont
  {Hoekstra}(2007)}]{Yurkin1}%
  \BibitemOpen
  \bibfield  {author} {\bibinfo {author} {\bibfnamefont {M.~A.}\ \bibnamefont
  {Yurkin}}\ and\ \bibinfo {author} {\bibfnamefont {A.~G.}\ \bibnamefont
  {Hoekstra}},\ }\href@noop {} {\bibfield  {journal} {\bibinfo  {journal} {J.
  Quant. Spectrosc. Radiat. Transf.}\ }\textbf {\bibinfo {volume} {106}},\
  \bibinfo {pages} {558} (\bibinfo {year} {2007})}\BibitemShut {NoStop}%
\bibitem [{\citenamefont {Moskalensky}\ and\ \citenamefont
  {Yurkin}(2021)}]{Yurkin3}%
  \BibitemOpen
  \bibfield  {author} {\bibinfo {author} {\bibfnamefont {A.~E.}\ \bibnamefont
  {Moskalensky}}\ and\ \bibinfo {author} {\bibfnamefont {M.~A.}\ \bibnamefont
  {Yurkin}},\ }\href@noop {} {\bibfield  {journal} {\bibinfo  {journal} {Rev.
  Phys.}\ }\textbf {\bibinfo {volume} {6}},\ \bibinfo {pages} {100047}
  (\bibinfo {year} {2021})}\BibitemShut {NoStop}%
\bibitem [{\citenamefont {Rytov}\ \emph {et~al.}(1989)\citenamefont {Rytov},
  \citenamefont {Kravtsov},\ and\ \citenamefont {Tatarskii}}]{Rytov}%
  \BibitemOpen
  \bibfield  {author} {\bibinfo {author} {\bibfnamefont {S.}~\bibnamefont
  {Rytov}}, \bibinfo {author} {\bibfnamefont {Y.}~\bibnamefont {Kravtsov}},\
  and\ \bibinfo {author} {\bibfnamefont {V.}~\bibnamefont {Tatarskii}},\
  }\href@noop {} {\emph {\bibinfo {title} {Principles of Statistical
  radiophysics, Vol. 3: Elements of Random Fields}}}\ (\bibinfo  {publisher}
  {Springer, Berlin},\ \bibinfo {year} {1989})\BibitemShut {NoStop}%
\bibitem [{\citenamefont {Polder}\ and\ \citenamefont {Hove}(1971)}]{Polder}%
  \BibitemOpen
  \bibfield  {author} {\bibinfo {author} {\bibfnamefont {D.}~\bibnamefont
  {Polder}}\ and\ \bibinfo {author} {\bibfnamefont {M.~V.}\ \bibnamefont
  {Hove}},\ }\href@noop {} {\bibfield  {journal} {\bibinfo  {journal} {Phys.
  Rev. B}\ }\textbf {\bibinfo {volume} {4}},\ \bibinfo {pages} {3303} (\bibinfo
  {year} {1971})}\BibitemShut {NoStop}%
\bibitem [{\citenamefont {Wang}\ \emph {et~al.}(2018)\citenamefont {Wang},
  \citenamefont {Wang}, \citenamefont {Jakob},\ and\ \citenamefont
  {Xu}}]{Wang}%
  \BibitemOpen
  \bibfield  {author} {\bibinfo {author} {\bibfnamefont {H.}~\bibnamefont
  {Wang}}, \bibinfo {author} {\bibfnamefont {L.}~\bibnamefont {Wang}}, \bibinfo
  {author} {\bibfnamefont {D.}~\bibnamefont {Jakob}},\ and\ \bibinfo {author}
  {\bibfnamefont {X.}~\bibnamefont {Xu}},\ }\href@noop {} {\bibfield  {journal}
  {\bibinfo  {journal} {Nature Commun.}\ }\textbf {\bibinfo {volume} {9}},\
  \bibinfo {pages} {2005} (\bibinfo {year} {2018})}\BibitemShut {NoStop}%
\bibitem [{\citenamefont {Komiyama}(2019)}]{Komiyama}%
  \BibitemOpen
  \bibfield  {author} {\bibinfo {author} {\bibfnamefont {S.}~\bibnamefont
  {Komiyama}},\ }\href@noop {} {\bibfield  {journal} {\bibinfo  {journal} {J.
  Appl. Phys.}\ }\textbf {\bibinfo {volume} {125}},\ \bibinfo {pages} {010901}
  (\bibinfo {year} {2019})}\BibitemShut {NoStop}%
\bibitem [{\citenamefont {Kajihara}\ \emph {et~al.}(2011)\citenamefont
  {Kajihara}, \citenamefont {Kosaka},\ and\ \citenamefont
  {Komiyama}}]{Kajihara2}%
  \BibitemOpen
  \bibfield  {author} {\bibinfo {author} {\bibfnamefont {Y.}~\bibnamefont
  {Kajihara}}, \bibinfo {author} {\bibfnamefont {K.}~\bibnamefont {Kosaka}},\
  and\ \bibinfo {author} {\bibfnamefont {S.}~\bibnamefont {Komiyama}},\
  }\href@noop {} {\bibfield  {journal} {\bibinfo  {journal} {Opt. Express}\
  }\textbf {\bibinfo {volume} {19}},\ \bibinfo {pages} {7695} (\bibinfo {year}
  {2011})}\BibitemShut {NoStop}%
\bibitem [{\citenamefont {Wilde}\ \emph {et~al.}(2006)\citenamefont {Wilde},
  \citenamefont {Formanek}, \citenamefont {Carminati}, \citenamefont {Gralak},
  \citenamefont {Lemoine}, \citenamefont {Joulain}, \citenamefont {Mulet},
  \citenamefont {Chen},\ and\ \citenamefont {Greffet}}]{DeWilde}%
  \BibitemOpen
  \bibfield  {author} {\bibinfo {author} {\bibfnamefont {Y.~D.}\ \bibnamefont
  {Wilde}}, \bibinfo {author} {\bibfnamefont {F.}~\bibnamefont {Formanek}},
  \bibinfo {author} {\bibfnamefont {R.}~\bibnamefont {Carminati}}, \bibinfo
  {author} {\bibfnamefont {B.}~\bibnamefont {Gralak}}, \bibinfo {author}
  {\bibfnamefont {P.-A.}\ \bibnamefont {Lemoine}}, \bibinfo {author}
  {\bibfnamefont {K.}~\bibnamefont {Joulain}}, \bibinfo {author} {\bibfnamefont
  {J.-P.}\ \bibnamefont {Mulet}}, \bibinfo {author} {\bibfnamefont
  {Y.}~\bibnamefont {Chen}},\ and\ \bibinfo {author} {\bibfnamefont {J.-J.}\
  \bibnamefont {Greffet}},\ }\href@noop {} {\bibfield  {journal} {\bibinfo
  {journal} {Nature}\ }\textbf {\bibinfo {volume} {444}},\ \bibinfo {pages}
  {740} (\bibinfo {year} {2006})}\BibitemShut {NoStop}%
\bibitem [{\citenamefont {Babuty}\ \emph {et~al.}(2013)\citenamefont {Babuty},
  \citenamefont {Joulain}, \citenamefont {Chapuis}, \citenamefont {Greffet},\
  and\ \citenamefont {Wilde}}]{Babuty}%
  \BibitemOpen
  \bibfield  {author} {\bibinfo {author} {\bibfnamefont {A.}~\bibnamefont
  {Babuty}}, \bibinfo {author} {\bibfnamefont {K.}~\bibnamefont {Joulain}},
  \bibinfo {author} {\bibfnamefont {P.-O.}\ \bibnamefont {Chapuis}}, \bibinfo
  {author} {\bibfnamefont {J.-J.}\ \bibnamefont {Greffet}},\ and\ \bibinfo
  {author} {\bibfnamefont {Y.~D.}\ \bibnamefont {Wilde}},\ }\href@noop {}
  {\bibfield  {journal} {\bibinfo  {journal} {Phys. Rev. Lett.}\ }\textbf
  {\bibinfo {volume} {110}},\ \bibinfo {pages} {146103} (\bibinfo {year}
  {2013})}\BibitemShut {NoStop}%
\bibitem [{\citenamefont {Jones}\ and\ \citenamefont {Raschke}(2012)}]{Jones}%
  \BibitemOpen
  \bibfield  {author} {\bibinfo {author} {\bibfnamefont {A.}~\bibnamefont
  {Jones}}\ and\ \bibinfo {author} {\bibfnamefont {M.}~\bibnamefont
  {Raschke}},\ }\href@noop {} {\bibfield  {journal} {\bibinfo  {journal} {Nano
  Lett.}\ }\textbf {\bibinfo {volume} {12}},\ \bibinfo {pages} {1475} (\bibinfo
  {year} {2012})}\BibitemShut {NoStop}%
\bibitem [{\citenamefont {O'Callahan}\ \emph {et~al.}(2014)\citenamefont
  {O'Callahan}, \citenamefont {Lewis}, \citenamefont {Jones},\ and\
  \citenamefont {Raschke}}]{O'Callahan}%
  \BibitemOpen
  \bibfield  {author} {\bibinfo {author} {\bibfnamefont {B.}~\bibnamefont
  {O'Callahan}}, \bibinfo {author} {\bibfnamefont {W.}~\bibnamefont {Lewis}},
  \bibinfo {author} {\bibfnamefont {A.}~\bibnamefont {Jones}},\ and\ \bibinfo
  {author} {\bibfnamefont {M.}~\bibnamefont {Raschke}},\ }\href@noop {}
  {\bibfield  {journal} {\bibinfo  {journal} {Phys. Rev. B}\ }\textbf {\bibinfo
  {volume} {89}},\ \bibinfo {pages} {245446} (\bibinfo {year}
  {2014})}\BibitemShut {NoStop}%
\bibitem [{\citenamefont {Herz}\ and\ \citenamefont {Biehs}(2019)}]{Herz2}%
  \BibitemOpen
  \bibfield  {author} {\bibinfo {author} {\bibfnamefont {F.}~\bibnamefont
  {Herz}}\ and\ \bibinfo {author} {\bibfnamefont {S.-A.}\ \bibnamefont
  {Biehs}},\ }\href@noop {} {\bibfield  {journal} {\bibinfo  {journal}
  {Europhys. Lett.}\ }\textbf {\bibinfo {volume} {127}},\ \bibinfo {pages}
  {44001} (\bibinfo {year} {2019})}\BibitemShut {NoStop}%
\bibitem [{\citenamefont {Ott}\ \emph {et~al.}(2021{\natexlab{b}})\citenamefont
  {Ott}, \citenamefont {An}, \citenamefont {Kittel},\ and\ \citenamefont
  {Biehs}}]{ottthgg}%
  \BibitemOpen
  \bibfield  {author} {\bibinfo {author} {\bibfnamefont {A.}~\bibnamefont
  {Ott}}, \bibinfo {author} {\bibfnamefont {Z.}~\bibnamefont {An}}, \bibinfo
  {author} {\bibfnamefont {A.}~\bibnamefont {Kittel}},\ and\ \bibinfo {author}
  {\bibfnamefont {S.-A.}\ \bibnamefont {Biehs}},\ }\href@noop {} {\bibfield
  {journal} {\bibinfo  {journal} {Phys. Rev. B}\ }\textbf {\bibinfo {volume}
  {104}},\ \bibinfo {pages} {165407} (\bibinfo {year}
  {2021}{\natexlab{b}})}\BibitemShut {NoStop}%
\bibitem [{\citenamefont {Eckhardt}(1984)}]{Eckhardt}%
  \BibitemOpen
  \bibfield  {author} {\bibinfo {author} {\bibfnamefont {W.}~\bibnamefont
  {Eckhardt}},\ }\href@noop {} {\bibfield  {journal} {\bibinfo  {journal}
  {Phys. Rev. A}\ }\textbf {\bibinfo {volume} {29}},\ \bibinfo {pages} {1991}
  (\bibinfo {year} {1984})}\BibitemShut {NoStop}%
\bibitem [{\citenamefont {Herz}\ \emph {et~al.}(2020)\citenamefont {Herz},
  \citenamefont {Kathmann},\ and\ \citenamefont {Biehs}}]{Herz4}%
  \BibitemOpen
  \bibfield  {author} {\bibinfo {author} {\bibfnamefont {F.}~\bibnamefont
  {Herz}}, \bibinfo {author} {\bibfnamefont {C.}~\bibnamefont {Kathmann}},\
  and\ \bibinfo {author} {\bibfnamefont {S.-A.}\ \bibnamefont {Biehs}},\
  }\href@noop {} {\bibfield  {journal} {\bibinfo  {journal} {Europhys. Lett.}\
  }\textbf {\bibinfo {volume} {130}},\ \bibinfo {pages} {44003} (\bibinfo
  {year} {2020})}\BibitemShut {NoStop}%
\bibitem [{\citenamefont {Kattawar}\ and\ \citenamefont
  {Eisner}(1970)}]{Eisner}%
  \BibitemOpen
  \bibfield  {author} {\bibinfo {author} {\bibfnamefont {G.}~\bibnamefont
  {Kattawar}}\ and\ \bibinfo {author} {\bibfnamefont {M.}~\bibnamefont
  {Eisner}},\ }\href@noop {} {\bibfield  {journal} {\bibinfo  {journal} {Appl.
  Opt.}\ }\textbf {\bibinfo {volume} {9}},\ \bibinfo {pages} {2685} (\bibinfo
  {year} {1970})}\BibitemShut {NoStop}%
\bibitem [{\citenamefont {Webber}(2013)}]{Webber}%
  \BibitemOpen
  \bibfield  {author} {\bibinfo {author} {\bibfnamefont {J.}~\bibnamefont
  {Webber}},\ }\href@noop {} {\bibfield  {journal} {\bibinfo  {journal} {Meas.
  Sci. Technol.}\ }\textbf {\bibinfo {volume} {24}},\ \bibinfo {pages} {027001}
  (\bibinfo {year} {2013})}\BibitemShut {NoStop}%
\bibitem [{\citenamefont {Weng}\ \emph {et~al.}(2018)\citenamefont {Weng},
  \citenamefont {Komiyama}, \citenamefont {An}, \citenamefont {Yang},
  \citenamefont {Chen}, \citenamefont {Biehs}, \citenamefont {Kajihara},\ and\
  \citenamefont {Lu}}]{Weng}%
  \BibitemOpen
  \bibfield  {author} {\bibinfo {author} {\bibfnamefont {Q.}~\bibnamefont
  {Weng}}, \bibinfo {author} {\bibfnamefont {S.}~\bibnamefont {Komiyama}},
  \bibinfo {author} {\bibfnamefont {Z.}~\bibnamefont {An}}, \bibinfo {author}
  {\bibfnamefont {L.}~\bibnamefont {Yang}}, \bibinfo {author} {\bibfnamefont
  {P.}~\bibnamefont {Chen}}, \bibinfo {author} {\bibfnamefont {S.-A.}\
  \bibnamefont {Biehs}}, \bibinfo {author} {\bibfnamefont {Y.}~\bibnamefont
  {Kajihara}},\ and\ \bibinfo {author} {\bibfnamefont {W.}~\bibnamefont {Lu}},\
  }\href@noop {} {\bibfield  {journal} {\bibinfo  {journal} {Science}\ }\textbf
  {\bibinfo {volume} {360}},\ \bibinfo {pages} {775} (\bibinfo {year}
  {2018})}\BibitemShut {NoStop}%
\end{thebibliography}%


%

\end{document}